# Cyanothiazole Copper(I) Complexes: Uncharted Materials with Exceptional Optical and Conductive Properties

Karolina Gutmańska[a], Agnieszka Podborska[b], Tomasz Mazur[b], Andrzej Sławek[b], Ramesh Sivasamy[b], Alexey Maximenko[c], Łukasz Orzeł[d], Janusz Oszajca[d], Grażyna Stochel[d], Konrad Szaciłowski[b,e]* Anna Dołęga[a]*

[a] *Gdansk University of Technology, Chemical Faculty, Department of Inorganic Chemistry, Narutowicza 11/12, 80-233 Gdańsk, Poland*

[b] *AGH University of Krakow, Academic Centre of Materials and Technology, al. Mickiewicza 30, 30-059 Kraków, Poland*

[c] *National Synchrotron Radiation Centre SOLARIS, Jagiellonian University, ul. Czerwone Maki 98, Kraków 30-392, Poland*

[d] *Jagiellonian University in Krakow, Faculty of Chemistry, Gronostajowa 2, Kraków, 30-387 Krakow Poland*

[e] *University of the West of England, Unconventional Computing Lab, Bristol BS16 1QY, United Kingdom*

**Abstract**

Cyanothiazoles, small and quite overlooked molecules, possess remarkable optical properties that can be fine-tuned through coordination with transition metals. In this study, we investigate a promising application of cyanothiazoles, where their combination with copper(I) iodide forms a new class of complexes exhibiting outstanding optical properties. X-ray crystallography of copper(I) iodide complexes with isomeric cyanothiazoles revealed key structural features, such as π-π stacking, hydrogen bonding, and rare halogen···chalcogen I···S interactions, enhancing stability and reactivity. Advanced spectroscopy and computational modeling allowed precise identification of spectral signatures in FTIR, NMR, and UV-Vis spectra. Fluorescence studies, along with XANES synchrotron analyses, highlighted their unique thermal and electronic properties, providing a solid foundation for further research in the field.

## Introduction

Thiazole-based compounds (**tzs**), characterized by the presence of an electron-donating sulfur (-S-) and an electron-accepting nitrogen (-C=N-) within their five-membered aromatic ring, exhibit a unique combination of donor and acceptor properties.[1] These compounds may find broad applications across various fields, including organic chemistry,[2] medicine,[3] and advanced material technologies.[4] Their electron-accepting ability is particularly valued, primarily due to the electron-withdrawing nitrogen atom in the imine group (C=N) within the ring structure.[4b] This feature significantly influences the interaction between the highest occupied molecular orbital (HOMO) and the lowest unoccupied molecular orbital (LUMO), leading to a reduction in the bandgap energy.[5] TZs are most commonly small molecules or conjugated polymers, which have been employed as n-type or p-type semiconductors in organic field-effect transistors (OFETs).[6] They have demonstrated impressive electron and hole mobilities.[2, 7] They have been utilized as electron donors in solution-processed organic photovoltaic devices (OPVs) and as electroluminescent or electron-transporting materials in organic light-emitting diodes (OLEDs).[8] In sensor applications, thiazole compounds stand out for their ability to detect metal ions and other analytes at trace levels. This is achieved through interactions between the nitrogen atom or a substituent group and heavy metals, which induce changes in optical properties, such as color shifts or fluorescence alterations.[9]

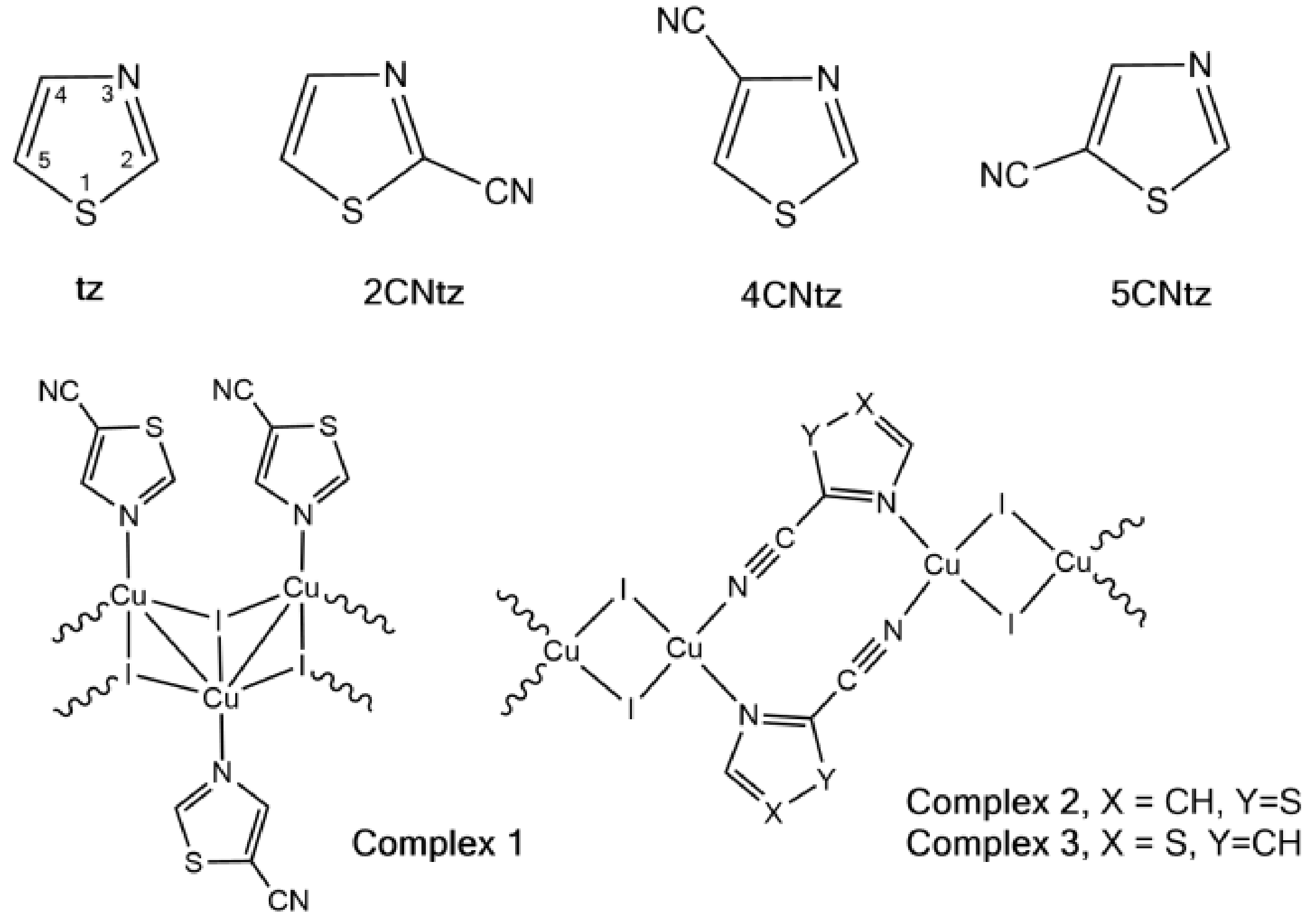

**Scheme I** Thiazole (tz), cyanothiazoles (**CNtz**) and their CuI complexes **1** – **3** studied within this work.

The presence of -C≡N group in the studied monocyanothiazoles and their complexes (**Scheme I**) has a significant impact on the chemical and physical properties of the thiazole ring, making these compounds particularly attractive both from a theoretical research perspective and for potential applications. The nitrile group, being a strong deactivator of the aromatic system, has the ability to withdraw electrons from the thiazole ring. Through inductive and mesomeric effects, it reduces the electron density on the heteroatoms of the ring, such as sulfur and nitrogen. As a result, these heteroatoms become weaker electron donors, limiting their ability to bind metals inside the thiazole ring.[10] The decrease in electron donation from the ring may simultaneously increase the preference of metals like $Ag^+$ or $Cu^+$ to coordinate with the nitrogen atom of the nitrile group, which we observed within the studied complexes (**Scheme I**). Exploiting these simple relationships, we aimed to design a novel, previously unexplored class of hybrid materials based on copper iodide and cyanothiazole isomers with unusual optical properties.

## Results and discussion

### Syntheses and molecular structures of CuI complexes with cyanothiazoles (CNtz)

Due to the remarkable ability of CuI to form various polymorphic modifications, which may be thermodynamically unstable,[11] the synthesis of well-defined complexes of cyanothiazole with copper(I) iodide presents a certain research challenge. Fortunately, in the reaction between CuI and **5CNtz**, only one stable polymeric complex, **1**, was successfully obtained. For **2CNtz**, regardless of the molar ratio of reagents or reaction parameters, an intermediate compound is initially produced, which quickly transforms into a stable form referred to as complex **2**. It is suggested that this intermediate is a so-called "disappearing or vanishing polymorph", characterized by an extremely short lifetime, rendering detailed experimental investigations virtually impossible.[12] In the case of complex **3**, depending on the synthesis method employed, two different forms can be obtained, which can transform into one another. The discussed synthetic pathways are illustrated in **Scheme II**. The processes highlight the complexity of systems based on copper(I) iodide, while also underscoring their research potential. The further studies focused exclusively on stable polymorphic forms, which may find application in future technological solutions.

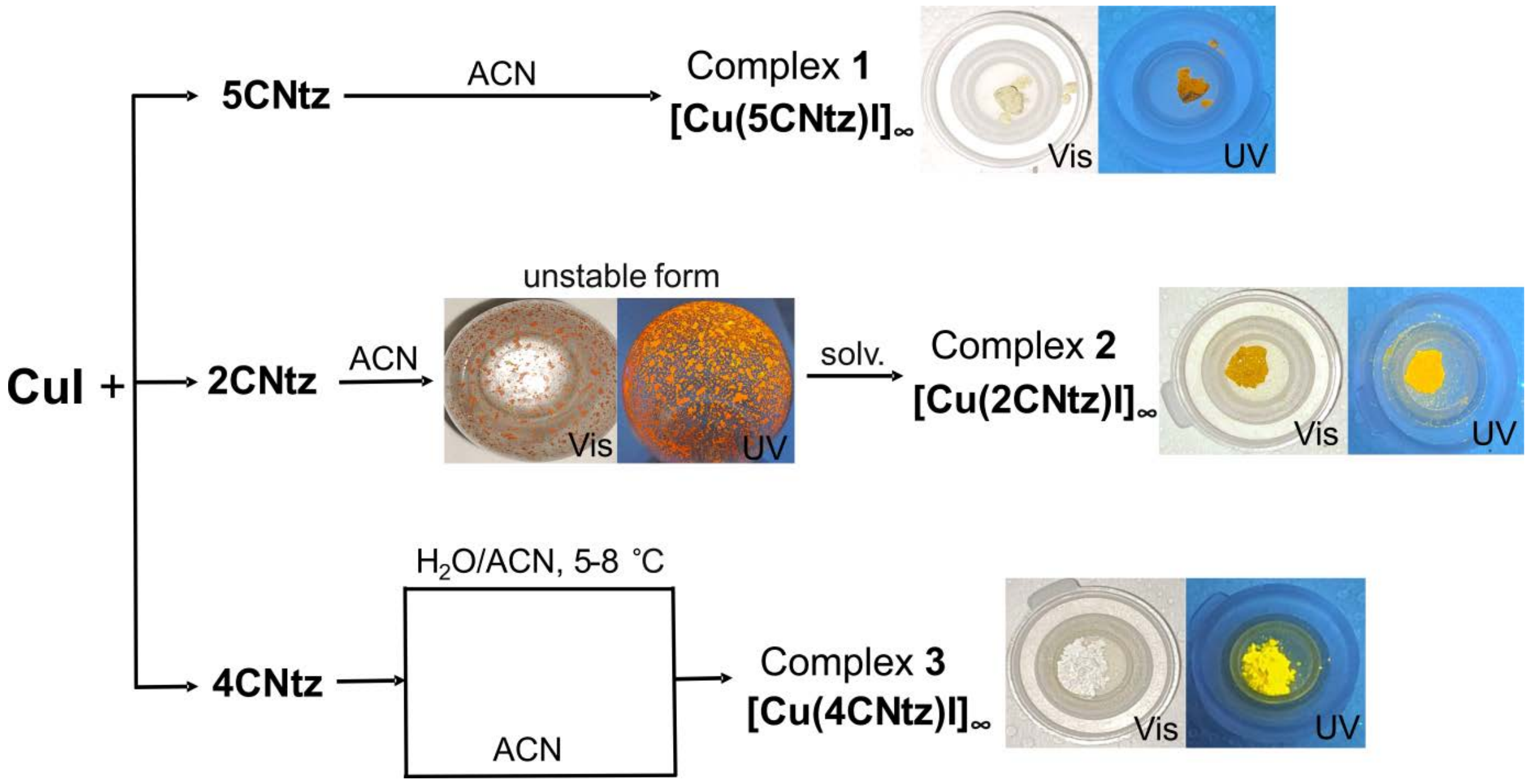

**Scheme II** Reactions of CuI with CNtz isomers. The formation of unstable form of **2** (polymorph?) is indicated. The details are described in the Experimental Section.

The crystal and refinement data for **1** - **3** are collected in Table S1. The details of the molecular geometry can be found in Tables S2 – S5. The symmetrically independent unit of each of the complexes **1** - **3** contains CuI fragment and one molecule of cyanothiazole. The basic synthons form 1D coordination polymers, which crystallize in the triclinic system, space group *P*-1. Copper(I) cations in each compound have a coordination number equal to 4 (CN=4) and exhibit a distorted tetrahedral geometry. This is due to the formation of $Cu_2I_2$ rings by the bridging iodide anions, which distort the local geometry around the metal atoms. The calculated geometrical parameters $\tau_4/\tau_4'$ confirm a slightly distorted tetrahedral geometry, with values of -0.89/0.89 for **1**, 0.90/0.98 for **2**, and -0.94/0.90 for **3**, respectively.[13] Another common feature of the complexes **1** - **3** is the center of symmetry, which is located in the middle of the resulting $Cu_2I_2$ ring. Among the three compounds, however, there are two distinct types of polymeric systems. Compound **1** features $Cu_2I_2$ rings sharing a common edge, forming a zigzag $(Cu_2I_2)_\infty$ ladder decorated with monodentate 5CNtz ligands linked to copper(I) *via* the nitrogen atom (**Figure 1a**). In contrast, compounds **2** and **3** consist of discrete $Cu_2I_2$ rings connected into one-dimensional polymers through two bridging 2CNtz or 4CNtz ligands, respectively, as illustrated by complex **2** in **Figure 1b**. Both **2** and **3** crystallize with the similar parameters of unit cell in the same crystal system (Table S1).

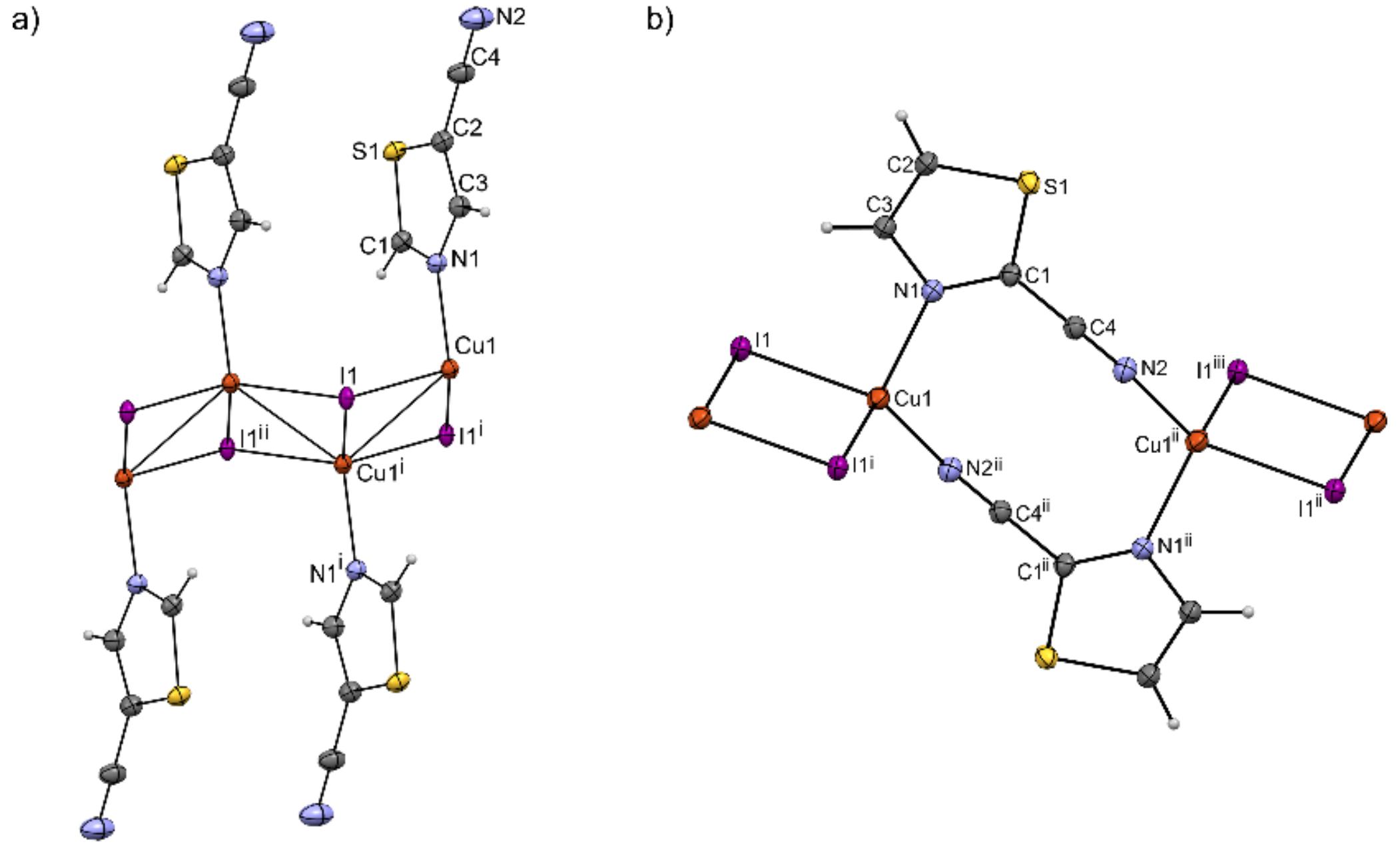

**Figure 1** Molecular structures of CuI complexes with cyanothiazole isomers **5CNtz** and **2CNtz**. The fragments of the polymeric ribbons are depicted: a) complex **1**, symmetry operations: $^{i}$: 1-x, 1-y, -z; $^{ii}$:-x, 1-y, -z; b) complex **2**; symmetry operations: $^{i}$:2-x, 2-y, 2-z; $^{ii}$:2-x, 1-y, 1-z; $^{iii}$: x, -1+y, -1+z. Thermal ellipsoids at 50% probability level.

The most distinct compound among **1** – **3** is compound **1**, in which coordination of copper cations by the nitrile group is not observed. This compound forms a typical copper(I) iodide chain in the "staircase" or “ribbon” form, consisting of copper and iodine atoms with the general formula $[Cu_xI_x]_n$.[14] Within the chain, cuprophilic interactions are observed, with a distance of approximately 2.75 Å. These Cu—Cu interactions cause the chain to deform, adopting a staircase shape. The 5TzCN molecules coordinate to copper atoms through the nitrogen atom of the thiazole ring, positioning themselves perpendicularly to the $[Cu_xI_x]_n$ chain at a distance of Cu1—N1 2.050(4) Å (**Figure 1a**). The interchain contacts in compound **1** are illustrated in **Figure 2**. The stabilization of the crystal structure is based on the several types of weak interactions. We indicate the presence of weak hydrogen bonds; the contacts that do not exceed the sum of van der Waals radii are C3–H3···$N2_{(1-x,1-y,1-z)}$ with the length of 3.413(7) Å, and C1–H1···$I1_{(1+x,\ -1+y,z)}$ with the length of 3.749(4) Å (see also Table 5S). Up to 0.2 Å exceeding the sum of the van der Waals radii, one can also find chalcogen···halogen contacts between sulfur and iodine atoms: S1···$I1_{(1+x,-1+y,z)}$ of 3.968(1) Å and S1···$I1_{(x,-1+y,z)}$ of 3.979(1) Å as well as chalcogen···pnictogen proximity N2···$S1_{(2-x,-y,1-z)}$ of 3.541(6) Å. Interestingly, no interchain π···π stacking interactions were observed in this system, which typically dominate the stabilization of thiazole-based molecules.[4b] The bond lengths, angles and short intermolecular contacts for complexes **1** - **3** are collected in Tables S2 – S5.

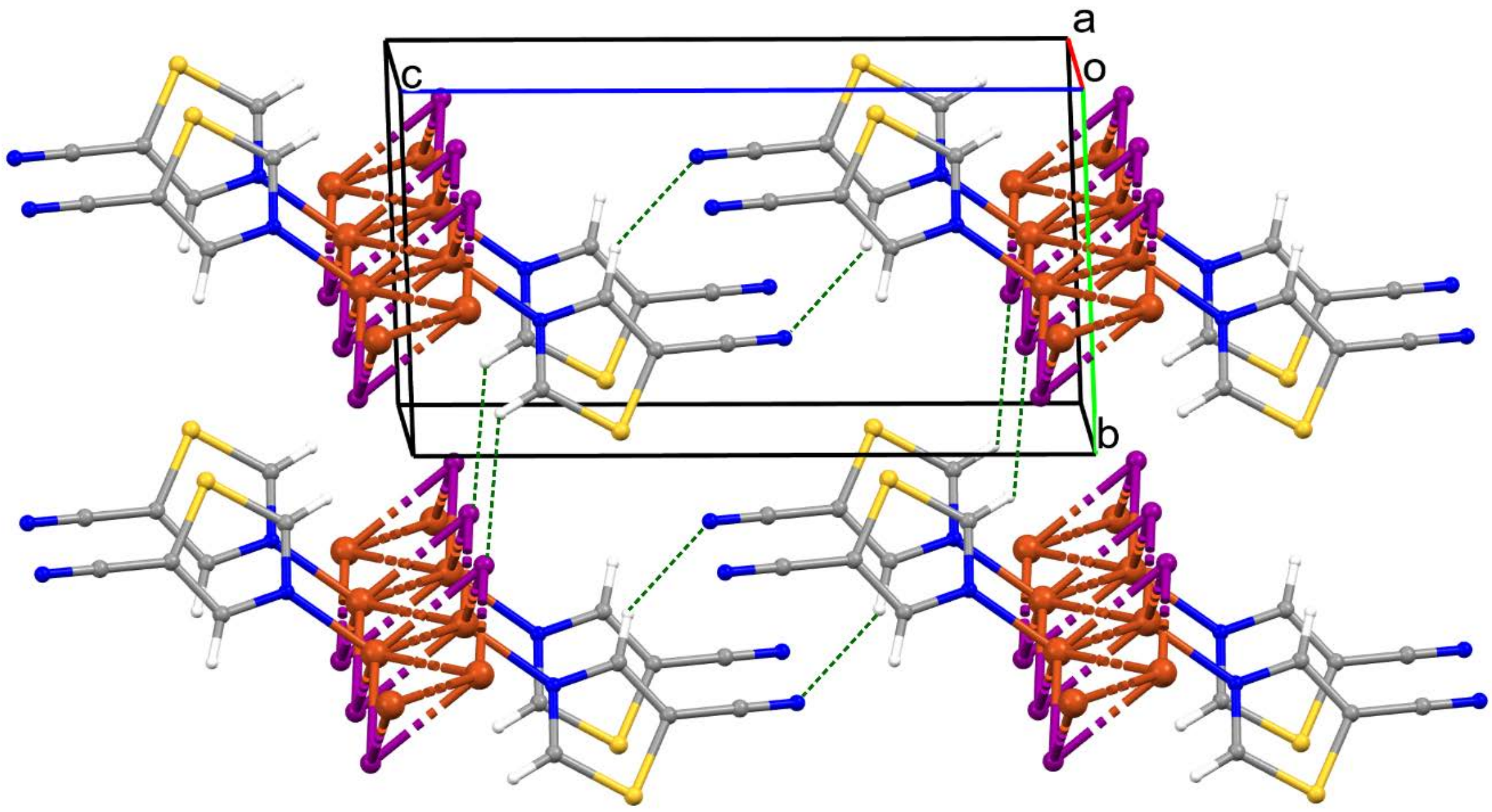

**Figure 2** Crystal packing of compound **1** viewed along axis *a*; C1–H1···I1 and C3–H3···N1 interactions are indicated as green dashed lines.

In compounds **2** and **3,** a characteristic element is the presence of the dimeric unit $Cu_2I_2$, which is present in about 57% of all described and synthesized copper(I) complexes[14a]. In both compounds the Cu1 atom is coordinated by an N1 atom derived from the thiazole ring, an N2 atom derived from the nitrile group and two iodine ions (**Figure 1b**). The second characteristic feature is the formation of a deformed ring including Cu atoms and CNtz isomers which is practically identical for **2** and **3** (see **Figure S1** for the overlay of the molecular structures of **2** and **3**). The Cu···Cu distances within these two rings: $Cu_2I_2/Cu_2(CNTz)_2$ are 3.1733(5)/5.3415(7) Å in compound **2** and 3.083(3)/5.356(3) Å in compound **3**. Due to the relatively high electron density at the iodine atoms, this ring is aligned perpendicularly to the $Cu_2I_2$ cluster. In addition, in both compound **2** and **3** the Cu1–$N2_{(C≡N)}$ bond with the nitrile group (1.943(2)Å in **2**/1.949 Å in **3**) is shorter than for Cu–$N1_{(Tz}$ $_{ring)}$ (2.064(2) Å in **2**/2.072(8) Å in **3**), which suggests large delocalization of the charge in cyanothiazoles and their complexes with copper(I). The phenomenon will be further discussed in the paragraph devoted to DFT calculations.

The chains of compounds **2** and **3** adopt similar mutual orientations with the set of analogous intermolecular interactions illustrated in **Figure 3** for complex **2**. Important stabilizing interactions between the chains in **2** and **3** are face-to-face π···π stacking interactions between the thiazole rings. The $Cg_{(Tz\ ring)}\cdots Cg_{(Tz\ ring)}$ distances are 3.783 Å for compound **2** and 3.882 Å for **3**, where Cg is the centroid of thiazole ring (Table S4). In **Figure 3** we indicate two types of C–H···I in compound **2**: C2–H2···$I1_{(-1+x,y,z)}$ of 3.776(2) Å and C3–H3···$I1_{(-1+x,y,z)}$ of 3.793(2) Å. In compound **3** the hydrogen bond lengths are C1–H1···$I1_{(1-x,1-y,2-z)}$ of 3.71(2) Å and C3–H3···$I1_{(1-x,1-y,1-z)}$ of 3.78(2) Å (Table S5).

Similar to **1** we have found relatively short chalcogen···halogen contacts between sulfur and iodine atoms in **2** and **3**. Halogen–chalcogen I···S interactions are rather uncommon and are not widely known and described in the literature.[15] These I···S distances that do not exceed the sum of van der Waals radii are: S1···$I1_{(-1+x,-1+y,-1+z)}$ 3.7143(7) Å for compound **2**, and S1···$I1_{(-1+x,-1+y,-1+z)}$/S1···$I1_{(1-x,2-y,1-z)}$ 3.632(2) Å/3.665(4) Å for compound **3**.

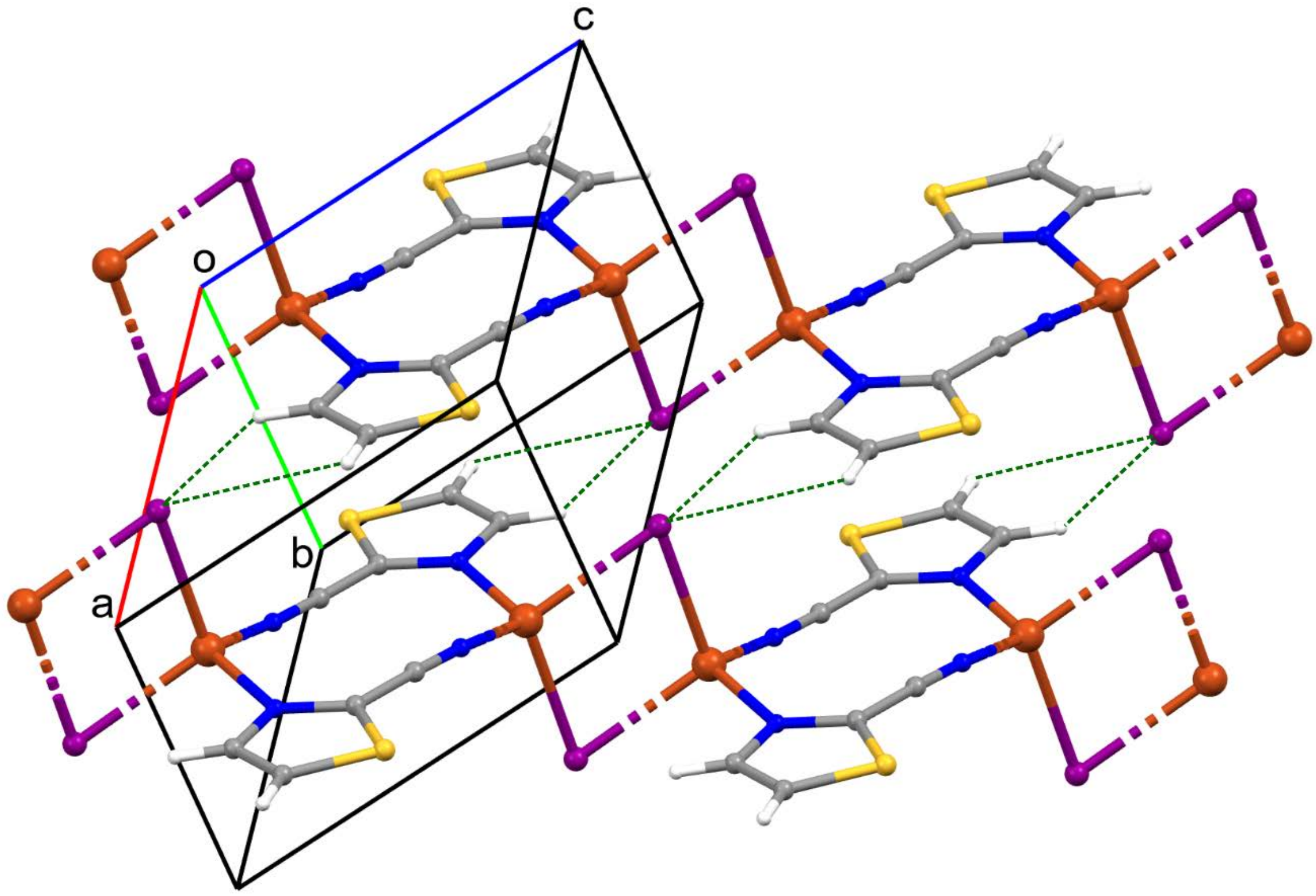

**Figure 3** Crystal packing of compound **2** indicating C–H···I interactions as green dashed lines.

In order to confirm and clarify the nature of the bonds and interactions in the obtained coordination compounds, a Hirshfeld surface (HS) analysis was performed, and fingerprint plots were generated.[16] These plots provide a two-dimensional mapping of the point distribution on the surface as a function of the distance parameters $d_i$ and $d_e$. All fingerprint plots for the presented compounds, which illustrate all interactions within the polymer, are characterized by a spread of points covering the ranges of $d_i$ and $d_e$ from 0.9 to 2.8 Å, respectively. The results of the Hirshfeld analysis are presented in **Figures 4** and **5**. Hirshfeld surfaces calculated for selected fragments of the polymeric chains are presented in Figure S2 while Table S6 collects the detailed fingerplots for the specific contacts on these HS.

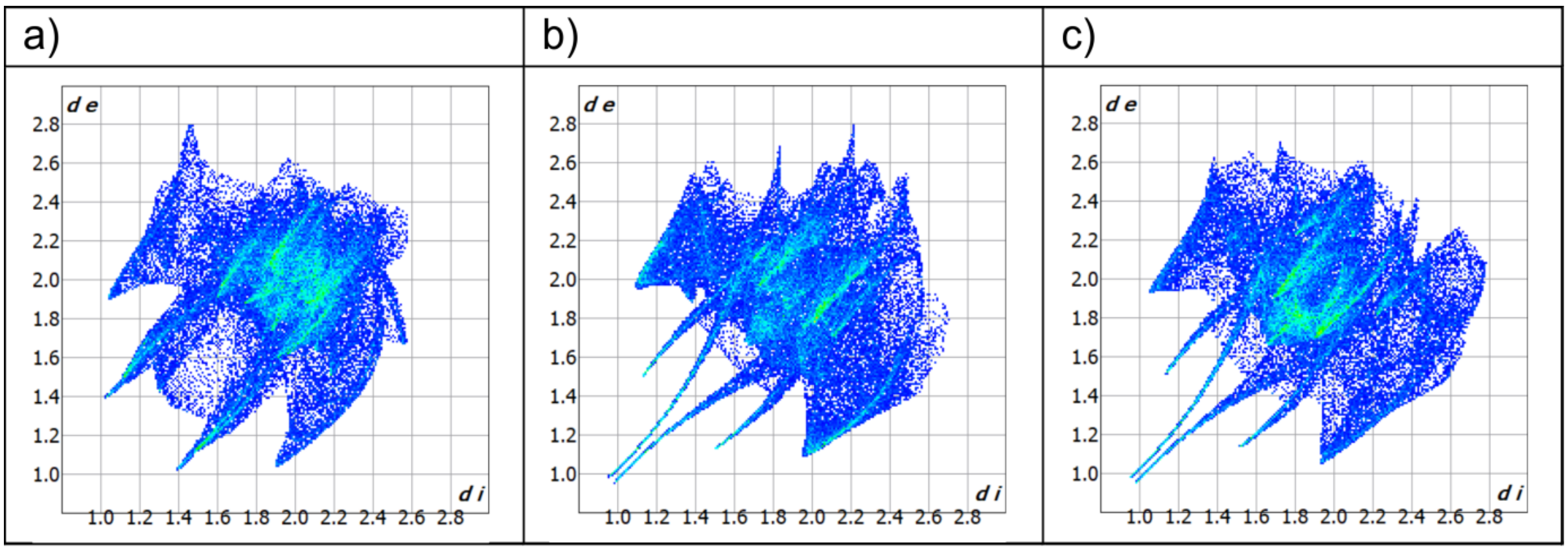

**Figure 4.** The all-interactions fingerprint plot for compound: a) **1**, b) **2** and c) **3**. Analysis performed with the use of CrystalExplorer.[16]

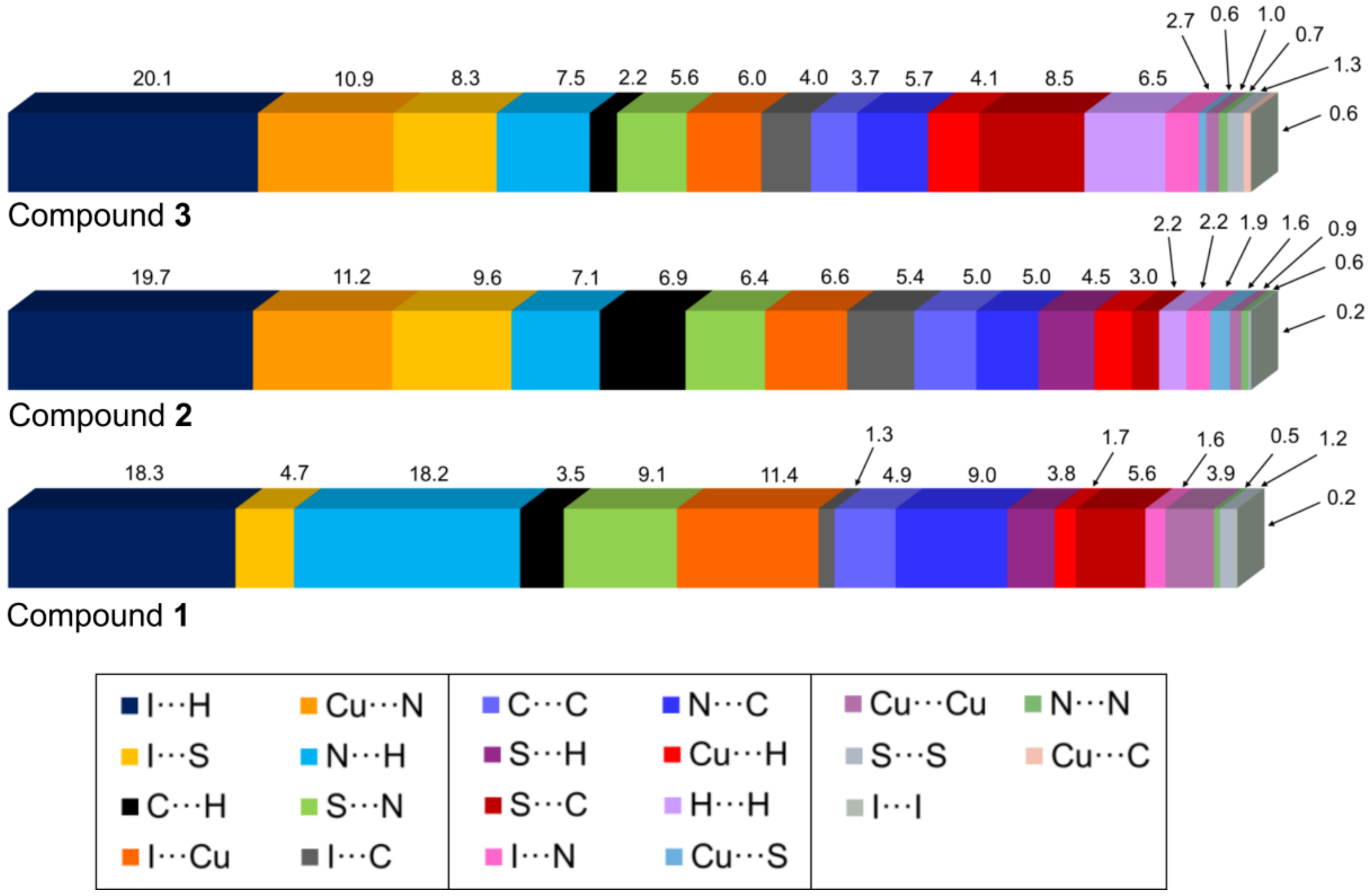

**Figure 5.** The contributions of various types of contacts on the Hirshfeld surfaces of the complexes **1** - **3**. Analysis performed with the use of CrystalExplorer.[16]

The dominant interactions with the greatest impact on structural stabilization are C-H···I which constitute approximately 20% of all interactions (percentage of the HS). The fingerprint plots exhibit a subtle shape that is shifted towards longer distances, with the $d_i$ and $d_e$ values averaging between 1.8 and 2.8 Å. This suggests that long-range interactions play a crucial role in the studied compounds. For compound **1**, C–H···N interactions are also significant, occurring at a comparable proportion of 18.2%. These interactions are characterized by sharp, distinct shapes, contrasting with the more diffuse and subtle shapes observed in compounds **2** and **3**. The previously discussed atypical I···S interactions are marked by a sharp, narrow, and slender peak in the upper region of the fingerprint plot. These interactions are most prevalent in compounds **2** (9.6%) and **3** (8.3%), while they are less pronounced in compound **1** (4.7%). Additionally, C–H···π (C···H contacts) and π···π (C···C, N···C, S···C contacts) interactions significantly contribute to the stabilization of the structures, collectively accounting for approximately 20% of the total interactions in each compound. Here some differences between **2** and **3** are encountered (**Figure 5** and Table S6).

**FTIR ATR spectroscopy**

The analysis of infrared (FTIR) spectra of the obtained compounds enabled the identification of characteristic band shifts, allowing for a detailed examination of copper(I) coordination polymers with cyanothiazoles. Due to their structural similarity, the three polymers will be discussed collectively, while the FTIR spectra are presented in Figures S3 – S20. In all FTIR spectra, a band shift of approximately +20 $cm^{-1}$ relative to the substrate was observed, attributed to symmetric and asymmetric stretching vibrations of C–H bonds within the thiazole ring.[17] Noticeable differences in band positions arise from the simplicity of the organic molecule, where even minor modifications within the thiazole ring result in shifts or splitting of these bands in the spectra (Figures S4,S5,S10,S11,S16,S17). For the nitrile group, the observed shift is minimal and insufficient to unequivocally confirm copper(I) coordination *via* the -C≡N group. In contrast, in complexes with $Ag^+$ ions, shifts in the nitrile group bands are significantly more pronounced.[18] The greater shifts in the FTIR bands for the nitrile group (-C≡N) coordinating to $Ag^+$ ions compared to $Cu^+$ ions arise from fundamental differences in their chemical and physical properties. $Ag^+$ ions are characterized by a larger ionic radius and higher polarizability, resulting in more diffuse and pronounced interactions with the ligand, which significantly alter the electronic distribution within the nitrile group. In the case of $Cu^+$, the smaller ionic radius and a stronger covalent component in the bond with the nitrile group limit its influence on the ligand's electronic structure. Additionally, $Cu^+$ exhibits the ability for back-donation—partial electron transfer from the nitrile group to the metal's d orbitals—which stabilizes the complex and reduces the electron deficiency in the nitrile group. This results in less significant band shifts compared to $Ag^+$, for which back-donation effects are negligible.[19] Further analysis of FTIR spectra of Cu(I) compounds with nitriles revealed that in complexes **2** and **3** the intensity of nitrile stretching mode is suppressed by coordination to the $Cu^+$ cations. This phenomenon can be attributed to the changes in the charge distribution on the ligand due to coordination to the metal ion.

Differences between the compounds become more apparent in the region around 1500 $cm^{-1}$. For compound **1**, a shift from 1493 $cm^{-1}$ to 1505 $cm^{-1}$ of the band attributed to stretching vibrations of the C=N group within the thiazole ring was observed.[20] In compounds **2** and **3**, such a shift does not occur, which can be explained by differences in molecular structure between the complexes. For compounds **2** and **3**, only minor shifts in bands corresponding to C=C stretching vibrations are observed in the range of 1464–1473 $cm^{-1}$, characteristic of skeletal vibrations of the thiazole ring.[21] Aromatic C–N bond vibrations appear in the range of 1300–1350 $cm^{-1}$.[22] In compound **1**, the shifts are minimal, whereas for compounds **2** and **3**, they are more pronounced, with values of 1317 to 1309 and $cm^{-1}$.

In spectroscopic studies, bands attributed to C–H bending vibrations were observed in the range of 1130–1113 $cm^{-1}$.[23] The C–S stretching vibrations typically appear in lower frequency regions, below 700 $cm^{-1}$.[22a, 24] , and for the analyzed compounds, these bands occur in the range of 536–507 $cm^{-1}$.

The analysis of bands in the fingerprint region allows for the identification of common features among the studied compounds. This is evident in the shifts of bands characteristic of the deformational vibrations of the thiazole ring and the N=C–H group in the 900–700 $cm^{-1}$ region. Significant shifts in these bands are a direct effect of copper coordination to the aromatic nitrogen atom.

**Optical and electronic properties– experimental and theoretical studies**

Diffuse reflectance spectra of studied complexes dispersed in barium sulfate matrix has been analysed according to Tauc approach. Diffuse reflectance spectra were converted to the Kubelka-Munk function, defined as follows (Equation 1):

$$F(R) = \frac{(1-R)^2}{2R} \qquad \text{(Equation 1)}$$

where *R* is the reflectance. For powder samples dispersed in scattering media it is commonly assumed that F(*R*) is proportional to the absorption coefficient α.[25] Then the Tauc function has been applied to fit the linear fragment of the spectrum (Equation 2):[26]

$$(\alpha \cdot hv)^{\frac{1}{r}} = A(hv - E_g) \qquad \text{(Equation 2)}$$

where *A* is the proportionality constant independent of the photon energy, *h* is the Planck constant, *v* is the photon frequency, $E_g$ is the band gap, *α* is the absorption coefficient, and *r* is the exponent describing the nature of the band gap: *r* = ½ for direct and *r* = 2 for indirect transitions, respectively.[26] According to DFT models the band gap of Complex **2** is direct and the band gaps of complexes **1** (5CNtz-CuI) and **3** (4-CNtz-CuI) are indirect, however the energy dispersion of band edges is very low, indicating low mobility resulting from insufficient electron delocalization in the solid. Therefore, as in the case of amorphous (or almost amorphous) materials or molecular crystals is can be safely assumed that *r* = 1.[27]

The same approximation is commonly used for molecular crystalline materials as well as for ionic crystal with only a week covalent interaction between ionic species.[28] Despite the fact that DFT models predict direct and indirect band gaps, the *r* = 1 case provides the best fit for the spectra, which is consistent with relatively weak intermolecular interactions in covalent crystals. Therefore, the final equation that allows the determination of the band gap is derived as follows (Equation 3):

$$F(R) \cdot hv = A(hv - E_g) \qquad \text{(Equation 3)}$$

Spectra, along with corresponding linear fits are shown in **Figure 6**. Thus evaluated band gap energies amount to 2.49, 3.06 and 2.71 eV for 2-, 4-, and 5-cyanothiazole complexes **2**, **3** and **1**, respectively. These data consistently indicate decrease of band gap energy as compared to parent CuI semiconductor. It is justified by changes in the electronic composition bands, mainly the conduction band. Electron acceptor character of cyanothiazole ligand makes the energy of the conduction band significantly lower than in the parent material. Moreover, both 2- and 5-cyanothiazole complexes show distinct absorption maxima at 3.10 and 3.25 eV, respectively, whereas 4-cyanothiazole complex shows well-pronounced Urbach tail extending down to 2 eV. This spectral feature is responsi ble for a distinct yellow colour of the complex despite relatively high band gap energy. The absorption maxima have Gaussian envelopes - one dominating Gaussian band and a low energy sub-band), consistent with partially localized charge transfer character of the transition (Figure S21).[29] Band gap energy for 4-cyanothiazole derivative **3** is the highest among studied complexes. Whereas complex 5-cyanothiazole cannot be directly compared due to different coordination mode, the comparison with 2-cyanothiazole shows the difference of over 0.5 eV. This huge effect can be attributed to differences in ground state charge distribution in the free ligand. It seems that nitrile group in 4-cyanothiazole is 0.13 *e* more negative than in 2-cyanothiazole (*cf.* **Table 1**). This may be a reason for much less efficient back donation, which results in higher energy of antibonding orbitals in the complex.

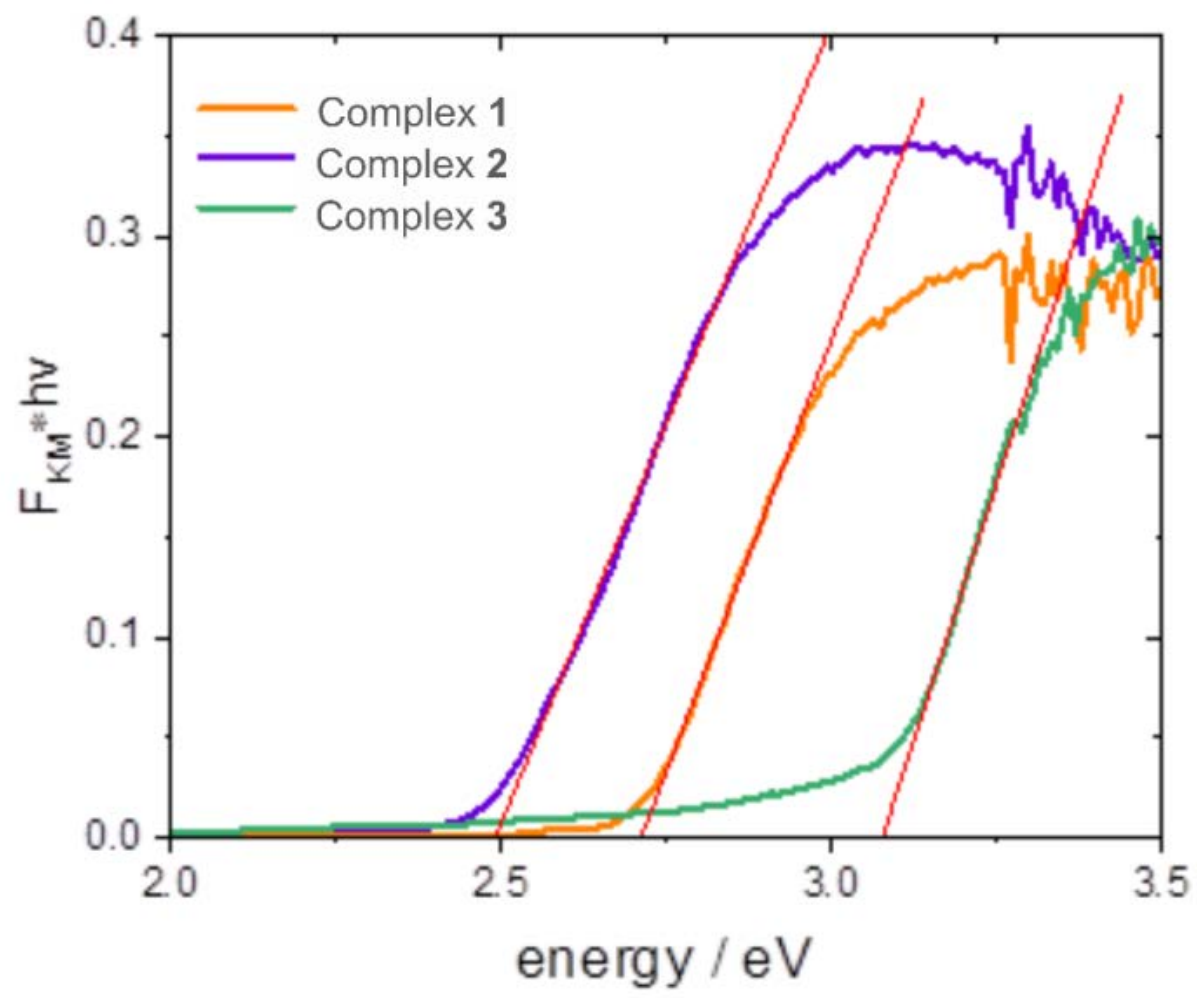

**Figure 6.** Diffuse reflectance spectra of three cyanothiazole-copper iodide complexes in the form of Kubelka-Munk function.

**Table 1**. Frontier orbitals, Mulliken point charges and the distribution of electrostatic potential for various isomers of cyanothiazole ligand as calculated using the DFT technique at the BVP86/DGDZVP level of theory.

| | **2CNtz** | **4CNtz** | **5CNtz** |
|---|---|---|---|
| HOMO | 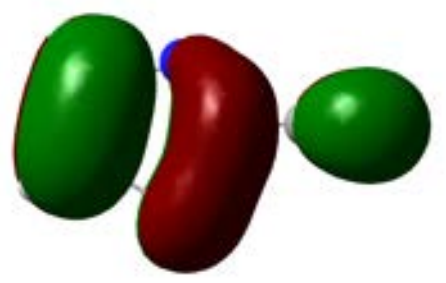 | 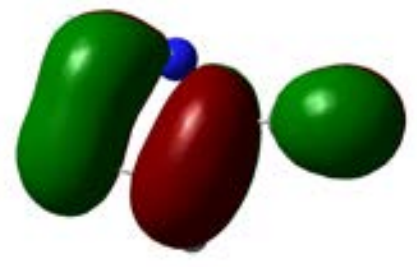 | 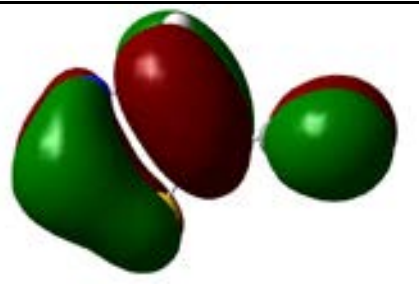 |
| LUMO | 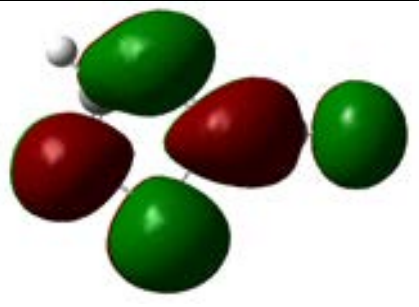 | 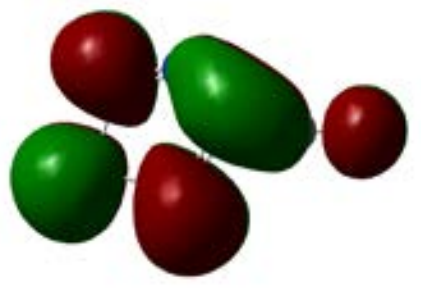 | 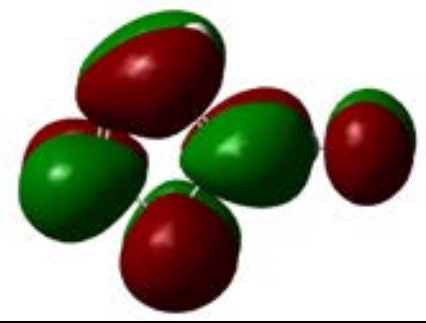 |
| ESP | 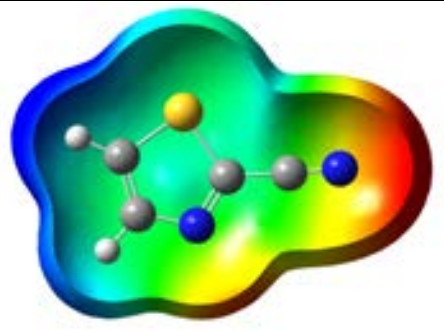 | 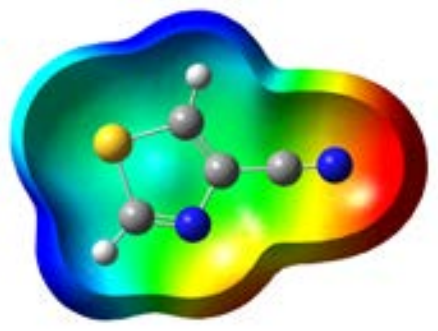 | 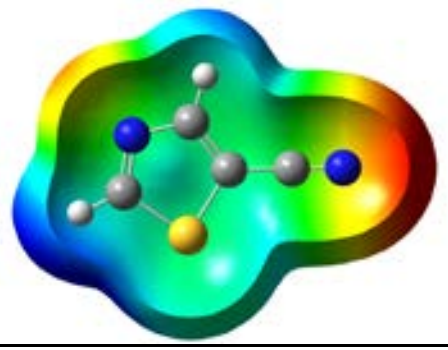 |
| Dipole moment / D | 5.17 | 5.26 | 3.39 |
| C≡N distance / Å | 1.153 | 1.153 | 1.155 |
| | | | |
| atom | Mulliken charges in a free ligand | | |
| C | 0.136 | 0.012 | 0.014 |
| N | -0.067 | -0.075 | -0.073 |
| | Mulliken charges in a complex | | |
| C | -0.030 | 0.012 | 0.010 |
| N | 0.112 | -0.085 | -0.129 |

All three studies cyanothizole isomers have very similar electronic structure (**Table 1**, **Figure 7**) with both HOMO and LUMO frontier orbitals fully delocalized over whole molecules. Both HOMO and LUMO have significant antibonding character with respect to C≡N bond in nitrile substituent. This has further consequences for interaction with Cu(I) centres. All HOMOs have almost the same energy, the LUMO of 4-cyanothiazole has slightly higher energy than other derivatives, but is seems not to have significant role. The largest negative charge can be observed in the vicinity of the nitrile group, in the case of 5-cyanothiazole also the lone electron pair at ring nitrogen significantly contributes to negative charge areas (**Table 1**). These results suggest that nitrile group should be the primary binding site for metal ions, for the 5-cyano isomer the ring nitrogen should be also taken into account. Therefore, two isomers, namely 2-cyanothiazole and 4-cyanothiazole form very similar complexes with copper iodide, based on $Cu_2I_2$ structural motifs with only small differences in their geometry. Surprisingly, the 5-cyano derivative forms complex based on infinite CuI chains. This may be a consequence of slightly different geometry of the ligand (the nitrile group and ring nitrogen atom are relatively more distant) as well as dipole moment of 5-cyanothiazole is also two times smaller than other ligands.

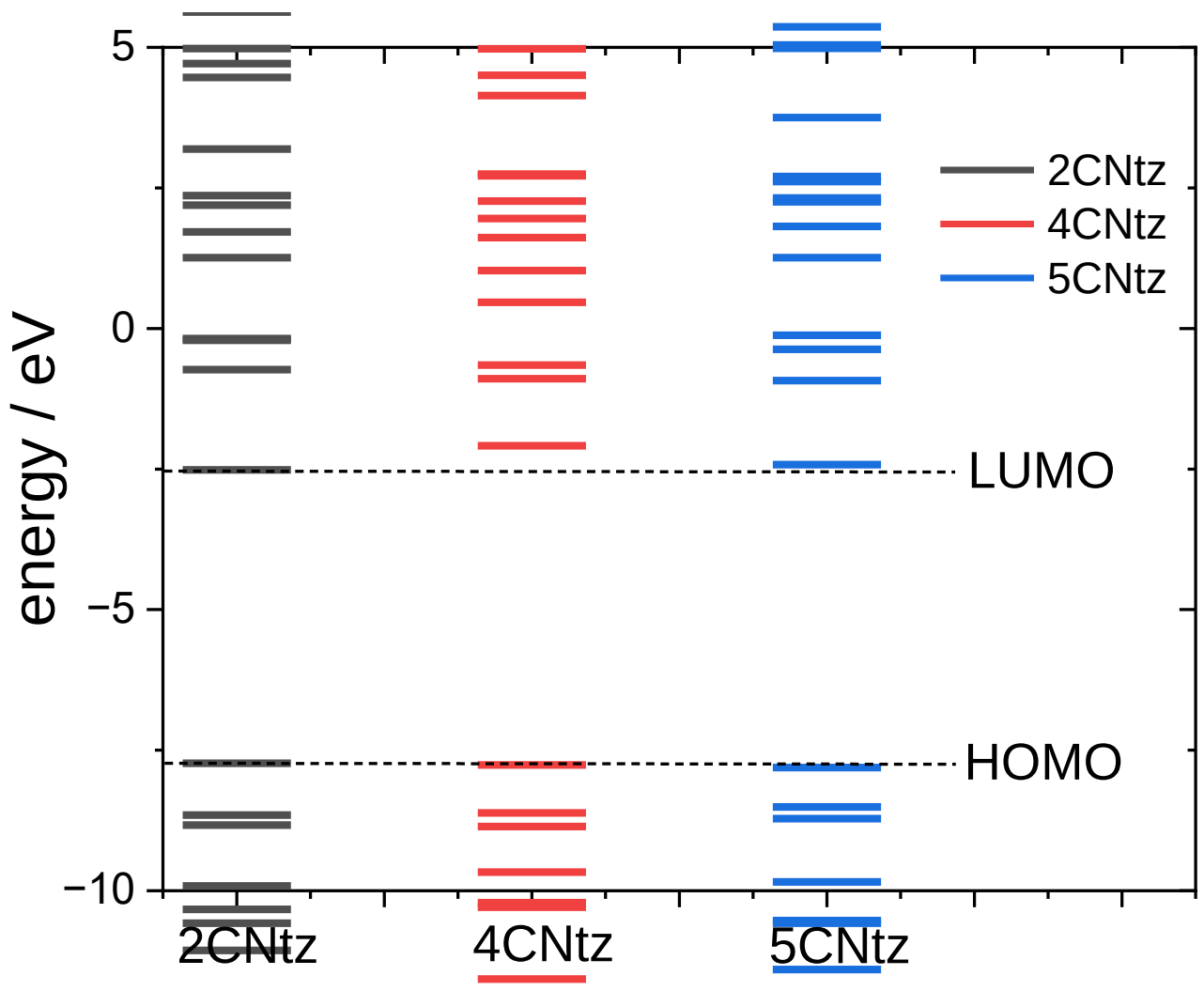

**Figure 7.** Electronic structure diagrams of cyanothiazols as calculated using the DFT technique at the B3LYP/TZVP level of theory.

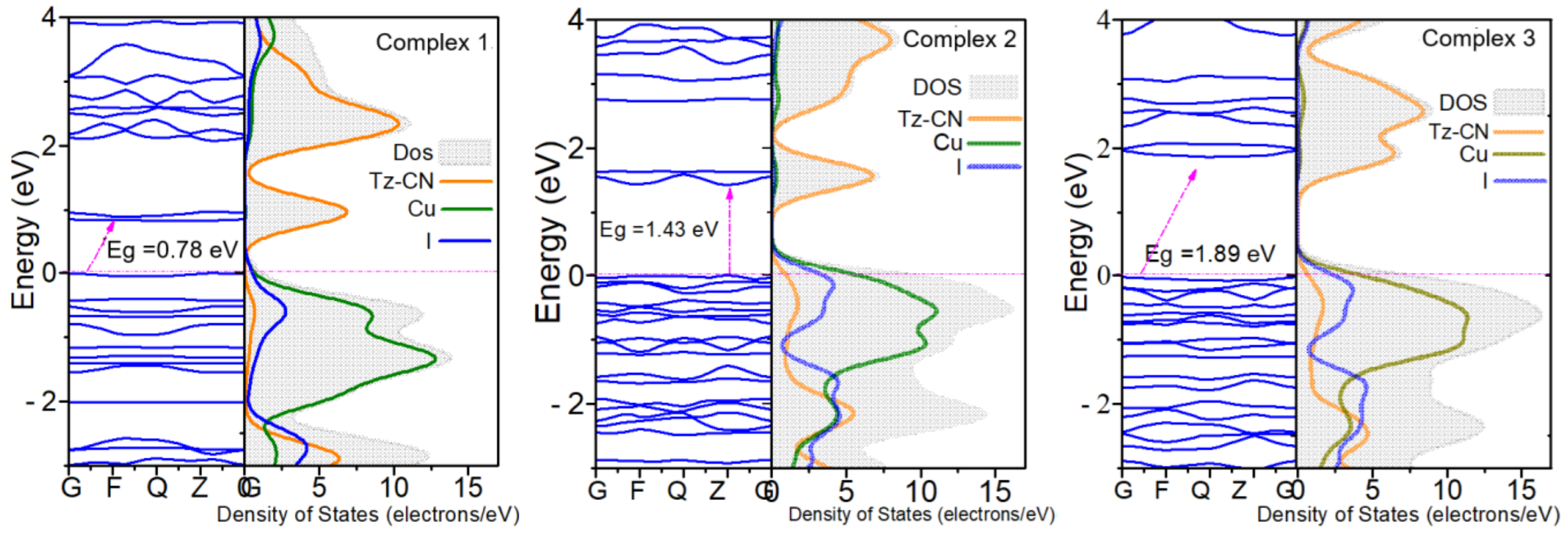

**Figure 8.** Band structure and density of states for complex **1** (**5CNtz**-CuI), complex **2** (**2CNtz**-CuI) and complex **3** (**4CNtz**-CuI) at DFT-MBD level of theory.

Despite very similar geometry and electronic structure of the three cyanothiazole ligands, the electronic structure of corresponding Cu(I) complexes with iodide ligands is significantly different. The fully relaxed, geometry-optimized structures of the **CNtz** crystals were used to calculate their electronic structures. As shown in **Figure 8,** the Valence Band Maximum (VBM) and the Conduction Band Minimum (CBM) are separated near the Fermi level, confirming that all the crystals exhibit semiconducting behavior. A closer examination of the band structure reveals that the VBM is close to the Fermi level and located at the Γ-point, while the CBM is at the Q-point, indicating a *p*-type indirect band gap (which is fully justified by the presence of copper(I) centres, which can be easily oxidized, thus yielding hoped as majority charge carriers). For the metal atoms with significant magnetic interactions, it is vital to apply spin-oribit coupling (SOC) to accurately capture their electronic properties as it modifies the electronic structure by splitting degenerate energy levels. In our examination we observed the band gap is effectively reduced compared to non-SOC calculations. This is due the crystal field effects splitting the d-orbitals into different energy levels, its further split with SOC. Also, SOC is creating hybridized orbitals by mixing the spin-up and spin-down states, that reduce the energy difference between the VBM and CBM, significantly narrowing the band gap (Figure S22). Calculating accurate band gaps in materials using the GGA-PBE-MBD method is challenging due to inherent errors. Hybrid functionals like HSE06 and PBE0, while more accurate, are computationally expensive. To address this, the DFT+U method has emerged as a promising alternative. It improves upon standard DFT functionals by introducing a correction term (the "Hubbard U") that accounts for the energy cost of having electrons in specific localized orbitals. This approach often achieves band gap predictions comparable to hybrid functionals close to the experimental value while being significantly less computationally demanding. In our study to obtain a bandgap value closely matching with the experimental data, we employed three different U values (3.5, 5.5, and 7.5 eV). The band gap predicted by the U7.5 functional closely approximates the experimental value. All calculated band structures are shown in the supporting information (Figures S23-S25). The bandgap values are listed in the **Table 2**. Further to analyze the electronic nature of the studied crystals we executed the atom-projected DOS calculations (Figure S26).

**Table 2**. Theoretical and experimental band gap data for cyanothiazole-copper iodide complex.

| **Compound** | **Theoretical band gap (GGA-PBE-D) / eV** | | | | | | **Experimental band gap / eV** |
|---|---|---|---|---|---|---|---|
| | **MB** | **SOC** | **U3.5** | **U5.5** | **U7.5** | **Type** | |
| **1** | 0.78 | 0.56 | 0.83 | 1.34 | 1.39 | Indirect | 2.71 |
| **2** | 1.42 | 1.30 | 1.60 | 1.83 | 1.95 | Direct | 2.49 |
| **3** | 1.89 | 1.77 | 2.25 | 2.41 | 2.53 | Indirect | 3.07 |

The valence band is composed mainly of 3d-copper orbitals hybridized with 5p orbitals of iodine with a significant contribution of organic ligand contribution at the bottom of the band, whereas lower part of the conduction band has also a significant contribution of occupied π orbitals of ligands. 2- and 4-cyano derivatives show also a small contribution of ligand π-orbitals to the top of the valence band. The bottom of the conduction band is built mainly from π-orbitals of the ligand, with a minor admixture of copper- and iodine-based orbitals. This contribution becomes slightly more significant in the upper part of the conduction band for the 5-cyano derivative. Such arrangement of energy levels indicates than organic ligands play a role of electron acceptors, thus decreasing the band gas of these materials, as compared to parent CuI material ($E_g$ = 3.1 eV).[30]

From the molecular orbital point of view **(Figure 9),** interactions between copper(I) ions and cyanothiazole ligands (2-, and 4-cyano isomers) are threefold: (i) coordination via lone electron pair at thiazole ring, (ii) coordination via lone pair at nitrile nitrogen and (iii) back donation from occupied $3d_{z^2}$ of copper, hybridized with $5p_x$ orbitals of iodine, to antibonding π-orbital of nitrile group. The latter bonding mode can be observed directly as a significant elongation of the C≡N bond on ligation with copper centre (~1.15 Å in the ligand, ~1.18 Å in complexes). Surprisingly, the same effect is observed for 5-cyano derivative, in which there is no direct interaction between Cu(I) centres and nitrile groups. In this particular case this effect may be attributed to electron flow from the Cu-I ribbon to the nitrile group via the ring. Analysis of Mulliken point charges confirms this interpretation: on ligation the total charge of the C≡N moiety increases from +0.041 to −0.119 e (**Table 1**)

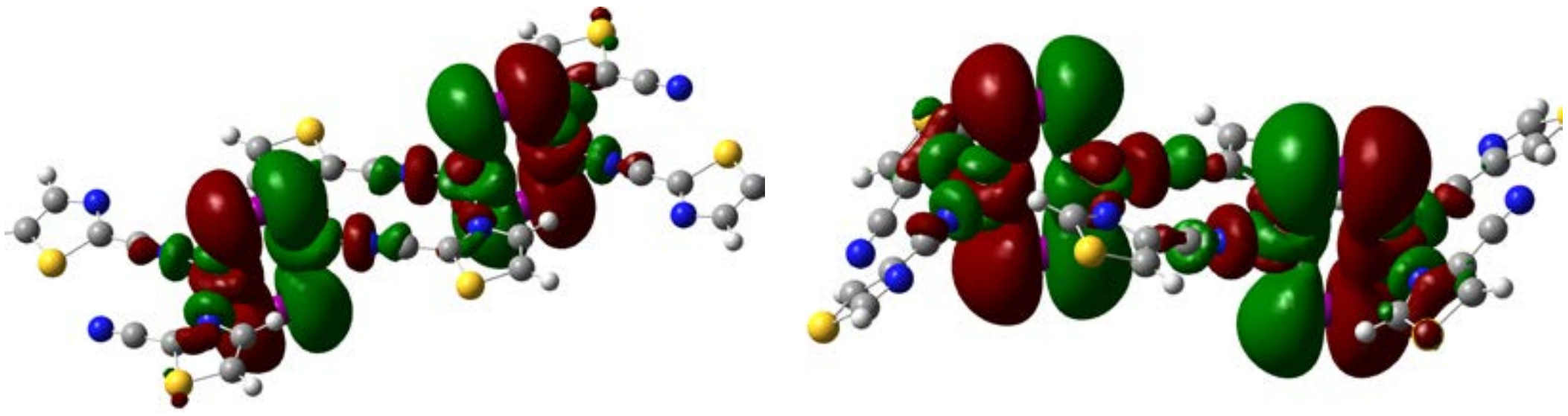

**Figure 9**. Contours of HOMO orbitals calculated for a relaxed-geometry fragment of polymeric chains of complex **2** (**2CNtz**-CuI, left) and **3** (**4CNtz**-CuI, right) complexes as calculated using DFT approach at the BVP86/DGDZVP level of theory.

It should be also noted that the energy dispersion of the top of valence and the bottom of the conduction band is rather small, which should result in relatively high effective masses of charge carriers, i.e. low electron and hole mobilities. It is fully justified, as any of structures under study does not provide significant π···π stacking interactions between neighbouring aromatic ligands, and the Cu-I polymeric ribbons in 5-cyano complex **1** extend only in one dimension. Therefore, as low mobility materials, copper iodide complexes with cyanothiazole ligands can find potential application as luminescent materials, catalysts or memristive materials, however they are not well-tailored for applications which require high mobility, e.g. field effect transistors.

**XAS analysis**

X-ray absorption spectroscopy (XAS) is an experimental technique that directly probes the electronic and structural nature of elements. In this work, XAS helped us to understand the nature of copper sites in the studied CuI complexes with cyanothiazoles compared to the reference CuI. We measured Cu K-edge spectra (**Figure 10**), where the sharp increase in the absorption – the edge – defines the threshold ionization energy. The value of the absorption edge usually increases with the oxidation state of the absorber element due to an increase in the core binding energy. The formal oxidation state of copper in all the studied materials is +1. Hence, the edge values are very close to each other and are 8982.05 eV for CuI reference, 8981.71 eV for **1**, 8982.14 eV for **2**, and 8981.95 for **3**. They were determined as the maximum of the first derivative of the spectrum with respect to energy (**Figure 11**).

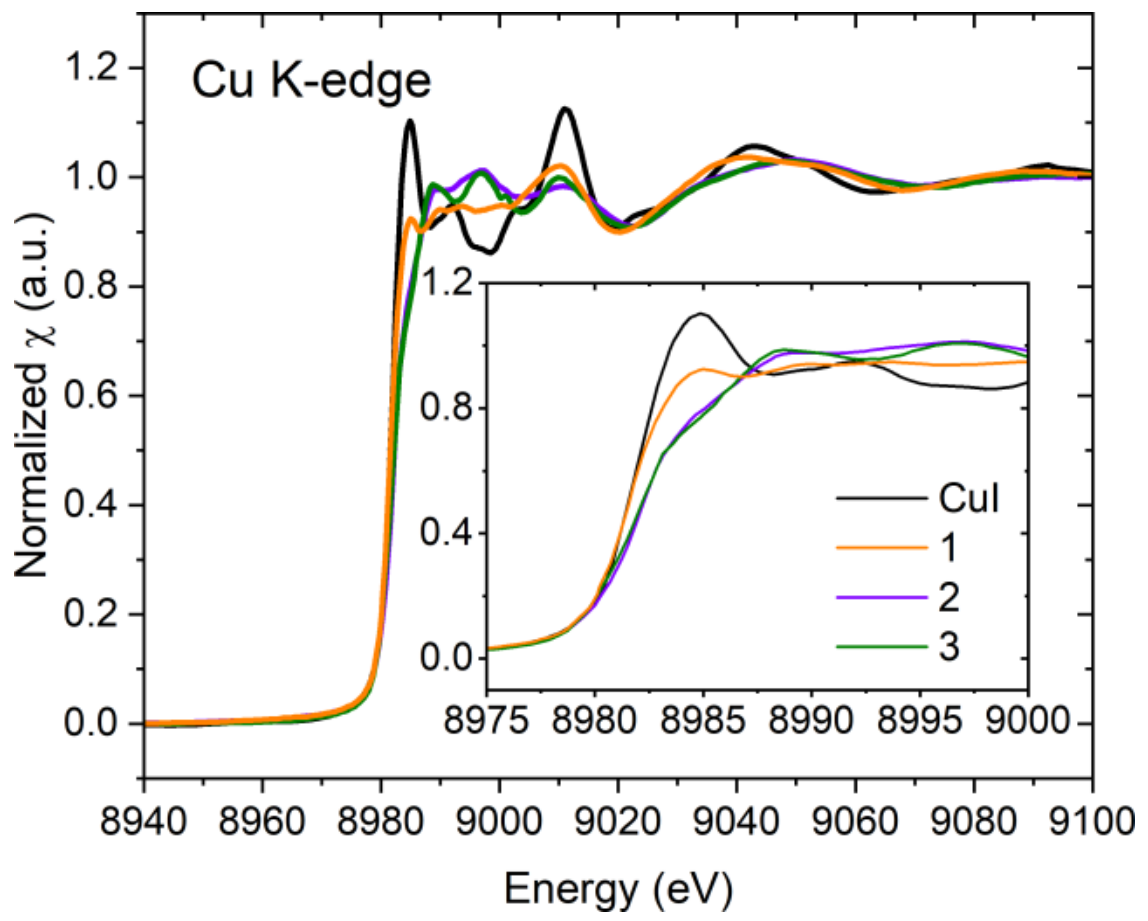

**Figure 10**. Experimental XAS Cu K-edge spectra for the CuI reference and its complexes with cyanothiazole isomers.

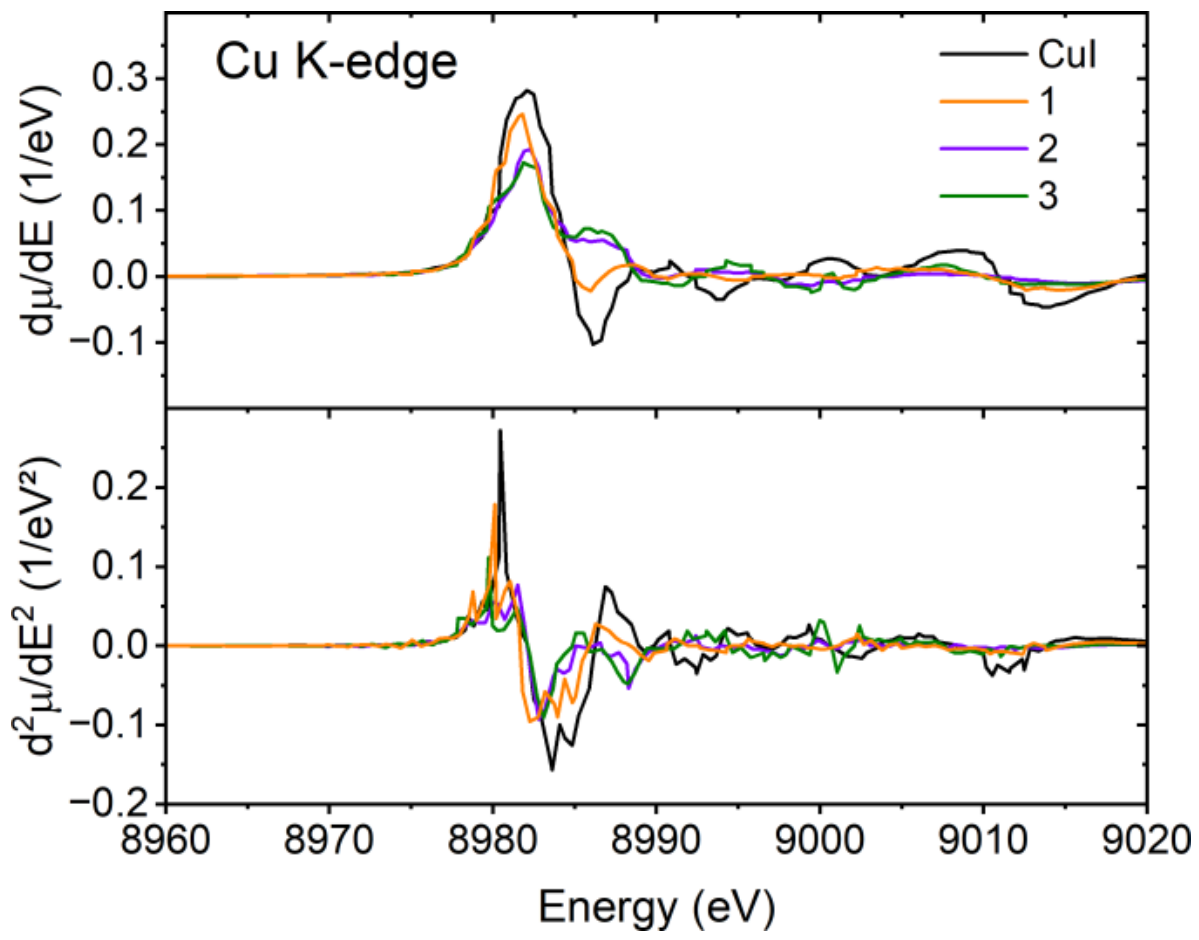

**Figure 11**. First (top) and second (bottom) derivatives with respect to energy of the experimental Cu K-edge XAS spectra for the CuI reference and the studied complexes with cyanothiazoles.

Above the edge lies the continuum consisting of a low-energy (up to 50–100 eV above the edge) X-ray absorption near-edge structure (XANES) part and a high-energy oscillatory extended X-ray absorption fine structure (EXAFS) part. XANES oscillations or their absence reflect the electronic and geometric characteristics of the molecular environment of the absorber. The basic electron configuration of copper is $[Ar]3d^{10}4s^1$, while the studied Cu(I) complexes with formal +1 oxidation state should reveal $[Ar]3d^{10}$. The main features in XANES result from the electronic transitions from 1s orbital, mainly to dipole-allowed 4p and quadrupole-allowed 4s or 3d orbitals. Depending on the element, the pre-edge part of XAS may also contain features of structural or electronic interest related to electronic transitions from 1s to higher bound states, the energies of which lie within the discrete part of the spectrum.

**Figure 10** shows Cu K-edge XANES for the studied complexes and CuI reference. Only for the reference CuI a distinct white line is observed, while spectra for the cyanothiazole complexes exhibit rather step-like course. This is related to the higher density of unoccupied states available for electronic transitions in CuI and, therefore, more localized absorption. The same spectral feature, but of much lower intensity, can be observed for 5-CNtz derivative. It is full justified, as the CuI ribbons offer much higher density of hybridized Cu and I states than other complexes. On the other hand, the iodide ions may be highly polarizable leading to significant electron redistribution. Such localized interaction may enhance the metal-ligand hybridization and create a more pronounced density of unoccupied states resulting in a high white line intensity. Salomon *et al.* studied XANES of several dozen Cu(I) and Cu(II) model complexes.[31] They also investigated the edge features of Cu(I) with ligand field theory. Since the 3d orbital in Cu(I) is fully occupied for Cu(I), the pre-edge peak was assigned to the dipole-allowed transitions, where the threefold degeneracy of $4p_{x,y,z}$ orbitals in free Cu(I) is split in the ligand field. Then, for linear 2-coordinate $D_{\infty h}$ complexes this would result in intense 1s → $4p_{x,z}$ pre-edge at lower energies and 1s → $4p_z$. For T-shaped 3-coordinate compounds ($C_{2v}$), this ends up in further degeneracy of $4p_{x,y}$, to obtain separate lower energy 1s → $4p_x$ pre-edge and low-intensity, higher energy 1s → $4p_z$ feature. In the 3-coordinate trigonal planar structure ($C_{3h}$), the $4p_y$ and $4p_z$ will be degenerated and shifted to a higher energy than $4p_x$, which is responsible for the pre-edge. Finally, in 4-coordinate tetrahedral ($T_d$) complexes, the 4p orbitals should be close to degenerate, but shifted to higher energy compared to free Cu(I), where the intensity of each 1s → $4p_i$ is reduced. It is worth noting that the Cu(I) complexes discussed here were characterised by N, O, and S ligation (or their mixtures), while in this work we study, the electron donors are nitrogen and iodide. Yet, the data on XAS for CuI or its complexes is scarce.[32]

Similarly as for the tetragonal 4-coordinated complexes reported by Salomon *et al.*,[31] we do not observe distinguish separate pre-edge peak at 8984, as for $Cu_2O$ (I).[33]

In Figures S27-S29 we show the DFT-calculated XAS spectra and density for states compared with experimental data for the studied CuI complexes. The calculated spectra are consistent with the experimental data. The main edge results from the dipole-allowable 1s → 4p transition. After deconvolution into separate molecular orbitals, the calculated PDOS splits in the lower-energy (< 8980 eV) and the higher-energy (> 8980 eV) edge features from different p-type orbitals. On the other hand, d-type orbitals contribute to higher-energy features (>8985 eV) and also would be responsible for the pre-edge, which is not present in the experimental spectra.

### Photoluminescence studies of CNTz-CuI complexes

All three complexes excited with UV light (375 nm, 3.59 eV), both at room and low temperature, emit yellow to orange, naked-eye visible photoluminescence, which is a typical behaviour of copper iodide derivatives (Figure S30).

Fluorescence spectra were measured for all compounds in a wide temperature range (6 - 325K). All studied compounds show dual emission (at low temperatures even three different emission components can be isolated), as shown in **Figures 12-14.** The emission maxima gradually shift to lower energies for **1** and **3** the latter one shows bathochromic shift of ca. 13 nm during 6 K – 325 K temperature scan, whereas **1** shows bathochromic shift of 58 nm. **(Figure 15)** At the same time emission intensity decreases with temperature, and the full width at half maximum increases with increasing temperature.

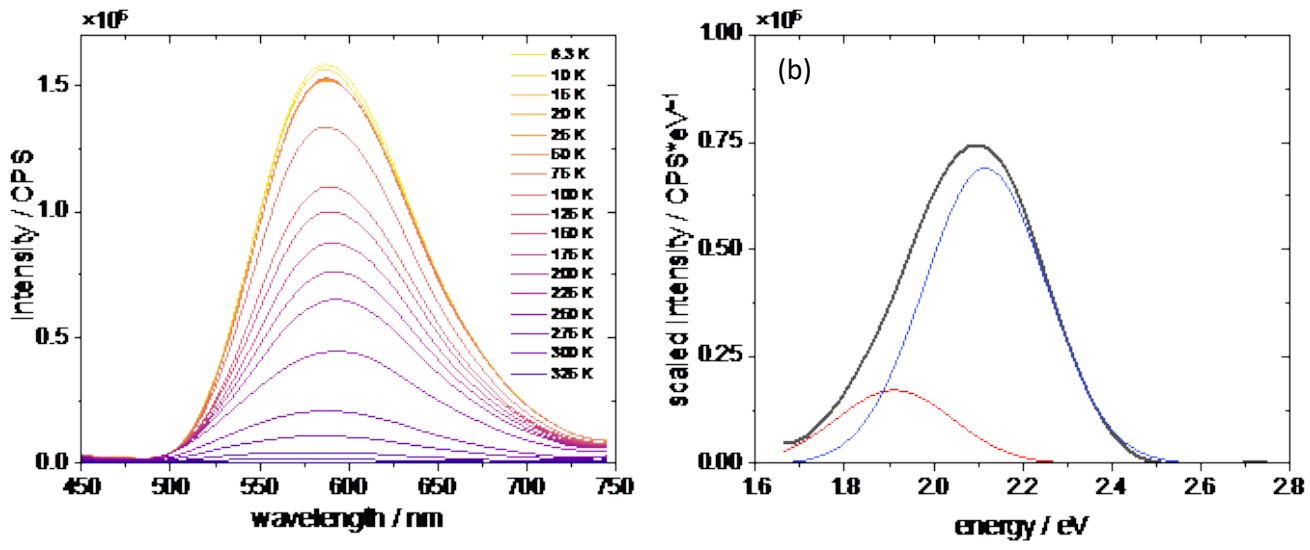

**Figure 12.** Emission spectra of complex **2** (a) along with deconvolution of the spectrum recorded at 6 K into Gaussian components (b).

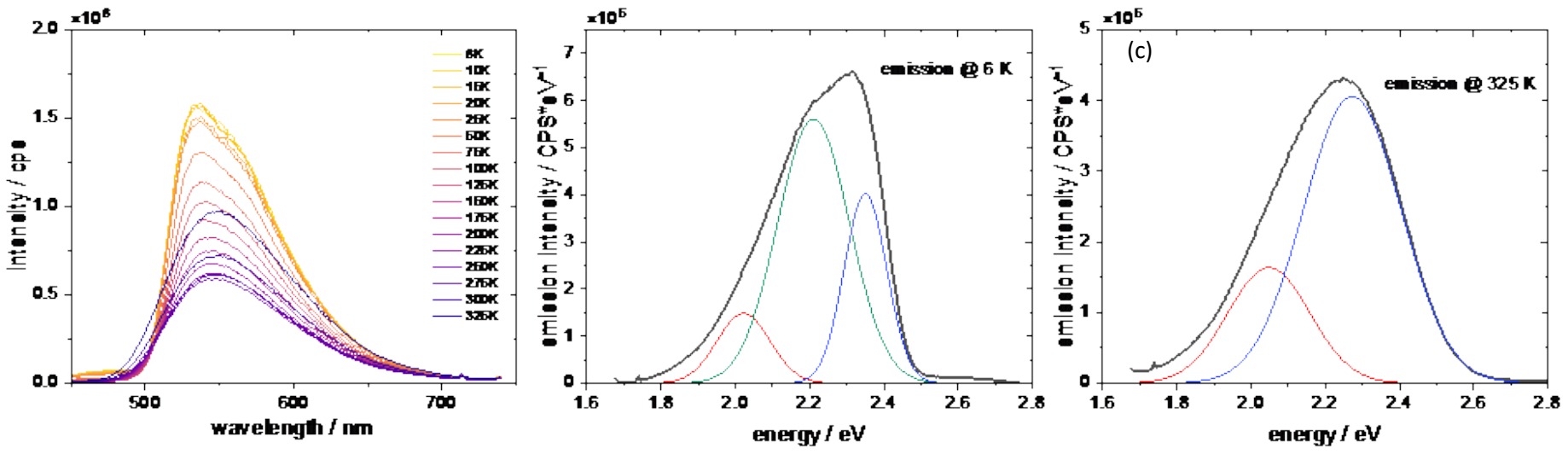

**Figure 13.** Emission spectra of complex **3** (a) along with deconvolution into Gaussian components of the spectra recorded at 6 K (b) and 325 K (c).

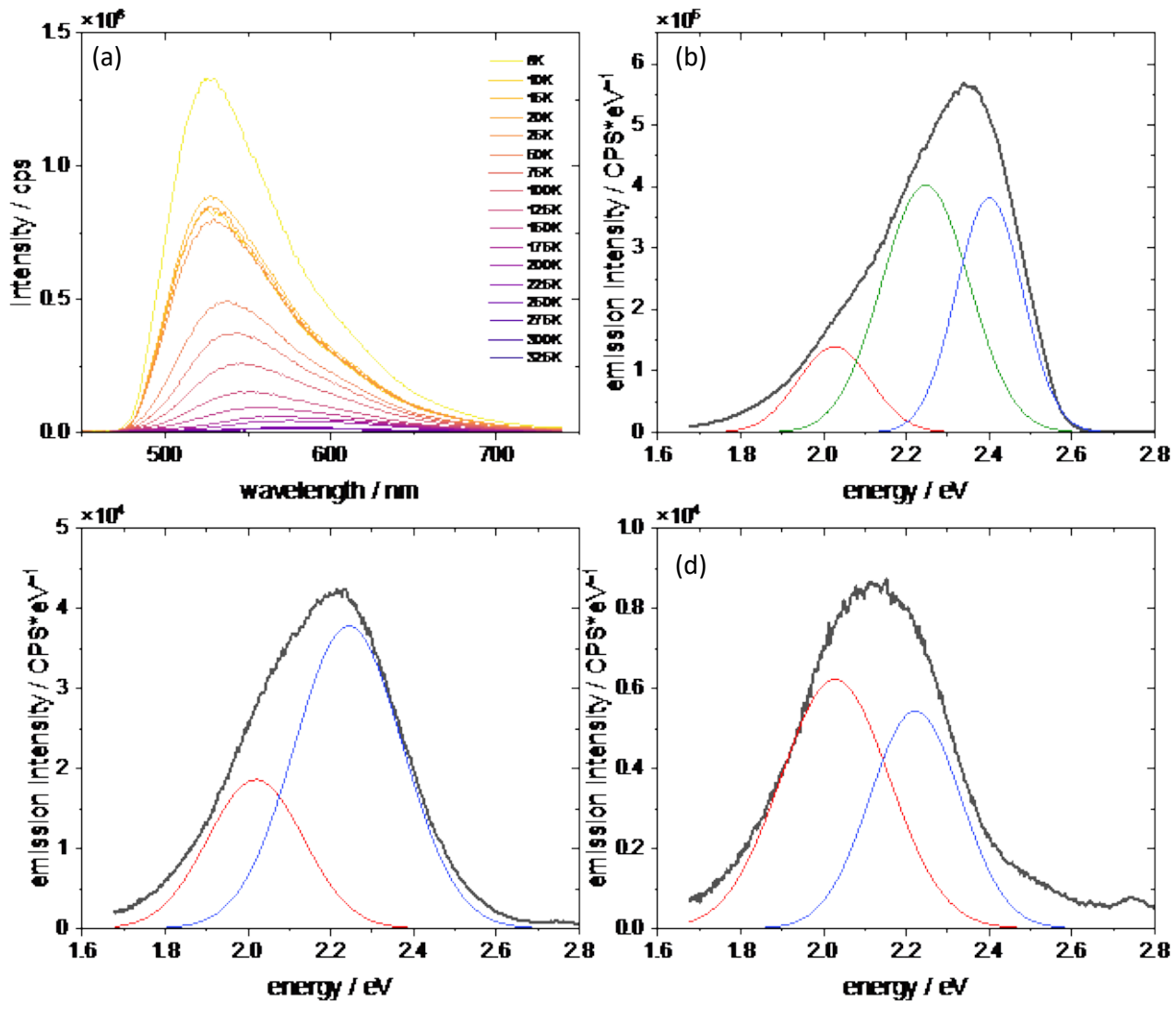

**Figure 14.** Emission spectra of complex **1** (a) along with deconvolution into Gaussian components of the spectra recorded at 6 K (b), 150 K (c) and 225 K (d).

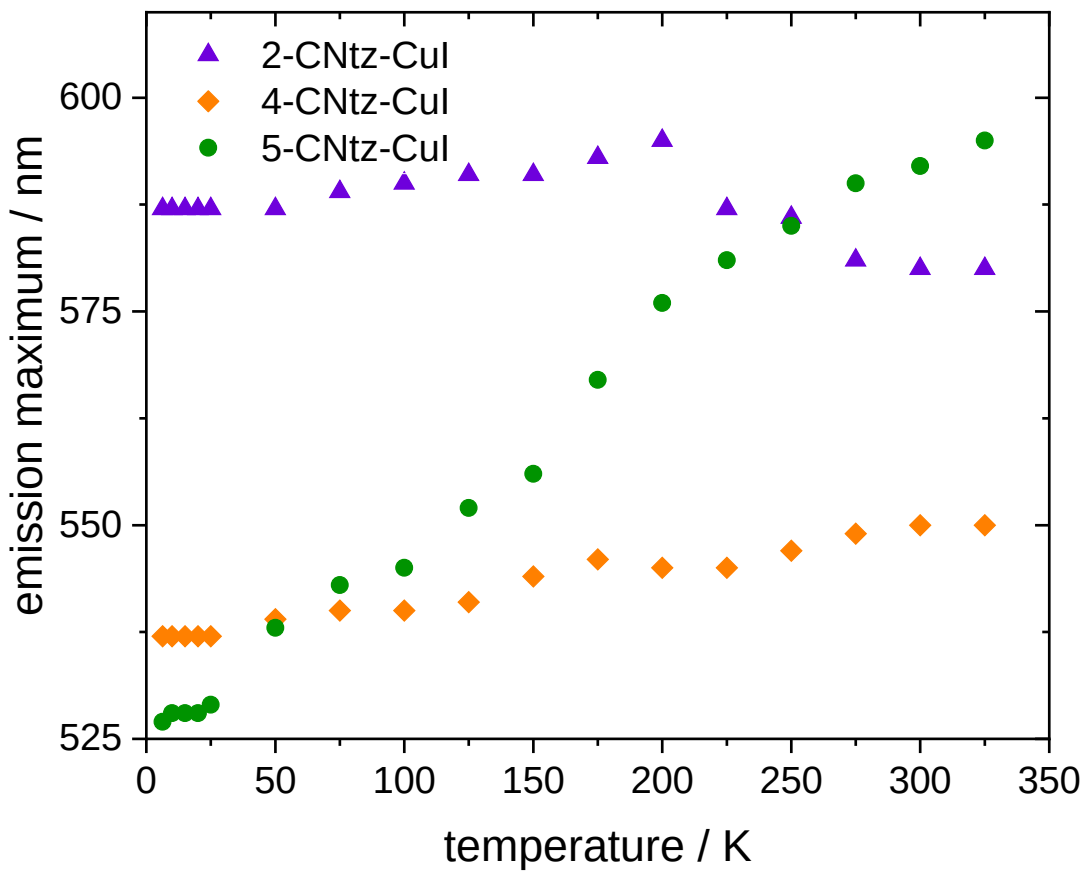

**Figure 15.** The changes in PL max peak for all complex in different temperatures.

The spectral data indicate the presence of at least two different emissive excited states. Analysis of electronic structure (density of states in particular) suggests three possibilities of low-lying excited states: MLCT (copper → cyanothiazole), MLCT (copper→iodide) (which can be also regarded as a local excited state (LE) involving hybridized orbitals of iodide and copper centres) and LLCT (iodide→cyanothiazole). The same set of transitions has been suggested for other copper iodide adducts with organic ligands.[34] The spectral data suggests that excitation to a higher excited state is followed by thermalization to low energy excited states of different charge transfer character and different internal reorganization (lower for LE state and much higher for MLCT state. The interchange between these states is possible via the thermally activated intersystem crossing. On this basis a tentative energy diagrams can be constructed **(Figure 16).** In the case of low interconversion energy barrier (**Figure 16a**) dual emission is not observed (or just a minor change of emission maximum can be observed). This is the consequence of an interplay between ground and excited state potential energy curve and also depends on reorganization energy. If the molecular displacement is small, then we should observe almost the same emission energy from both excited states, as shown in **Figure 16a**. This is the situation observed in the case of 2-CNtz-CuI. Despite that detailed shape analysis shows two distinct emission bands (**Figure 12b**) at 1.91 and 2.11 eV, which may correspond to two different excited states, which, due to low energy barrier are in the thermal equilibrium at any temperature. Similar situation, however with slightly larger special separation of relaxed excited states are observed in the case of **3**. Slightly different spatial arrangement of organic ligand may result in slightly higher activation barrier, therefore weak bathochromic shift is observed with increasing temperature.

More spectacular spectral changes can be observed for the 5-CNtz-CuI complex **1**. Large bathochromic shift of the emission peak corresponds to the systems with high activation energy between two emissive states. Furthermore, at low temperatures a third peak is clearly visible. Broadening of bands as well as decrease in intensity with increasing temperature is also clearly visible.

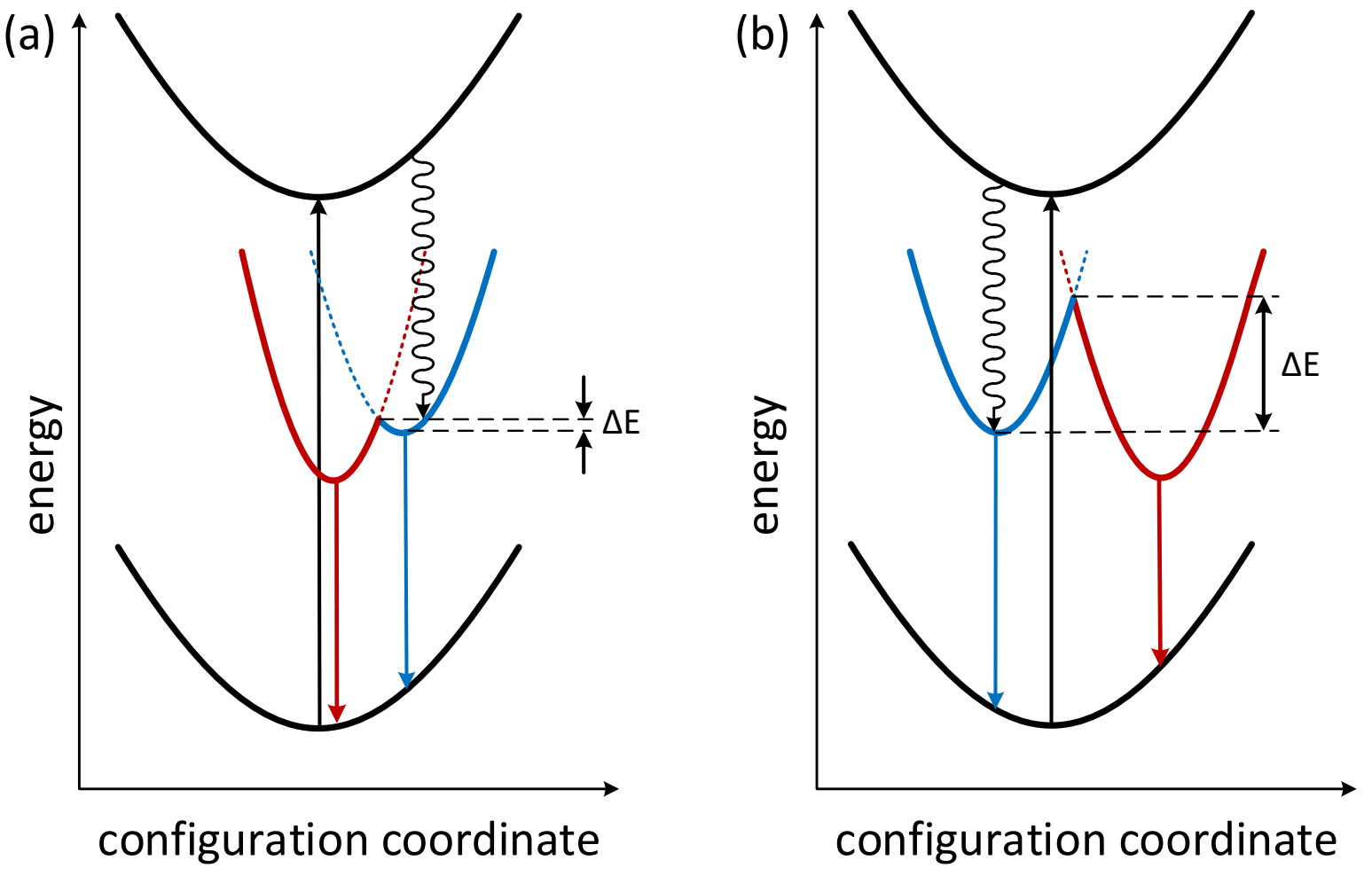

**Figure 16**. Schematic of photophysical processes with two self-trapped emitting states (red and blue line) with a) small and b) large potential barriers. Please note that energies of low energy emissive excited states are identical in (a) and (b), the only difference is the configurational coordinate (molecular displacement in the Frank-Condon state).

Broadening of bands may be used to evaluate the Huang-Rhys parameter, which is the measure pf photon-exciton coupling, as given by Equation 4:

$$FWHM = 2.36\sqrt{s}\hbar\omega_{photon}\sqrt{\coth\frac{\hbar\omega_{photon}}{2k_BT}} \qquad \text{(Equation 4)}$$

where FWHM is the full width at half maximum (derived from deconvolution of emission spectra into Gaussian components), $s$ is the Huang-Rhys factor and $\omega_{phonon}$ is the phonon energy. Fitting of Equation 4 to experimental data yields Huang-Rhys factors of 15.21, 12.52 and 39.8 for 2-CNtz-CuI, 4-CNtz-CuI and 5-CNtz-CuI, respectively. Photon energies amount 34.7, 19.8 and 13.0 meV, respectively. The latter, much higher value is justified by the specific structure of the complex – ribbon-like CuI core decorated by the 5-cyanothiazole ligands.

In conclusion, all cyanothiazole adducts with copper(I) iodide are luminescent materials with dual-type emission, however 2- and 4- isomers yield polymers with periodically arranged $Cu_2I_2$ fragments separated by organic ligands. Therefore internal reorganization energy is low, which is reflected in low activation barrier between two emissive excited states and dual emission features are barely visible. The 5- isomer in turn yields polymeric ribbons decorated with organic ligands. Therefore the MLCT transition is accompanied with much larger geometry displacement as compared to other polymers. As a consequence, the activation energy barrier between two emissive states is much higher due to increased internal reorganization energy and temperature-dependent dual emission is a dominating spectral feature of the 5-CNtz-CuI complex **1**.

## Electric measurements

For electrical measurements for information processing layered devices were prepared, and their current – voltage (I-V) characteristics were measured at room temperature. each of the compounds presented a hysteresis curve, however, with respect to behavior in the 1$^{st}$ quadrant, complex **2** and **3** were characterized by the clockwise switching and the complex **1** by the counter-clockwise switching (see **Figures 17a-c**). These characteristics influence directly the pulse behavior. For the first two compounds positive pulses decrease the conductivity, whereas the negative pulses increase conductivity. The incremental switching behavior patterns – either in a form of potentiation or depression are shown for complex **2** in **Figures 17e** and **h**. Depression behavior is registered above threshold of +2.2 V for switching pulses, whereas potentiation phenomena are available with only low values of switching pulse potentials – as low as -0.6 V. **Figures 17f** and **i**. shows results for complex **3**, for which sample the depression is possible with switching pulse potentials of +1.4 V, potentiation starts for -1V pulses. Sample complex **1** (**Figures 17d** and **g**) can be potentiated with pulses over +1 V (signal intensity varies throughout the sequence), depression can be achieved for pulses of -1.4 V or higher. In all cases gradual switching is reversible.

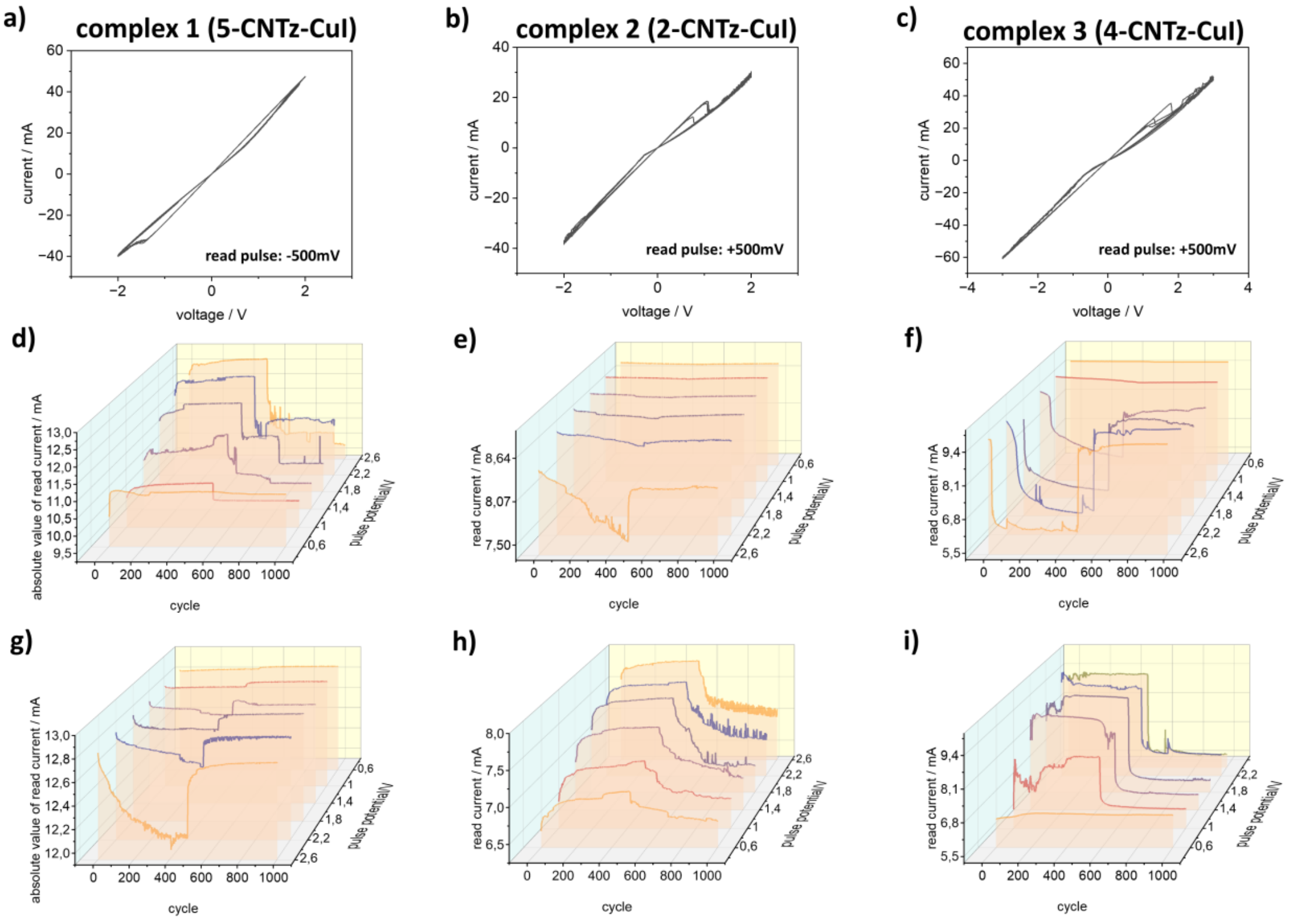

**Figure 17.** Basic electrical and plasticity behavior characterization. I-V characteristic for the a) complex **1** (**5CNtz**-CuI) b) complex **2** (**2CNtz**-CuI) and c) complex **3** (**4CNtz**-CuI). Potentiation-depression experiments: row d)-f) after conditioning samples in -3 V, followed by positive and negative set of switching pulses. Row g)-i) after conditioning samples in +3 V, followed by negative and positive set of switching pulses. In case of negative readout (complex **1**) values, the current is presented as absolute value, allowing easy comparison of the different behavior types for similar conditions.

These characteristics influence directly the pulse behavior. For the first two compounds positive pulses decrease the conductivity, whereas the negative pulses increase conductivity. Sample 5CNtz complex **1** (**Figure 17d** and **g**) can be potentiated with pulses over +1 V (signal intensity varies throughout the sequence), depression can be achieved for pulses of -1.4 V or higher. The incremental switching behavior patterns – either in a form of potentiation or depression are shown for 2CNtz complex **2** in **Figures 17e** and **h**. Depression behavior is registered above threshold of +2.2 V for switching pulses, whereas potentiation phenomena are available with only low values of switching pulse potentials – as low as -0.6V. Figure **17f** and **i** shows results for 4CNtz complex **3**, for which sample the depression is possible with switching pulse potentials of +1.4 V, potentiation starts for -1 V pulses. In all cases gradual switching is reversible.

Regarding electrical endurance (on-off test) and stability of the switching (retention tests) for the devices, complexes **1** (**Figures 18a** and **d**) and **2** (**Figure 18b** and **e**) can be switched in repeatable and predictable manner, with both LRS and HRS states stable for the period of 12h. Complex **3** (**Figure 18c** and **f**) however switches randomly over 1000 cycles and is characterized by instability of one of the states (over 1.5 h).

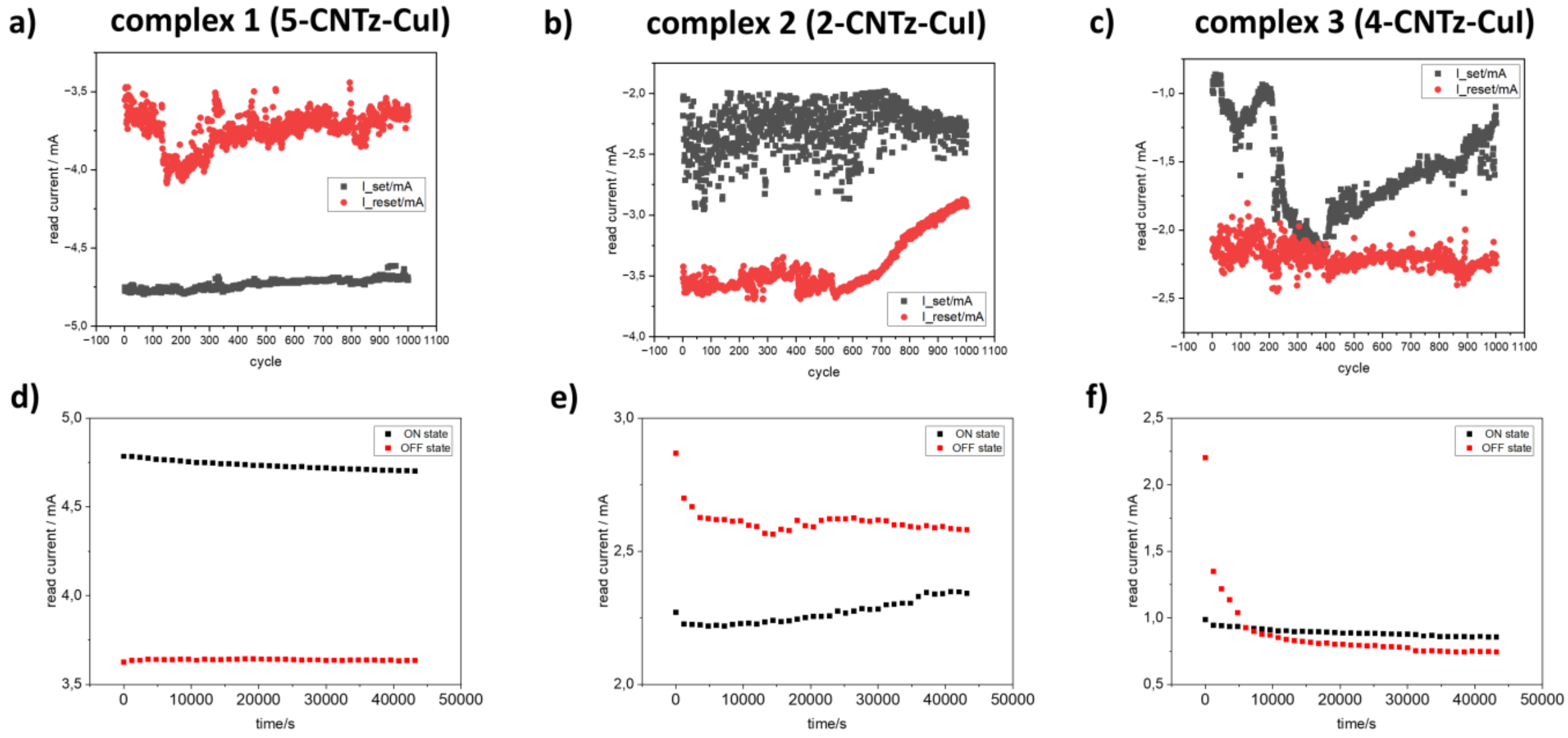

**Figure 18.** Device endurance and state retention tests. Results of the on-off switching (endurance) tests for: a) complex **1** (**5CNtz**-CuI), b) complex **2** (**2CNtz**-CuI) and c) complex **3** (**4CNtz**-CuI). Results of the retention tests for the HRS and LRS states for the same complexes: d) **1**, e) **2** and f) **3**.

## Conclusions

In this work, we synthesized simple yet previously unreported CuI complexes with cyanothiazole ligands. Moreover, we experimentally and theoretically demonstrated how the position of the CN group in the thiazole ligand impacts the structure of copper complexes and, consequently, influences their optical and electrical properties. The cyanothiazole adducts with copper(I) iodide form coordination polymers; **5CNtz** derivative (complex **1**) forms polymeric ribbons decorated with organic ligands and the complexes of **2CNtz** and **4CNtz** (complexes **2** and **3** respectively) feature periodically arranged $Cu_2I_2$ fragments separated by organic ligands. All compounds are luminescent materials with dual-type emission; the internal reorganization energy in **2** and **3** is low, which is reflected in low activation barrier between two emissive excited states and dual emission features are barely visible. The MLCT transition complex **1** with **5CNtz** is accompanied by much larger geometry displacement as compared to the other two polymers. The increased internal reorganization energy leads to a significantly higher activation energy barrier between the two emissive states, making temperature-dependent dual emission the predominant spectral feature.

The Cu K-edge XAS analysis confirmed the +1 oxidation state of the studied complexes. It also revealed a slightly lower degree of hybridization between Cu and I p-type orbitals in complexes **2** and **3** compared to complex **1**, with both exhibiting significantly reduced hybridization relative to the CuI reference.

The DFT results confirms that these complexes have the electronic properties typical of p-type semiconductors, implying potential applications in electronic devices that utilize hole-based conduction.

Regarding potential use of the compounds in information processing only complexes **1** and **2** can be switched between HRS/LRS in repeatable and predictable manner. Electrical characteristic of complex **3** makes it more of a candidate for volatile behavior or random number generation applications. Gradual switching profile of compound **2** would suggests it can be implemented in constructing material-based neural networks.

## EXPERIMENTAL

### Syntheses

*General Considerations*

The following chemicals were used as purchased: cooper(I) iodide CuI, 98.0%, WARCHEM; 5-cyanothiazole (5TzCN), 97%, Angene; 2-cyanothiazole (2TzCN), 99.24%, AmBeed; 4-cyanothiazole (4TzCN), 97%, Angene ;acetonitrile ACN, 99.9%.

*Synthetic protocols*

**Compound 1:** A solution of CuI (0.086g, 0.0452mmol), was prepared in 3 mL of ACN. Solid **5TzCN** (0.05g, 0.0452mmol) was added into the solution, and the reaction mixture was stirred until complete dissolution of the substrate. The resulting reaction mixture was subjected to crystallization by slow evaporation of the solvent under controlled conditions. After approximately 24 hours, well-defined yellowish crystals were obtained, yield: 86%, decomposition temperature above: 217.3 °C. Elemental analysis: calcd 15.98% C, 0.67% H,

9.32% N; found 15.97%C, 0.69%H, 9.33%N. FTIR ATR: 3092(m), 3072(vs), 2227(vs), 1505(m), 1375(m), 1308(s), 1231(vs0, 1114(s), 896(s), 820(s), 597(m) $cm^{-1}$.

**Compound 2** was synthesized using the same procedure as described for compound **1**, well-defined yellow crystals were obtained, yield: 68%, decomposition temperature above: 137.1 °C. Elemental analysis: calcd 15.98% C, 0.67% H, 9.32% N; found 15.96%C, 0.66%H, 9.34%N . FTIR ATR: 3098(w), 3082(vs), 2231(w), 1464(m), 1351(s), 1309(s), 1202(w), 1136(vs), 1051(s), 909(w), 885(m), 733(w), 770(vs), 702(w), 619(m), 524(m), 492(w) $cm^{-1}$.

The **unstable form of compound 2** was synthesized following the same procedure as described for compound **1**, with the modification that the solvent was quickly evaporated to yield a reddish powder. FTIR-ATR ($cm^{-1}$): 3103(s), 3112(s), 3084(vs), 2233(s), 1478(m), 1468(m), 1373(s), 1307(w), 1189(w), 1134(s), 1057(m), 883(w), 766(s), 753(s), 617(w), 532(w), 487(w) $cm^{-1}$.

**Compound 3** was synthesized using two methods described below, decomposition temperature above: 155.5°C. Elemental analysis: 15.98% C, 0.67% H, 9.32% N; found 15.99%C, 0.69%H, 9.30%N. FTIR-ATR: 3094(vs), 3073(vs), 1416(s), 1299(m), 1213(m), 1130(m), 986(vs), 8819m), 843(vs), 784(vs), 504(s) $cm^{-1}$.

<u>Method A:</u> A solution of CuI (0.086g, 0.0452mmol) was prepared in 3 mL of ACN. Solid 5TzCN (0.05g, 0.0452mmol) was then added to the solution, followed by the addition of **1** mL of distilled water. The reaction mixture was stirred until the substrate completely dissolved and left for crystallization in 5-8°C. A white crystalline product was subsequently obtained. Yield: 46%

<u>Method B:</u> The synthesis procedure was identical to that described for compound **1**, except the solvent was evaporated to 1 mL of the initial volume of the reaction mixture. This resulted in the formation of a crystalline yellowish powder. Yield: 52%

**Computational methods**

DFT modeling of single ligand molecules as well as fragments of polymeric chains has been performed using Gaussian 16 Revision C.01 software package[35] and post-processed and visualized using GaussView package.[36] For organic molecules the B3LYP hybrid functional[37] and the TZVP basis set[38] have been used, whereas metal complexes were modelled using the BVP86 hybrid functional[39] and the DGDZVP Gaussian-type double zeta basis set.[40]

The geometry optimization and the band structure and the density of states were obtained, were calculated using CASTEP (Cambridge Serial Total Energy package) code. The plane-wave basis set and pseudopotentials were used implemented with the PBE-GGA exchange-correlation functional. The calculations employed plane-wave basis set with pseudopotentials to carry out the PBE-GGA (Perdew-Burke-Ernzerhof Generalized Gradient

Approximation) exchange-correlation functional. Ion-electron interactions were described using the projected augmented wave (PAW) formalism. Non-covalent interactions were included through the DFT-MBD (Density Functional Theory with Many-Body Dispersion) relativistic correction method. To improve the accuracy of the electronic structure we employed the DFT-MBD+*U* method with GGA-PBE exchange-correlation functional. The $U_{eff}$ parameters is employed for the localized 3d electrons of the Cu. To ensure exact convergence, the periodic boundary conditions were set with the following parameters: a plane-wave cut-off energy of 580 eV, an energy convergence tolerance of $5 \times 10^{-7}$ eV/atom, a maximum force tolerance of 0.01 eV/Å, a maximum stress tolerance of 0.02 GPa, and a displacement tolerance of $5 \times 10^{-4}$ Å. A Monkhorst-Pack grid of 5 × 5 × 5 k-points was applied for all calculations.

XAS spectra were modelled using density functional theory (DFT) calculations performed with FDMNES software[41] using local spin density approximation. The structure for the calculations was previously DFT-optimized using CASTEP. The finite difference method was used for X-ray absorption fine structure[42] with dipole (Δl=±1) and quadrupole (Δl=0, ±2) trasitions and 9 Å cluster radius. Relativistic and spin-orbit coupling effects were neglected. Lorentzian convoluted spectra were presented.

**Electrical measurements**

Thin film samples were prepared according to following procedural steps. Firstly, all compounds were dissolved in the mixture of DMSO (typically 30mg of compound in 1mL of DMF) and mixed on magnetic stirrer at elevated temperatures (100°C). Substrates, ITO glass (Ossila, The Netherlands) were washed (water, isopropyl alcohol), dried and cleaned with $O_2$ plasma, then heated to 100°C. Thin films were deposited on hot substrates via spincoating technique, typically 3000 rpm for 30s and post-baked in 100°C on a hotplate for 25 min. Copper electrodes were thermally deposited afterwards, through shadow mask (Ossila, The Netherlands) with electrode dimensions 1.3 x 1.5 mm.

All of the I-V responses, state retention measurements, endurance tests and potentiation-depression tests were registered on SP-300 potentiostat (BioLogic, France) equipped with Instec TP102V Thermoelectirc Probe Station. The system was designed as two-terminal device, with working electrode (WE) connected to Cu electrodes and counter (CE) and reference (RE) electrodes were connected the ITO substrate.

Potentiation-depression tests were conducted two-way. Firstly the devices were switched to either high-resistance state (HRS) or low-resistance state (LRS). In the next step a set of 500 pulses of one polarity was followed by a set of 500 pulses of reversed polarity. Device state was read after each incremental switching – either by 200mV or -200mV. The change in reading voltage was to increase S/N ratio and was based on the shape of the hysteresis loop (see **Figure 17 a)-c)**).

The endurance test (on-off switching) was conducted by subjecting the devices to alternating extreme potential values (-2 V and +2 V) for 100 ms. This switching sequence was repeated 1000 times, with the device state measured at a reading voltage of either -200mV after each cycle.

State switching stability (state retention) was evaluated for both HRS and LRS. After applying a DC bias of either -2 V or +2 V for 1 s, the device state was measured at an arbitrary chosen reading voltage of +200 mV every 20 minutes over a period of 12 hours.

**Physicochemical methods**

FTIR spectra of the pure crystalline products were recorded using a Nicolet iS50 spectrometer equipped with a Specac Quest diamond ATR accessory. All FTIR spectra were collected and processed using OMNIC software.

Elemental CHNS analyses were performed on a Vario EI Cube Elemental Analyzer.

The melting points of the compounds were determined using a Stuart Scientific SMP3.

Solid-state diffuse reflectance spectra were recorded with a Perkin Elmer Lambda 365+ double-beam UV/Vis spectrometer, equipped with an integrating sphere. Barium sulfate ($BaSO_4$) was used as a reference blank for the reflectance spectra in the 200–1100 nm range, with a slit width of 5 nm and a scan speed of 489 nm/min. The Kubelka–Munk transformation was applied to the reflectance data to estimate the energy band gap ($E_g$). Data evaluation was carried out using the UV WinLab Data Processor and Viewer (Version 10.6.2).

Cu K-edge X-ray absorption spectra (XAS) were measured at the bending magnet ASTRA beamline at the SOLARIS National Synchrotron Radiation of Poland.[43] Powder samples were mixed with microcrystalline cellulose, ground with mortar and pestle, then pressed into thin pellets and put between Kapton tapes. The measurements were performed in transmission mode using an incident photon beam delivered by a modified Lemmonnier-type double-crystal monochromator equipped with Ge (220) crystals. For the monochromator energy calibration, we used Cu foil placed in the reference chamber, with the absorption edge at 8979 eV. Positions of the absorption edges were determined based on a maximum of the first derivative of the spectrum. The final spectra were merged from at least three consecutive scans. All spectra were processed using the Athena program from the Demeter software package.[44]

Photoluminescence spectra have been recorded on Fluorolog-3 Horiba Jobin Yvon spectrophotometer with high pressure xenon lamp. The samples were placed in a quartz soldered capillary. The emission spectra were recorded in the range of 450-740 nm with 375 nm excitation wavelength and resolution of 1 nm. The spectra were measured in a wide temperature range from 6-325 K with 25 K steps provided by the use of a closed-cycle helium cryostat (ARS Inc.) equipped with LakeShore 331 Temperature Controller.

**Crystallography**

The crystal structure data for compounds **1** - **3** were collected on an IPDS 2T dual-beam diffractometer (STOE & Cie GmbH, Darmstadt, Germany) at 120.0(2) K using Mo$K_{\alpha}$ (complexes **1** and **2**) and Cu$K_{\alpha}$ (complex **3**) radiation from a microfocus X-ray source (GeniX 3D Mo High Flux, Xenocs, Sassenage, France). Crystals were cooled with a Cryostream 800 open-flow nitrogen cryostat (Oxford Cryosystems). The crystallographic data are summarized in Table S1. Data collection and image processing for compounds **1** - **3** were carried out using X-Area 1.75. Intensity data were scaled with LANA (part of X-Area) to minimize differences in the intensities of symmetry-equivalent reflections (integration method). The structures were solved using the intrinsic phasing procedure implemented in SHELXT, and all non-hydrogen atoms were refined with anisotropic displacement parameters by the full matrix least squares method based on $F^2$, using the SHELX–2014 program package.[45] The Olex[46] and WingX[47] program suites were employed to prepare the final version of the CIF files. Mercury was used to prepare the figures.[48]

Hydrogen atoms were refined using isotropic model with $U_{iso}$(H) values fixed to be 1.2 times $U_{eq}$ of the carbon atoms to which they were attached.

**Acknowledgements**

The authors acknowledge the financial support from the Polish National Science Center within the OPUS programme (grant agreement No. 2022/47/B/ST4/00728). This research was partly supported by program “Excellence initiative–research university” for the AGH University of Science and Technology and Gdańsk University of Technology. The authors gratefully acknowledge Polish high-performance computing infrastructure PLGrid (HPC Center: ACK Cyfronet AGH) for providing computer facilities and support within computational grant no. PLG/2024/017405. The XANES measurements at ASTRA beamline were made under the provision of the Polish Ministry of Education and Science project: ‘Support for research and development with the use of research infrastructure of the National Synchrotron Radiation Centre SOLARIS’ under contract nr 1/SOL/2021/2. The further development of the ASTRA beamline for measuring at low photon energies was supported within the EU Horizon2020 programme (952148-Sylinda). The Faculty of Chemistry of the Jagiellonian University is the beneficiary of structural funds from the European Union, grant no. POIG. 02.01.00-12-023/08 "Atomic Scale Science for Innovative Economy (ATOMIN)".

## Table of contents

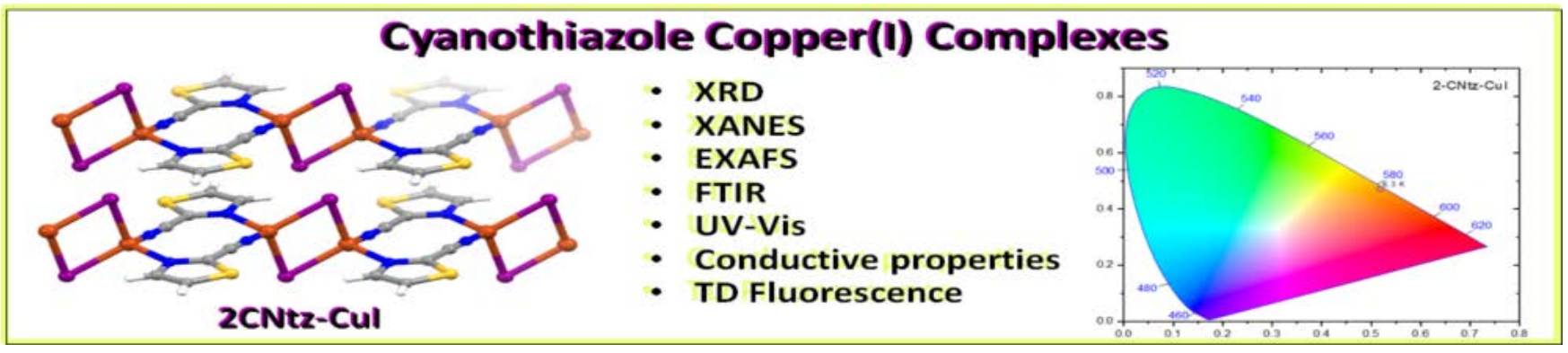

Cyanothiazoles, paired with copper(I) iodide, form complexes with luminescent properties. X-ray crystallography revealed two types of coordination polymers, while spectroscopic, synchrotron and electric measurements supported by DFT calculations studies highlighted their unique electronic features, offering a foundation for future research.

# Supporting information

## Cyanothiazole Copper(I) Complexes: Uncharted Materials with Exceptional Optical and Conductive Properties

Karolina Gutmańska[a], Agnieszka Podborska[b], Tomasz Mazur[b], Andrzej Sławek[b], Ramesh Sivasamy[b], Alexey Maximenko[c], Łukasz Orzeł[d], Janusz Oszajca[d], Grażyna Stochel[d], Konrad Szaciłowski[b,e]* Anna Dołęga[a]*

[a] *Gdansk University of Technology, Chemical Faculty, Department of Inorganic Chemistry, Narutowicza 11/12, 80-233 Gdańsk, Poland*

[b] *AGH University of Krakow, Academic Centre of Materials and Technology, al. Mickiewicza 30, 30-059 Kraków, Poland*

[c] *Jagiellonian University, National Synchrotron Radiation Centre SOLARIS, ul. Czerwone Maki 98, Kraków 30-392, Poland*

[d] *Jagiellonian University in Krakow, Faculty of Chemistry, Gronostajowa 2, Kraków, 30-387 Krakow Poland*

[e] *University of the West of England, Unconventional Computing Lab, Bristol BS16 1QY, United Kingdom*

**Table S1** Crystal and refinement data for **1** - **3**

| Compound | Complex 1 | Complex 2 | Complex 3 |
|---|---|---|---|
| **Empirical formula** | $C_4H_2CuIN_2S$ | $C_4H_2CuIN_2S$ | $C_4H_2CuIN_2S$ |
| **Formula weight (g mol$^{-1}$)** | 300.58 | 300.58 | 300.58 |
| **Wavelength (Å)** | 0.71073 | 0.71073 | 1.54186 |
| **Temperature (K)** | 120(2) | 120(2) | 120(2) |
| **Crystal system** | triclinic | triclinic | triclinic |
| **Space group** | *P* -1 | *P* -1 | *P* -1 |
| ***a*** **(Å)** | 4.1231(4) | 6.6261(6) | 6.4835(7) |
| ***b*** **(Å)** | 6.9179(6) | 7.6830(7) | 7.5760(9) |
| ***c*** **(Å)** | 12.9182(12) | 8.1333(8) | 8.3822(9) |
| ***α*** **(°)** | 92.993(7) | 115.757(7) | 114.401(8) |
| ***β*** **(°)** | 93.993(7) | 108.542(7) | 107.399(8) |
| ***γ*** **(°)** | 95.290(7) | 91.328(8) | 91.821(9) |
| **Volume (Å$^3$)** | 365.37(6) | 347.19(6) | 352.03(7) |
| ***Z*** | 2 | 2 | 2 |
| **Calculated density (g cm$^{-3}$)** | 2.732 | 2.875 | 2.836 |
| **Crystal size (mm)** | 0.188x 0.087x 0.031 | 0.178x 0.124x 0.05 | 0.214x 0.144x 0.069 |
| **Absorption coefficient (mm$^{-1}$)** | 7.405 | 7.793 | 40.82 |
| ***F*(000)** | 276 | 276 | 276 |
| **θ range (°)** | 2.962 to 29.159 | 2.986 to 29.164 | 6.172 to 67.131 |
| **Limiting indices** | -5≤h≤5 | -9≤h≤9 | -7≤h≤7 |
| | -9≤k≤9 | -10≤k≤10 | -8≤k≤8 |
| | -17≤l≤17 | -11≤l≤11 | -9≤l≤9 |
| **Reflections collected / unique/unique [*I>2σ(I)*]** | 4363, 1976, 1808 | 4680, 1864, 1803 | 2650, 1151, 1143 |
| **$R_{int}$** | 0.0228 | 0.0192 | 0.0411 |
| **Completeness to $θ_{max}$ (%)** | 99.2 | 99.6 | 91.7 |
| **Data / restraints / parameters** | 1976 / 0 / 82 | 1864 / 0 / 82 | 1151 / 0 / 82 |
| **Goodness-of-fit on $F^2$** | 1.048 | 1.074 | 1.122 |
| **Final R indices [*I>2σ(I)*]** | $R_1$ = 0.0288 | $R_1$ = 0.0188 | $R_1$ = 0.0599 |
| | $wR_2$ = 0.0751 | $wR_2$ = 0.0491 | $wR_2$ = 0.1725 |
| **R indices (all data)** | $R_1$ = 0.0324 | $R_1$ = 0.0195 | $R_1$ = 0.0601 |
| | $wR_2$ = 0.0774 | $wR_2$ = 0.0496 | $wR_2$ = 0.1727 |
| **Largest diff. peak and hole (e Å$^{-3}$)** | 1.001/-1.148 | 0.966/-0.67 | 1.729/-1.012, |
| **CCDC deposition number** | 2417242 | 2417243 | 2417244 |

## **Table S2** Selected bond lengths and short contacts in angstroms [Å] in complexes **1** - **3**

| **Bonds [Å]/Compound** | **Complex 1** | **Complex 2** | **Complex 3** |
|---|---|---|---|
| **Cu1—I1/I1$^{i/ii/iii/vi}$** | 2.6177(5)/2.6372(6)/2.6390(6) | 2.6709(4)/2.6357(4) | 2.6428(18)/ 2.6776(18)$^{i}$ |
| **Cu1—N1/N2$^{i/vii}$** | 2.050(3) | 2.0647(18)/1.943(2) | 2.072(11)/1.949(10)$^{i}$ |
| **S1—C1/C3/C2** | 1.707(4)/-/1.367(6) | 1.717(2) /1.701(2) | 1.701(12)/1.698(12)$^{i}$ |
| **N1—C1/C2/C3** | 1.311(5)/ -/1.366(5) | 1.324(3)/1.369(3) | 1.290(14)/1.372(14)$^{i}$ |
| **N2—C4** | 1.143(6) | 1.148(3) | 1.151(15) |
| **C2—C3/C4** | 1.367(6)/1.425(6) | 1.368(3) | 1.389(17)/1.434(15)$^{i}$ |
| **Cu1—Cu1$^{iv}$/ Cu1$^{v}$** | 2.7460(10)/2.7503(10) | - | - |
| **C4—C1** | - | 1.423(3) | - |
| **Cu···Cu$^{ii/vi}$** | - | 3.173 | 3.083 |
| **S···I$^{vii/ix}$** | - | 3.714 | 3.632/3.665 |

Symmetry operations: $^{i}$: 2-x, 1-y, 2-z; $^{ii}$: 1+x, y, z; $^{iii}$: 1-x,1-y, -z; $^{iv}$: 2-x,1-y, -z; $^{v}$:1-x, 1-y, -z; $^{vi}$:2-x,2-y,2-z; $^{vii}$:2-x,1-y,1-z; $^{viii}$:-1+x, -1+y, -1+z; $^{ix}$: 1-x, 1-y, 2-z

## **Table S3** Selected angles in degrees [°] in complexes **1** - **3**

| **Angles [°]/Compound** | **Complex 1** | **Complex 2** | **Complex 3** |
|---|---|---|---|
| **N1—C1—S1** | 115.5(3) | 115.1(2) | 117(1) |
| **C3—C2—C4** | 126.8(4) | 124(1) | 124(1) |
| **C3—C2—S1** | 110.7(3) | 110.7(2) | 109(1) |
| **C4—C2/C1—S1** | 122.4(3) | 122.6(2) | - |
| **C2/C3/C4—C1/C2/C3—N1** | 114.1(4) | 115.1(2)/122.2(2) | 116(1)/119(1) |
| **C1/C2/C4—C2/C4—N2** | 179.5(6) | 176.5(3) | 178(1) |
| **I1/I1$^{i/ii/iv}$—Cu1—N1/N1$^{vi}$/N2$^{vii}$** | 109.0(1)/104.72 | 101.59(6)/103.52(6)/ 110.82(7)/118.49(7) | 102.9(3)/114.8(3)/ 101.7(3)/112.3(3) |
| **N1—Cu1—Cu1$^{iii/iv}$** | 123.8(1)/ 119.3(1) | - | - |
| **I1/I1$^{iii/v/vii/viii}$—Cu1/Cu1$^{iii/iiiv}$—Cu1/Cu1$^{v/viii}$** | 58.89(1)/58.61(1)/58.13(1)/ 131.12(2)/58.55(1)/131.24(2) | - | - |
| **I1/I1$^{ii}$—Cu1/Cu1$^{i/iv/v}$—I1/I1$^{i/v/vi}$** | 103.37(2)/117.16(2)/117.02(2) | 106.55(1) | 109.18(7)/109.18(7) |
| **Cu1$^{iv}$—Cu1—Cu1$^{v}$** | 97.21(2) | - | - |
| **Cu1—I1/I1$^{i/ii/iv}$—Cu1$^{i/iv/v/vi/vii}$** | 62.84(2)/62.98(2)/103.37(2) | 73.45(1)/73.45(1) | 70.82(6)/70.82(6) |
| **C1—N1—C2/C3** | 111.1(3) | 109.9(2) | 109(1) |
| **C1/C2/C3/C4—N1/N2—Cu1/Cu$^{v/vii}$** | 122.4(3)/126.5(3) | 129.7(2)/119.7(2)/ 168.3(2) | 120.0(9)/130.9(9)/ 170(1) |
| **C1—S1—C2/C3** | 88.6(2) | 89.2(1) | 89.6(6) |
| **N1—Cu1—N2** | - | 114.22(9) | 114.8(4) |

Symmetry operations: $^{i}$: 2-x, 1-y, 2-z; $^{ii}$: 1+x, y, z; $^{iii}$: 1-x,1-y, -z; $^{iv}$: 2-x,1-y, -z; $^{v}$:1-x, 1-y, -z; $^{vi}$:2-x,2-y,2-z; $^{vii}$:2-x,1-y,1-z; $^{vii}$: -1+x,y,z

## **Table S4** π···π stacking interactions in complexes **1** - **3**

| **Compound** | **π···π stacking interactions** | **Distances between Cg-Cg [Å]** | **Cg definition** |
|---|---|---|---|
| 1 | $Cg_{Tz1}$···$Cg_{Tz2}$ | 4.123 | $Cg_{Tz1}$: N1–C1–S1–C2–C3<br>$Cg_{Tz2}$: N1i–C1i–S1i–C2i–C3i |
| 2 | $Cg_{Tz3}$···$Cg_{Tz4}$ | 3.783 | $Cg_{Tz1}$: N1–C3–C2–S1–C1<br>$Cg_{Tz2}$: N1$^{ii}$–C3$^{ii}$–C2$^{ii}$–S1$^{ii}$–C1$^{ii}$ |
| 3 | $Cg_{Tz4}$···$Cg_{Tz5}$ | 3.882 | $Cg_{Tz1}$: N1–C2–C3–S1–C1<br>$Cg_{Tz2}$: N1$^{ii}$–C2$^{ii}$–C3$^{ii}$–S1$^{ii}$–C1$^{ii}$ |

$Cg_{Tz}$ - centroid of the thiazole ring, Symmetry operations: $^{i}$:-1+x,y,z; $^{ii}$:1-x,1-y,1-z

**Table S5** Geometrical parameters of C–H···X hydrogen bonds in complexes **1** – **3**

| **Compound** | bond | ***D···A* [Å]** | **H···*A* [Å]** | ***D*–H [Å]** | ***ΔDHA* [°]** | **Type of interaction** |
|---|---|---|---|---|---|---|
| **1** | C1–H1···I1$^{i}$ | 3.749(4) | 3.0588 | 0.950 | 130.8 | |
| | C3–H3···N2$^{ii}$ | 3.413(7) | 2.558 | 0.950 | 140.0 | intermolecular |
| **2** | C3–H3···I1$^{iii}$ | 3.793(2) | 3.1721 | 0.950 | 124.6 | |
| | C2–H2···I1$^{iii}$ | 3.776(2) | 3.1677 | 0.950 | 123.5 | |
| | C2–H2···I1$^{iv}$ | 3.753(2) | 3.1364 | 0.950 | 124.2 | |
| **3** | C1–H1···I1$^{v}$ | 3.71(2) | 3.1321 | 0.95 | 120.89 | |
| | C3–H3···I1$^{vi}$ | 3.81(2) | 3.1005 | 0.95 | 129.95 | |

Symmetry operations: $^{i}$: 1-x, -1+y, z; $^{ii}$: 1-x, 1-y, 1-z; iii:-1+x, y,z; $^{iv}$:1-z,2-y, 1-z; $^{v}$:1-x,1-y,2-z; $^{vi}$:-1+x,y, -1+z

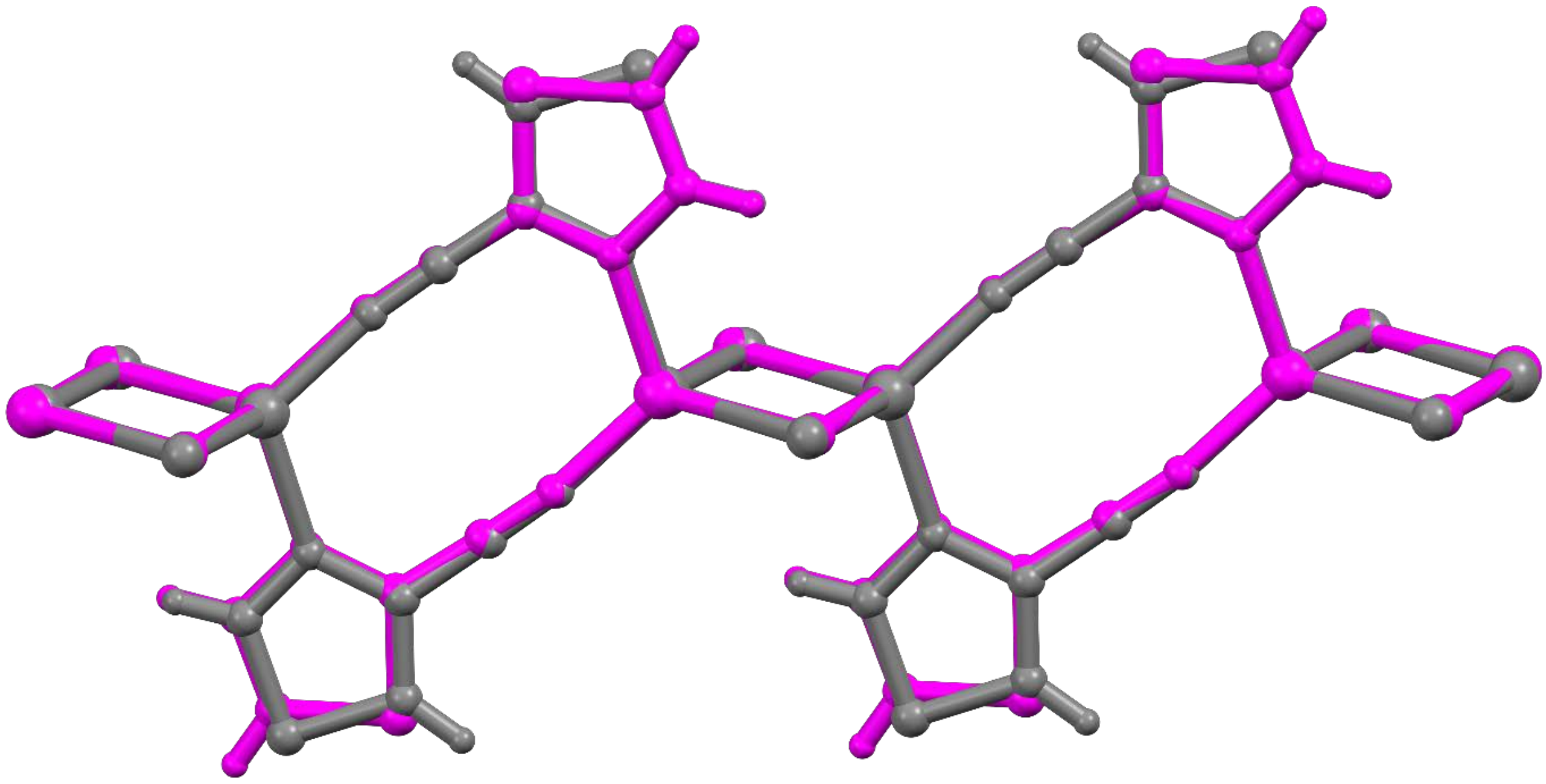

**Figure S1**. The overlay of molecular structures of **2** (magenta) and **3** (grey) illustrating the similar size of both rings in the studied complexes. The overlay prepared in the program Mercury.

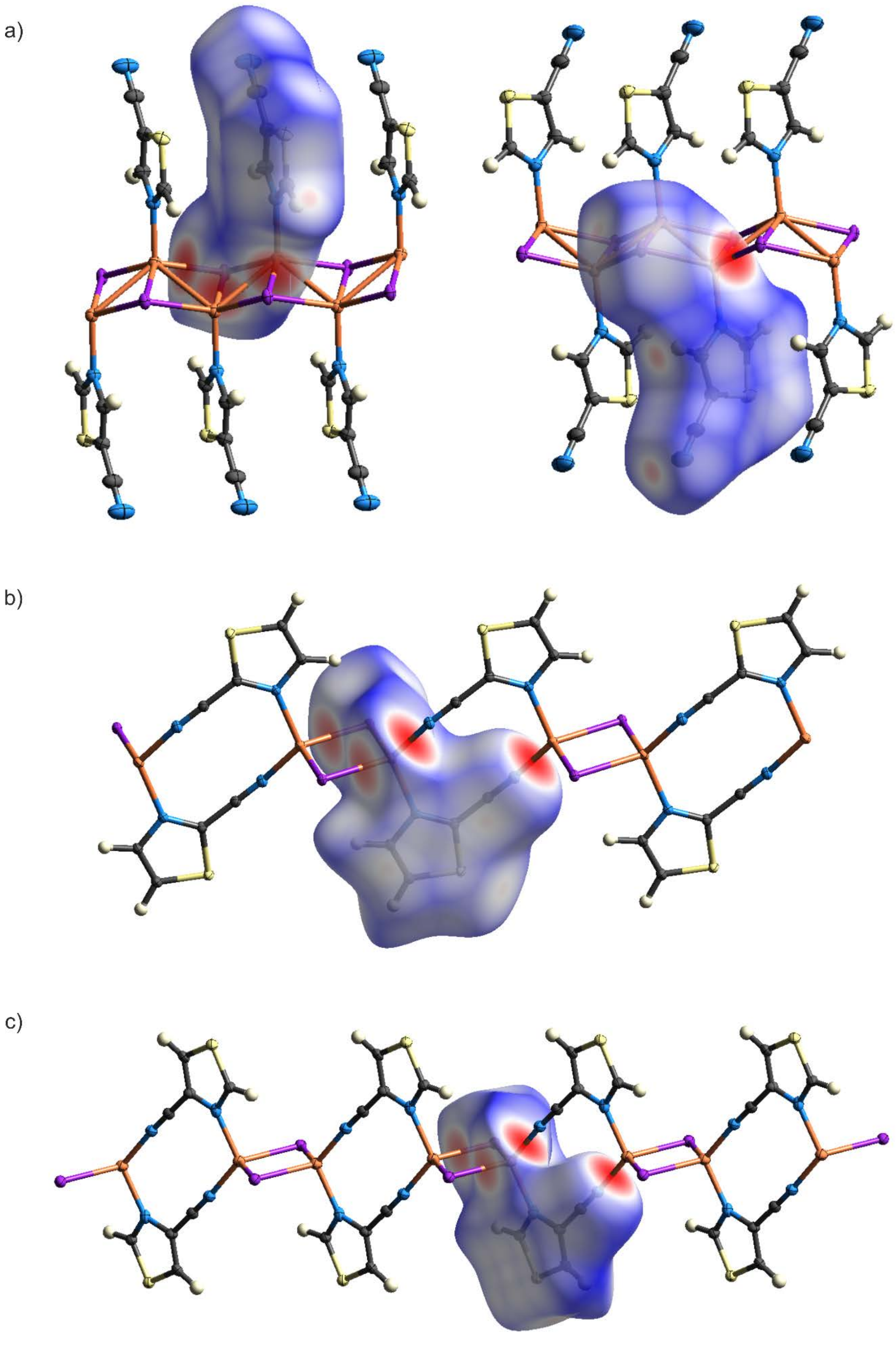

**Figure S2**. The fragments of **1** - **3** selected for the calculation of Hirshfeld surfaces and HSs: a) two views of the HS of complex **1**; b) HS of complex **2**; c) HS of complex **3**.

**Table S6** The fingerprint plots for specific intermolecular interactions in: a) **1**, b) **2** and c) **3**.

| **Compound 1** | **Compound 2** | **Compound 3** |
|---|---|---|
| **a)** C-H···I(H···I) | | |
| 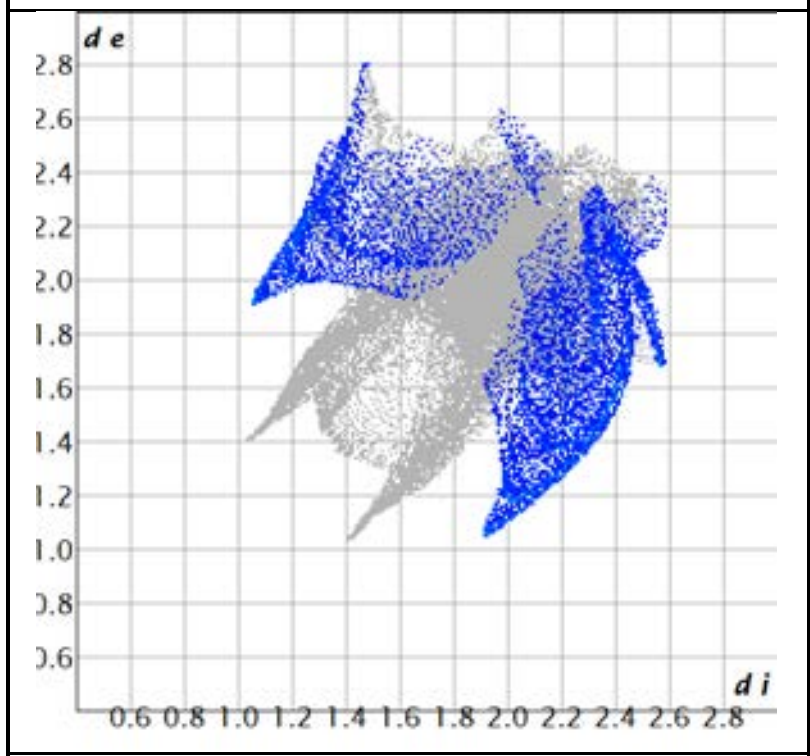 | 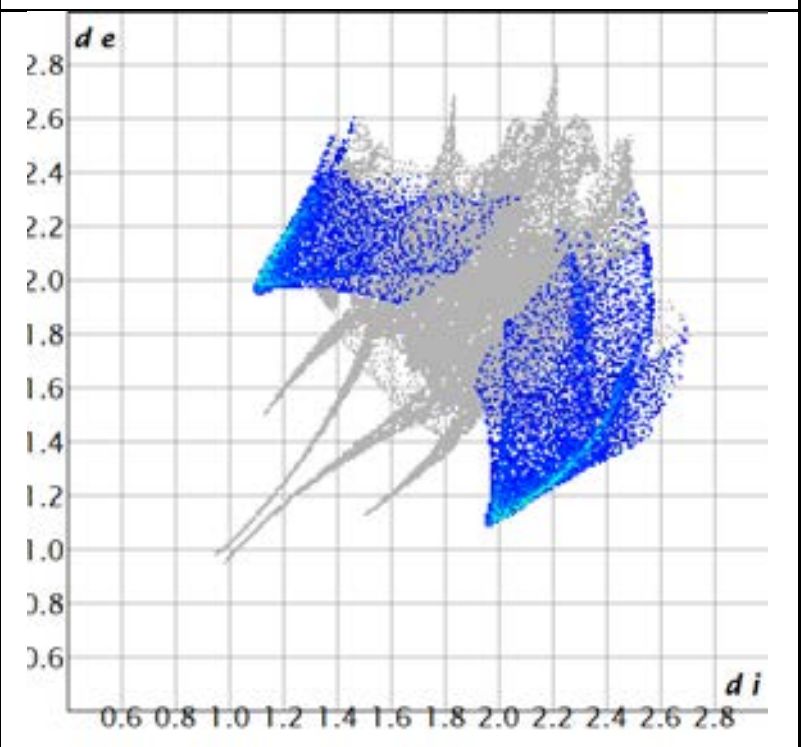 | 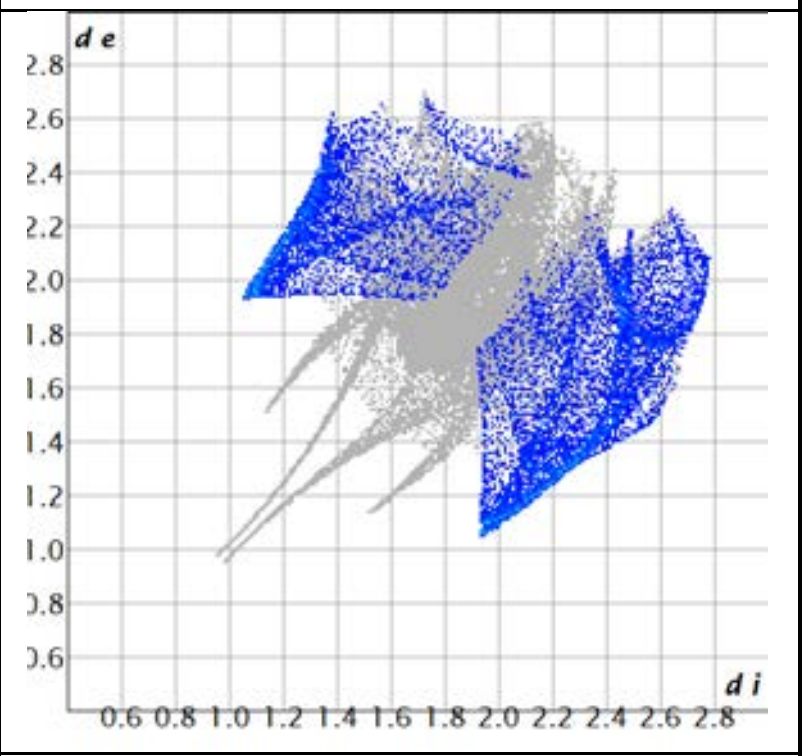 |
| b) C-H···N | | |
| 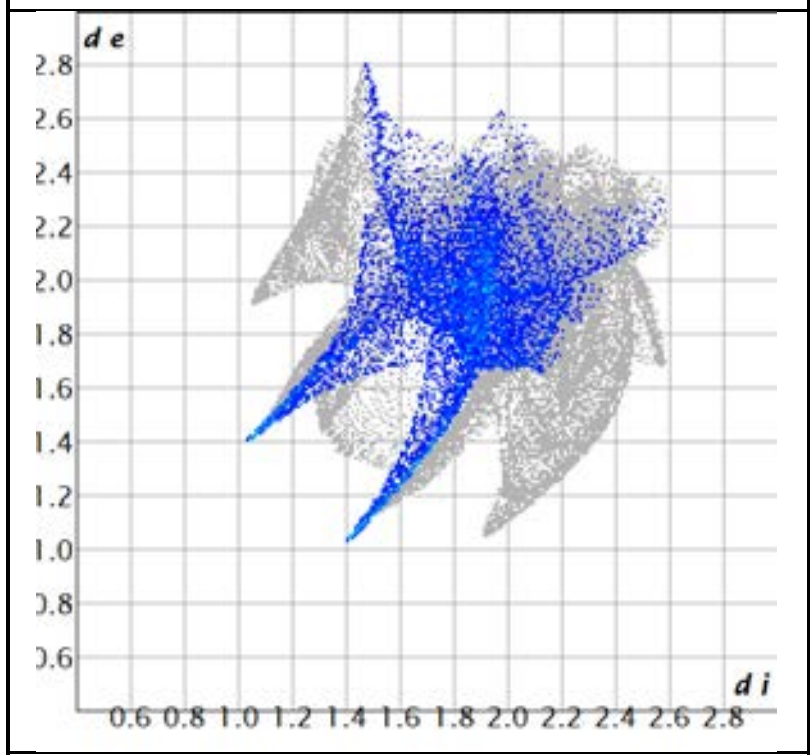 | 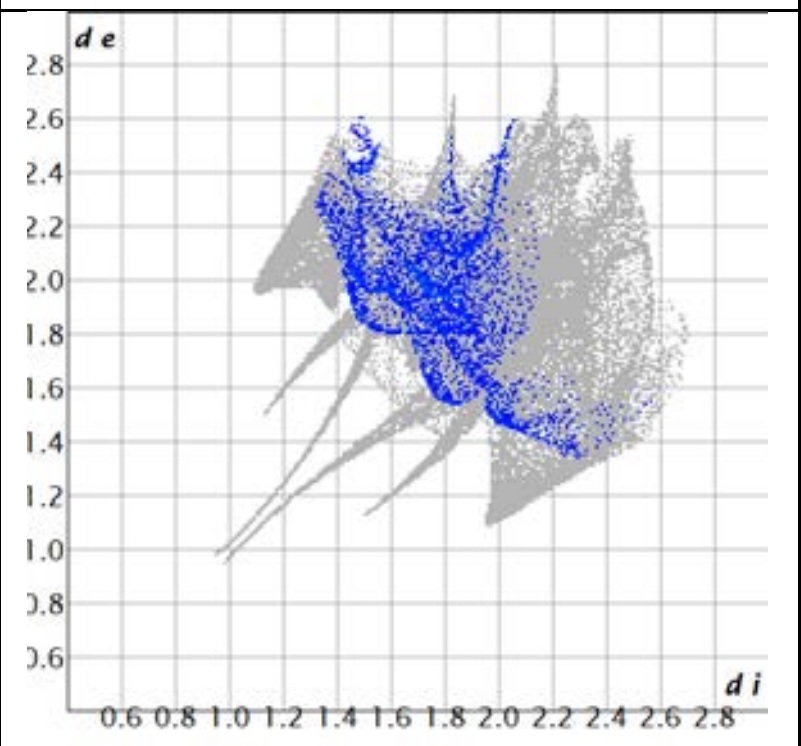 | 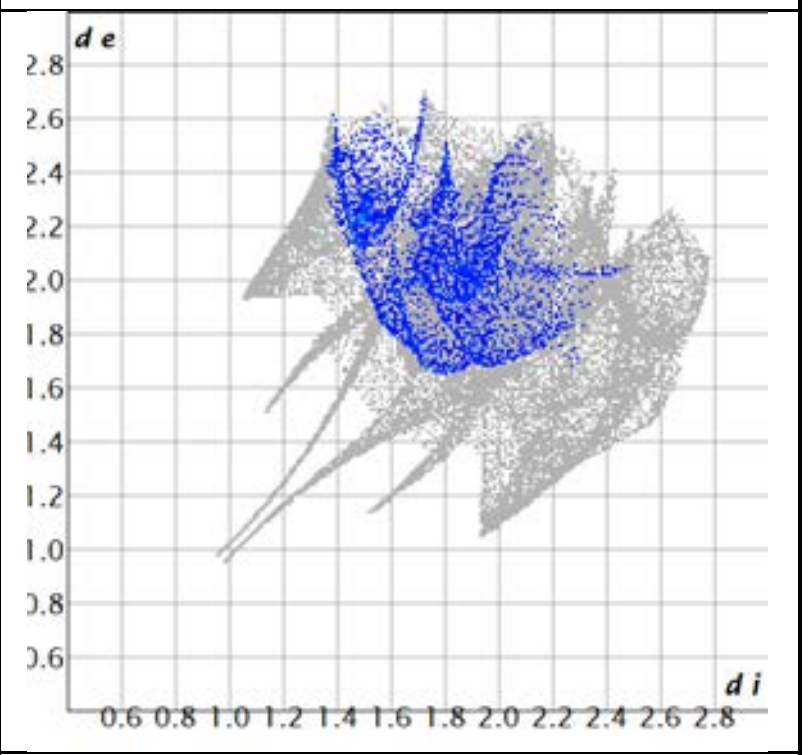 |
| c) I···S | | |
| 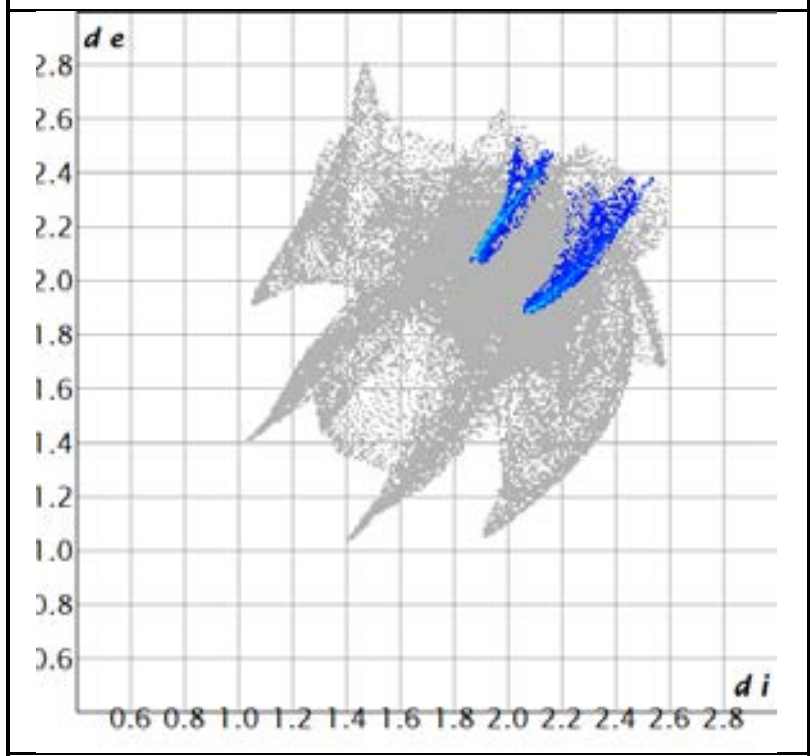 | 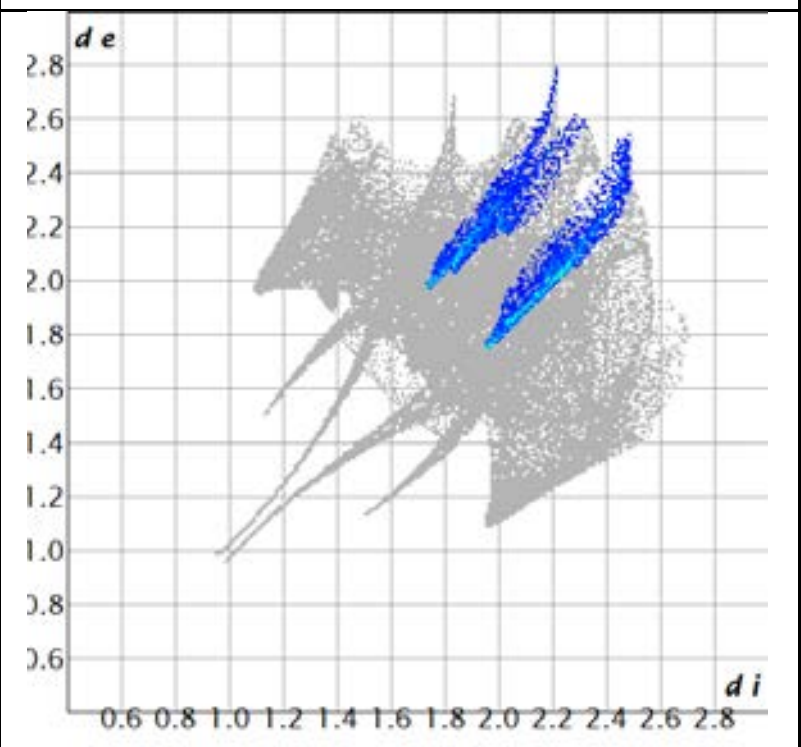 | 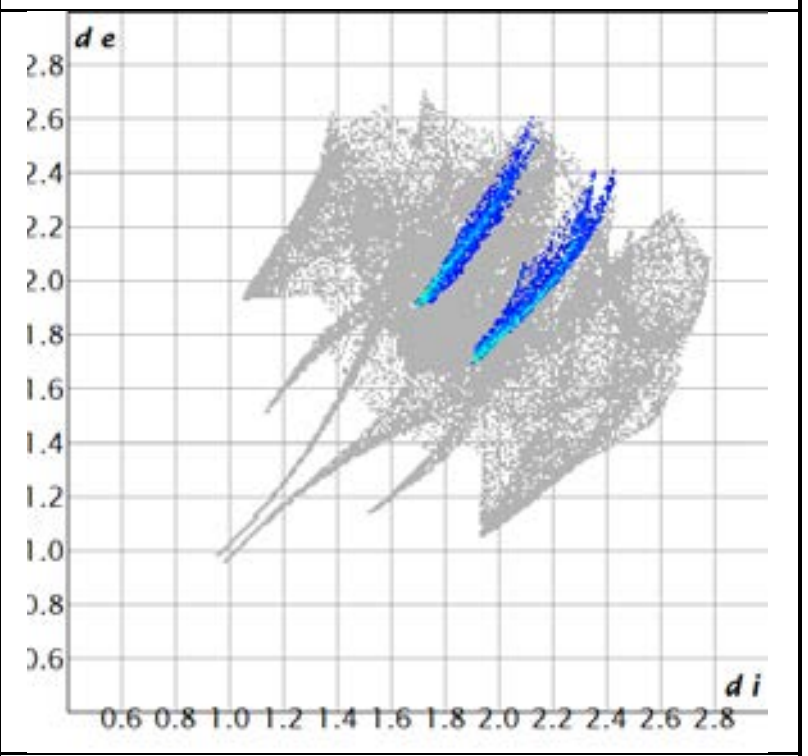 |
| d) C···H | | |
| 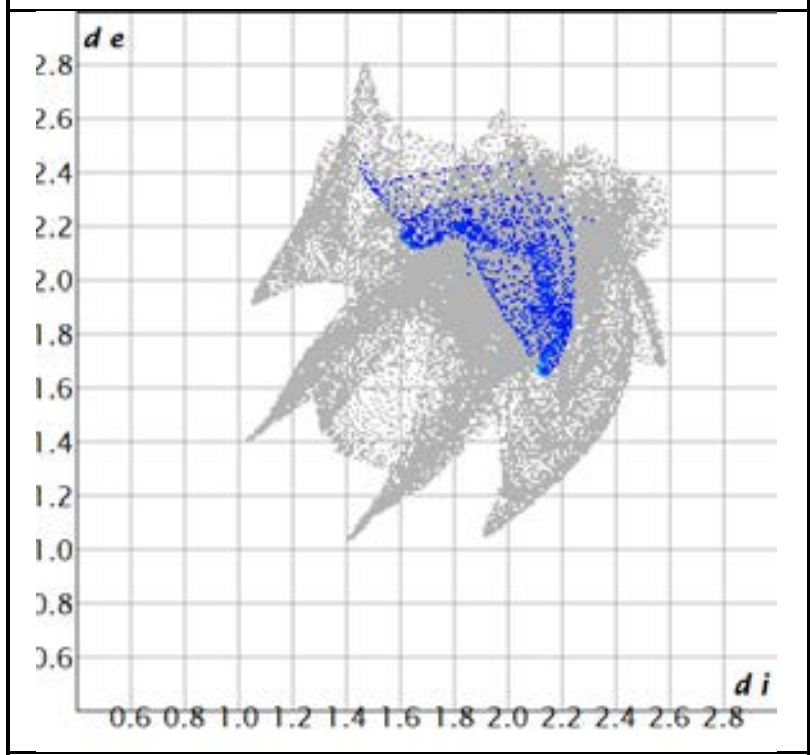 | 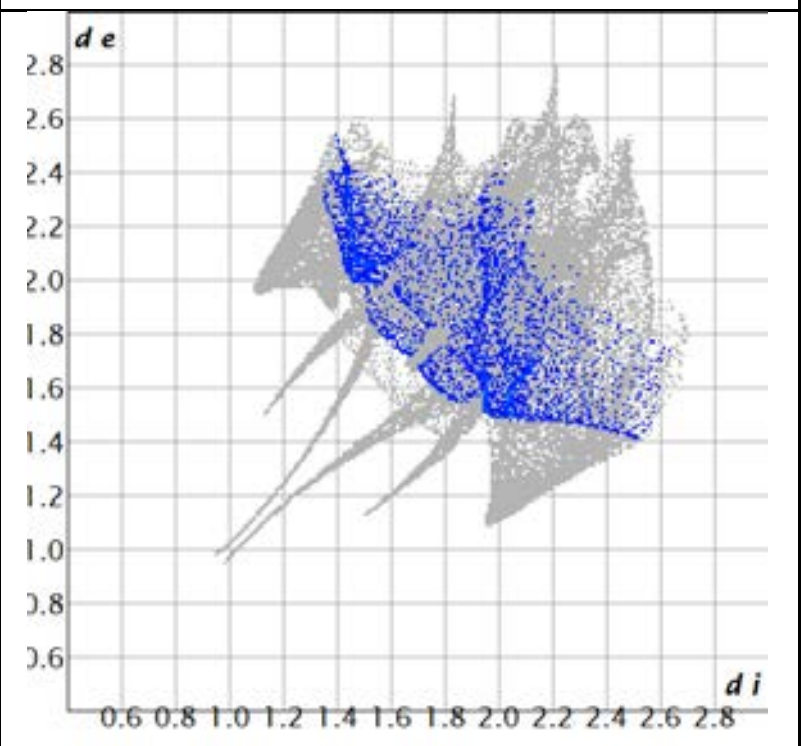 | 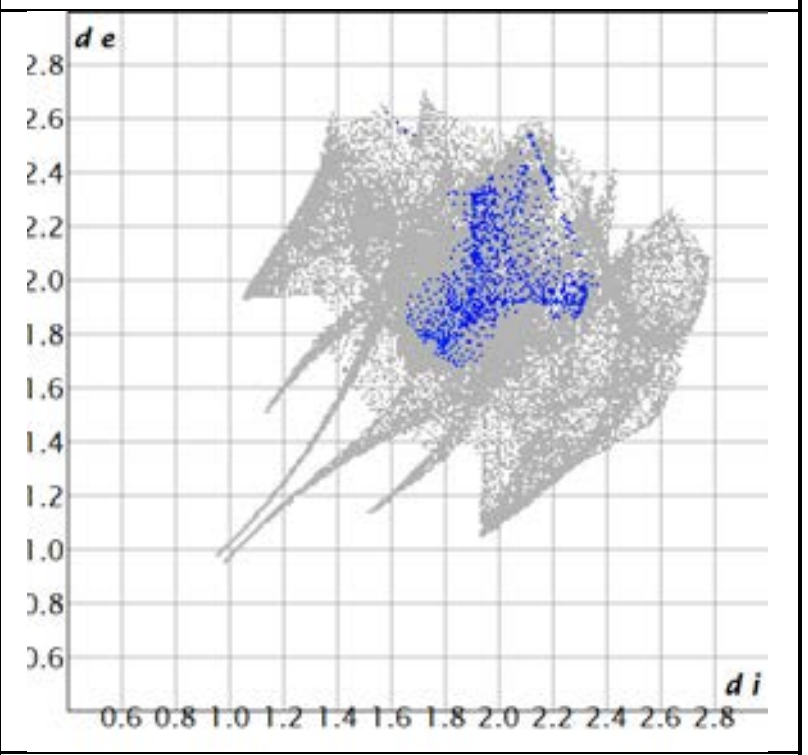 |
| e) C···C | | |

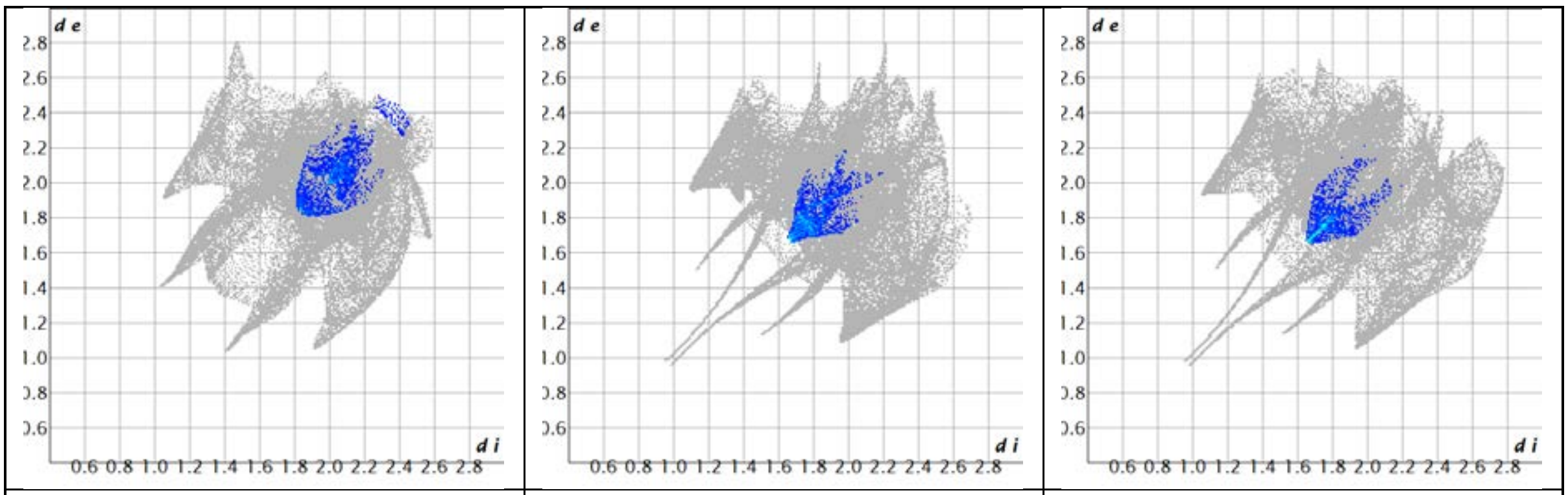

f) N···C

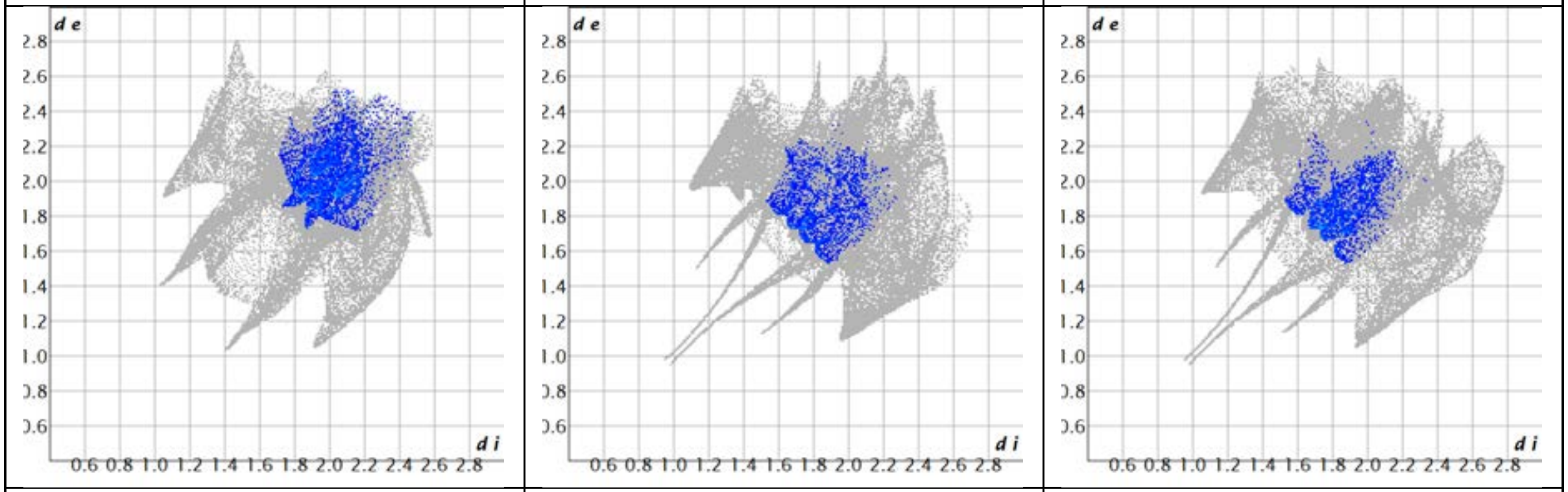

g) S···C

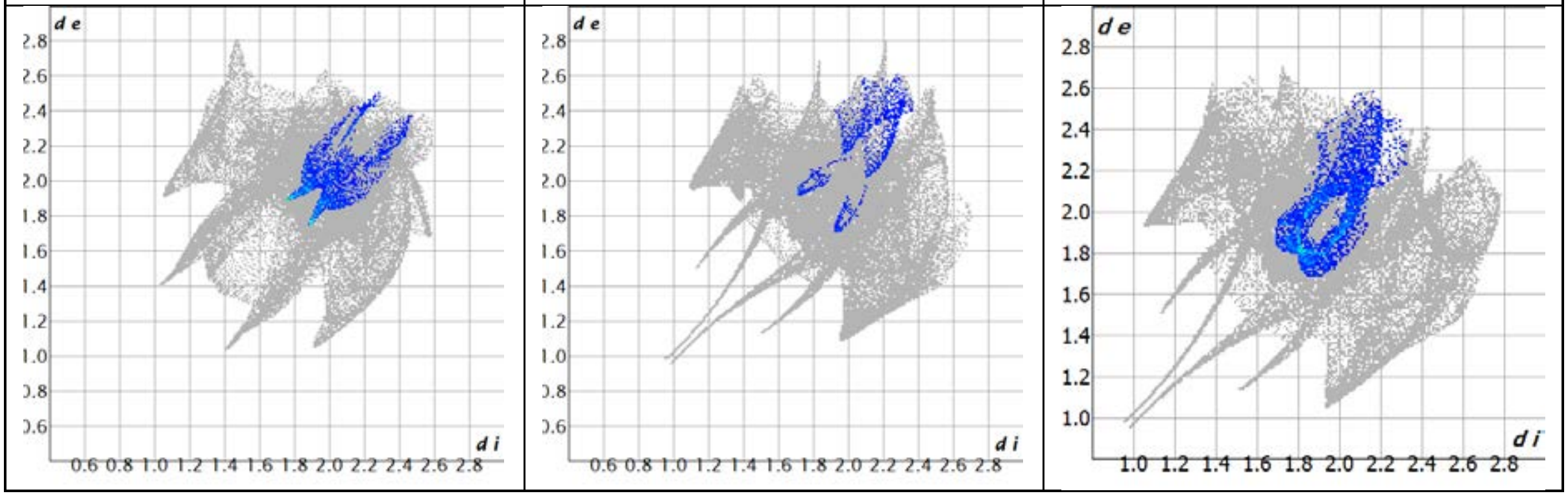

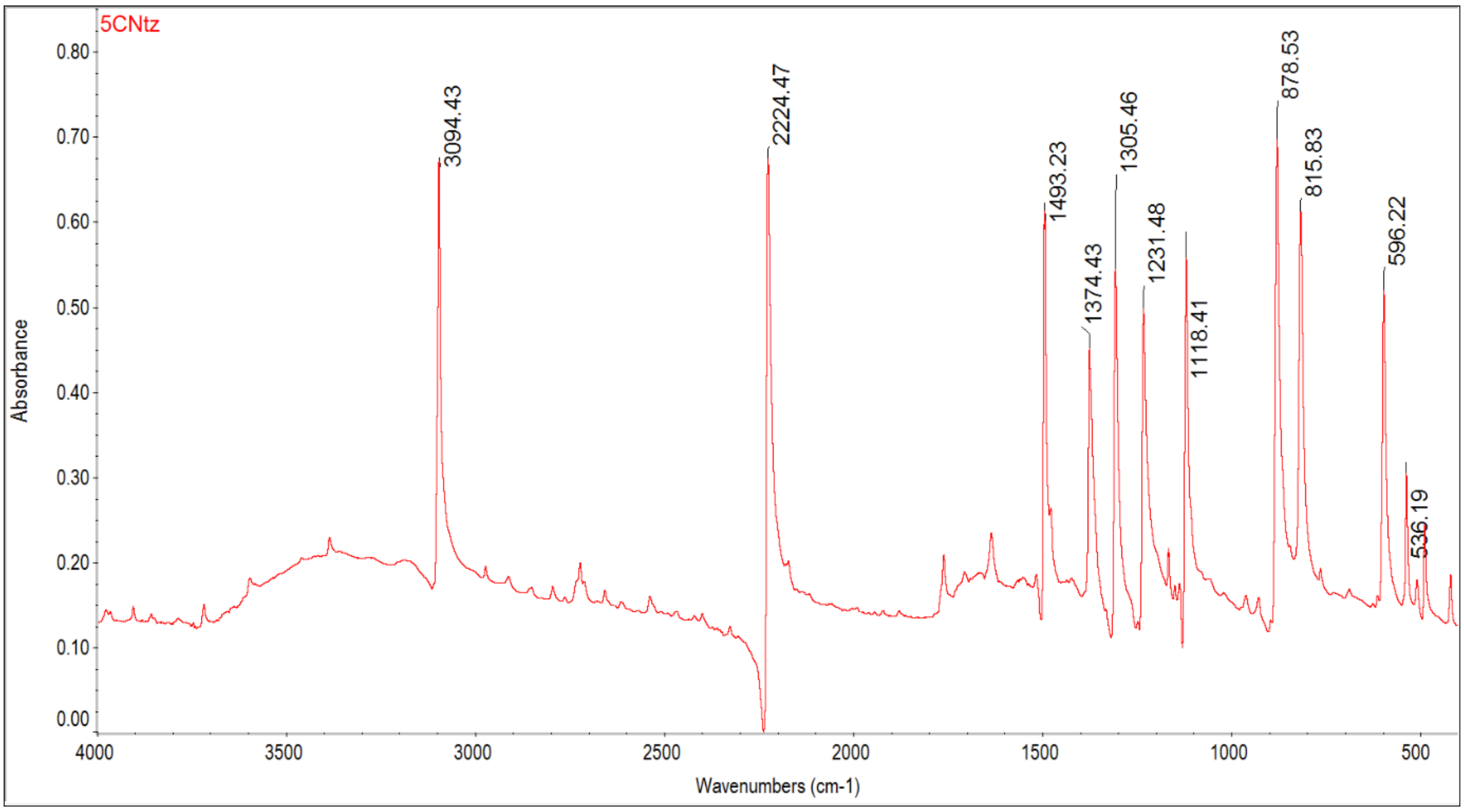

**Figure S3**. FTIR spectra of 5-cyanothiazole.

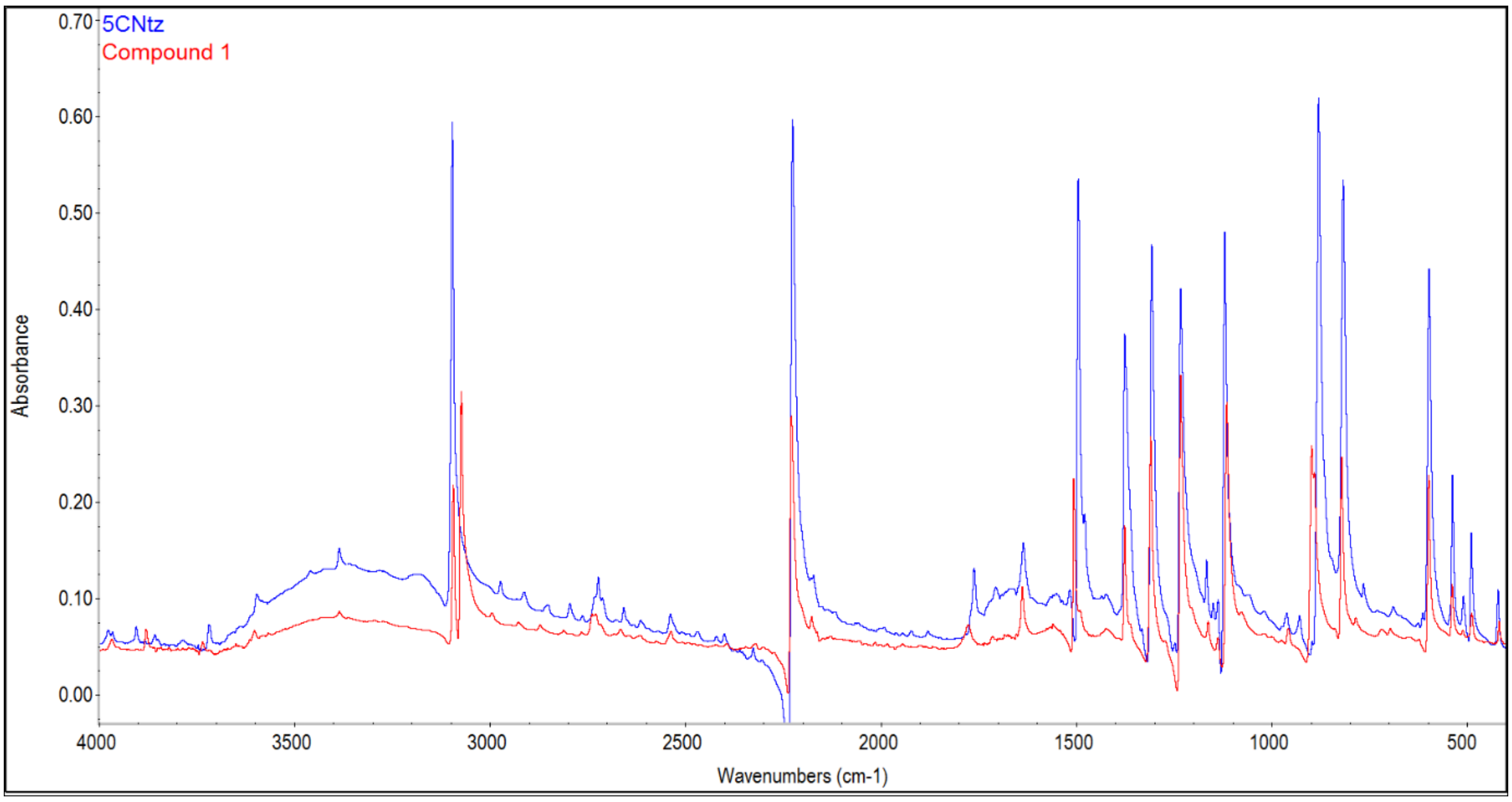

**Figure S4**. Superimposed FTIR spectra of compound **1** and **5CNtz**.

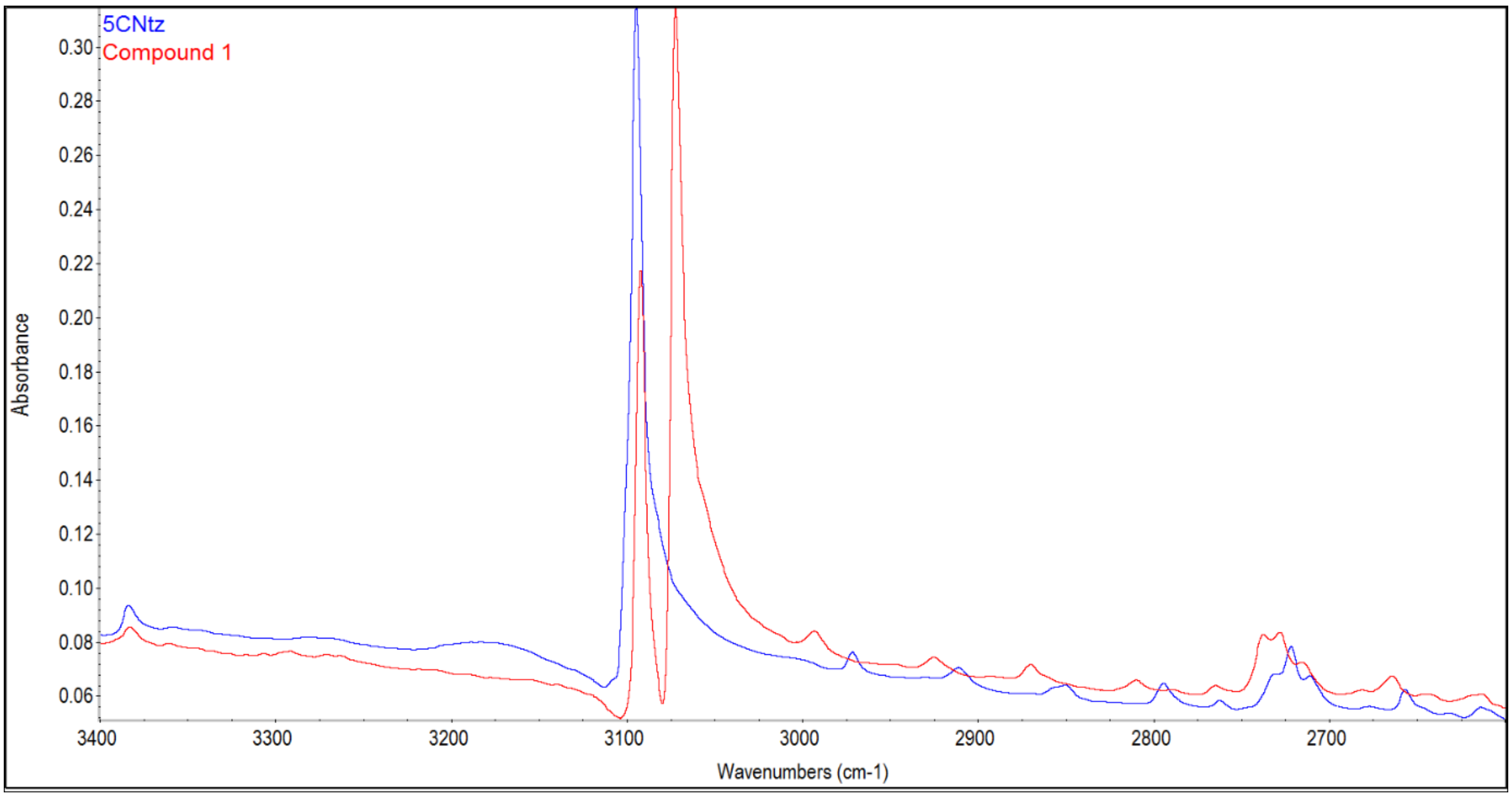

**Figure S5** Superimposed FTIR spectra of compound **1** and **5CNtz** in the range of 3400-2600 $cm^{-1}$.

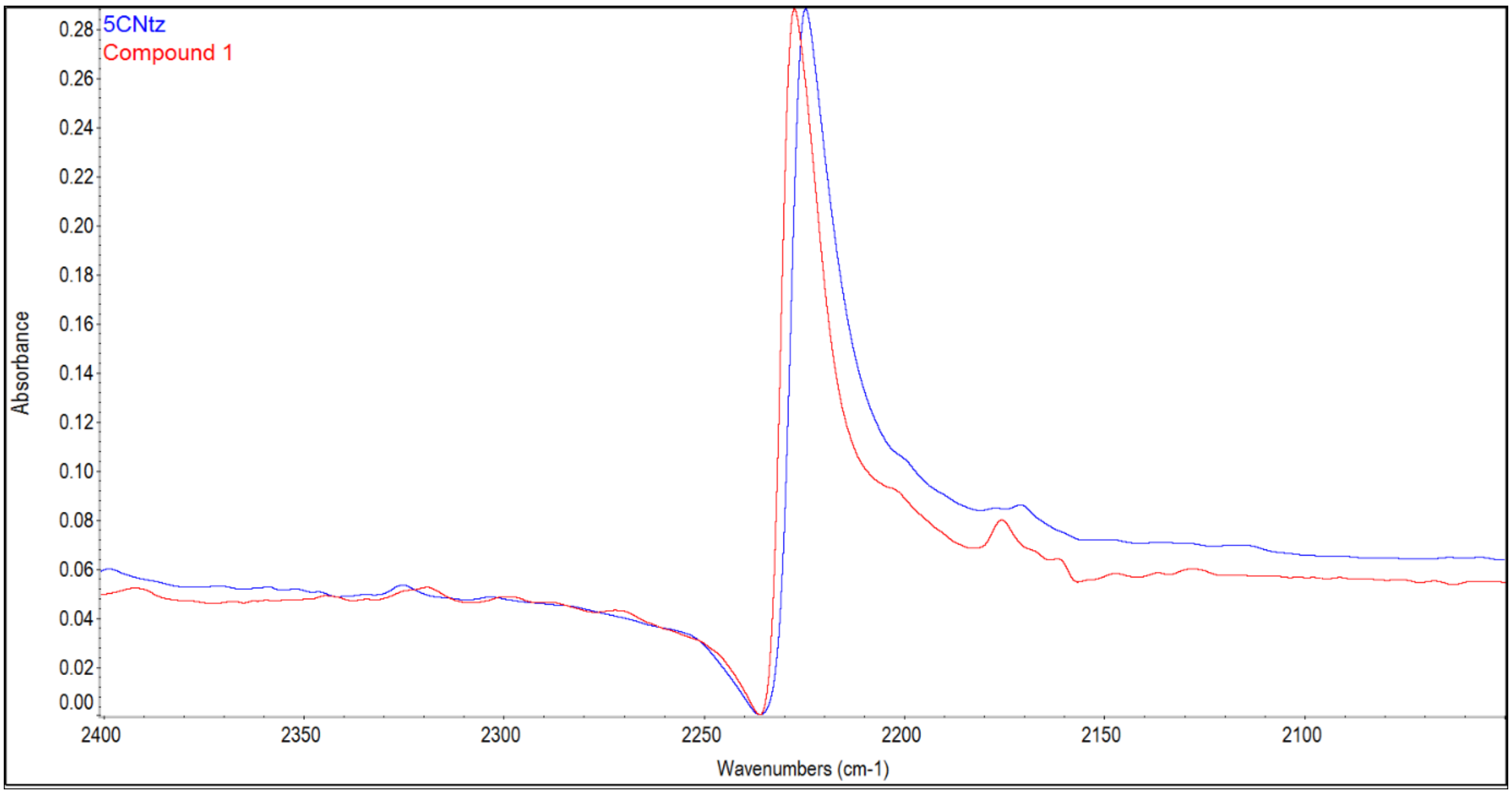

**Figure S6** Superimposed FTIR spectra of compound **1** and **5CNtz** at the range of 2400-2050 $cm^{-1}$.

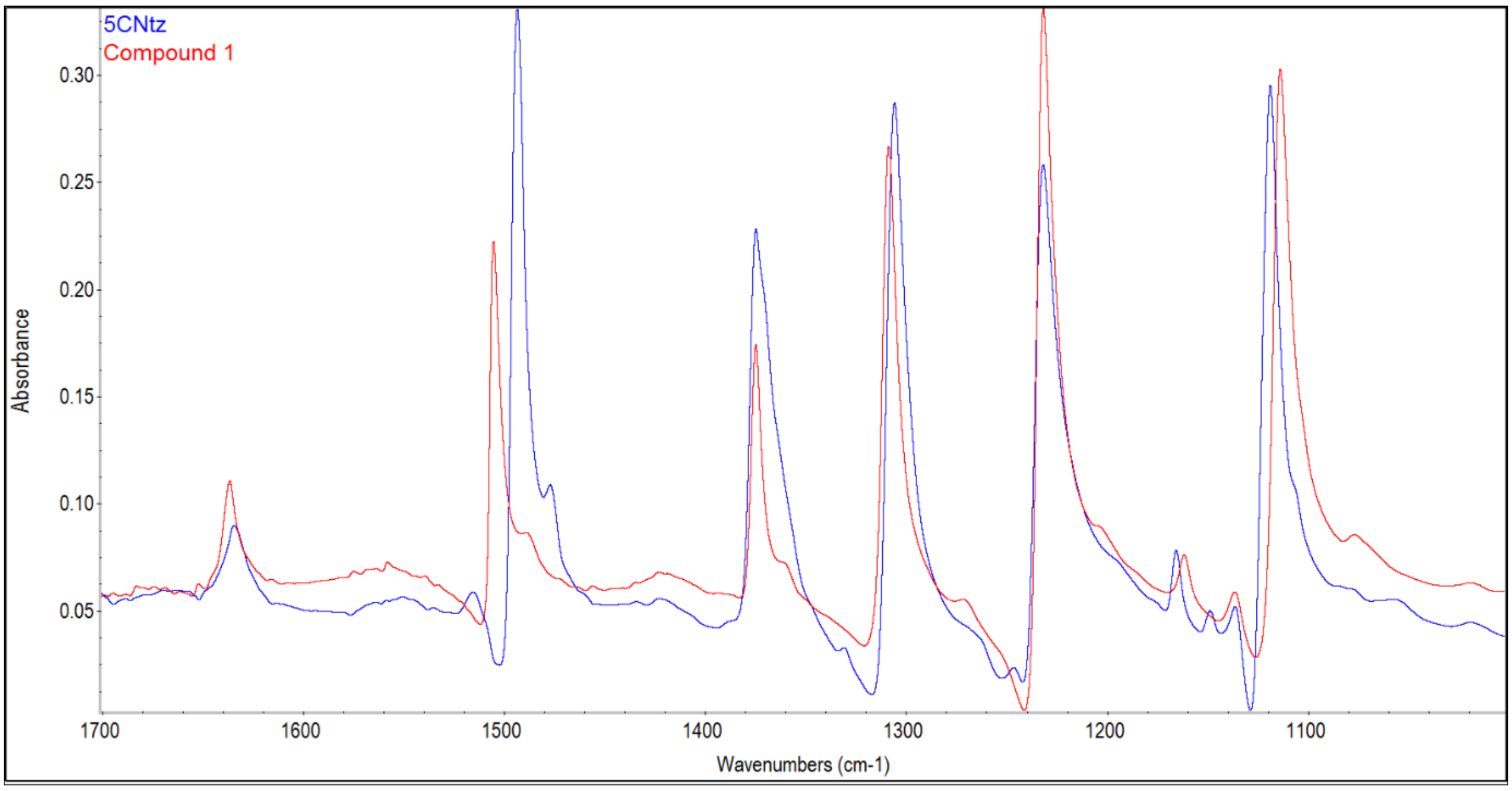

**Figure S7** Superimposed FTIR spectra of compound **1** and **5CNtz** in the range of 1700-1000 $cm^{-1}$.

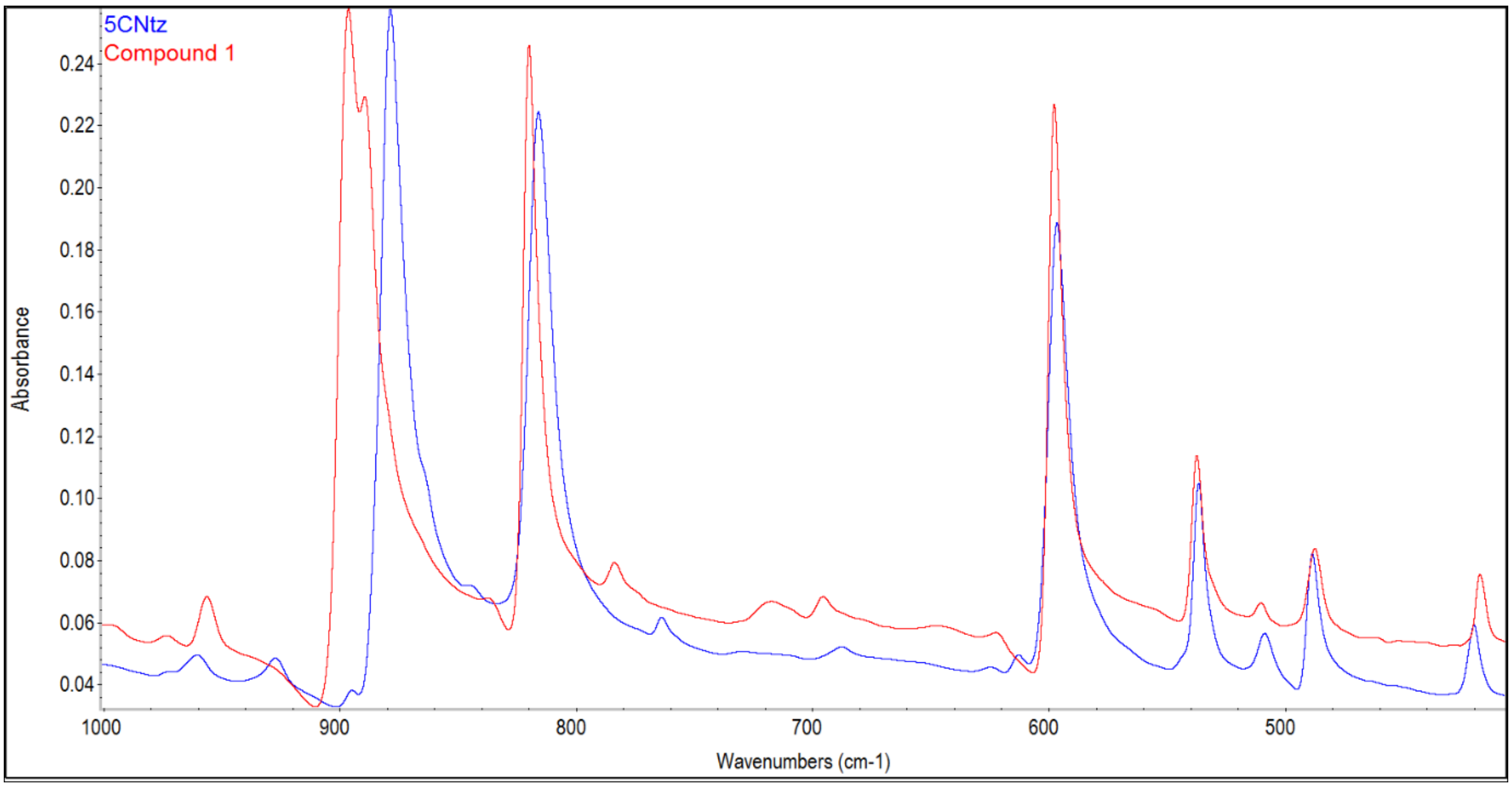

**Figure S8** Superimposed FTIR spectra of compound **1** and **5CNtz** in the range of 1000-400 $cm^{-1}$.

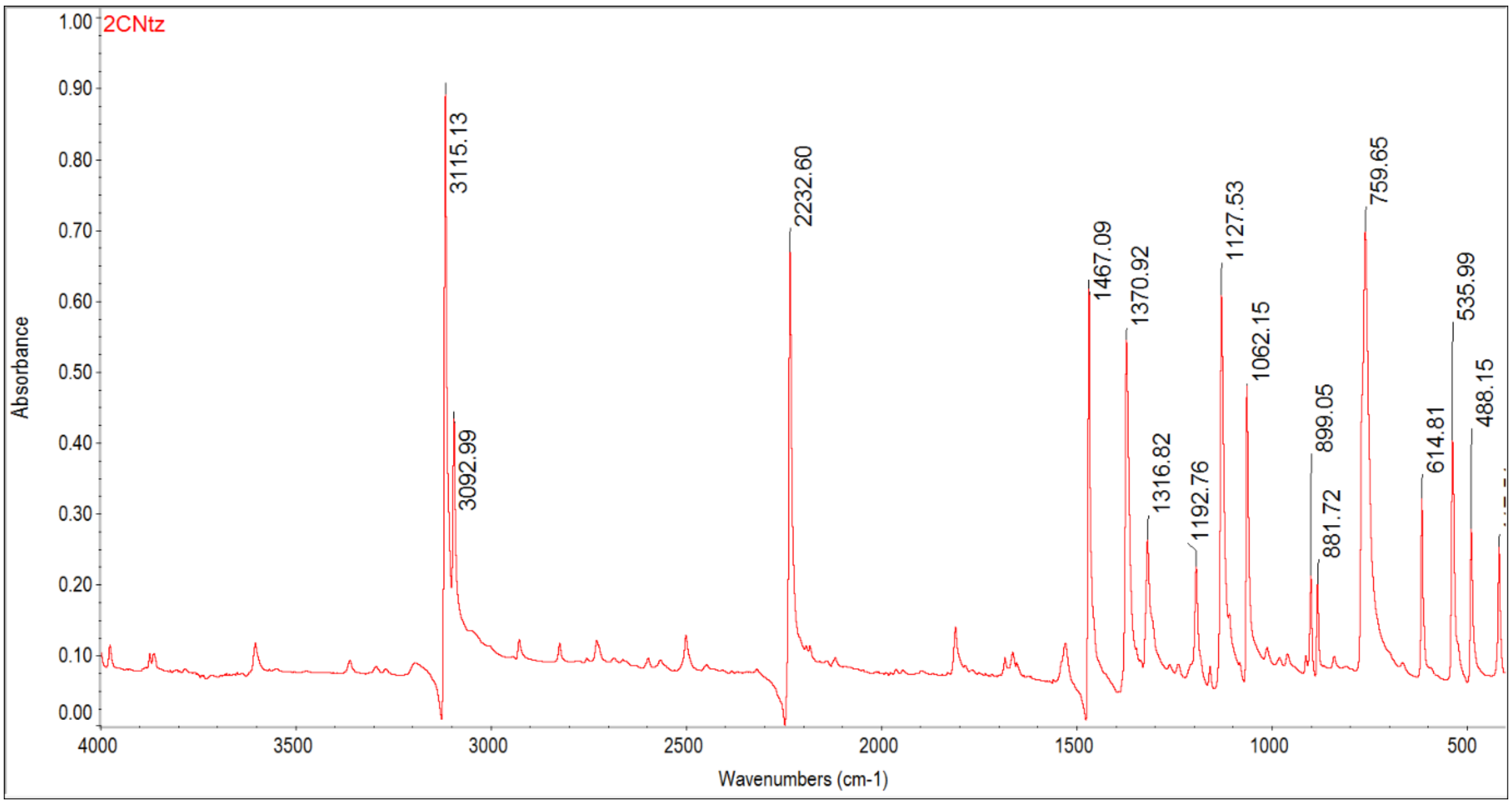

**Figure S9**. FTIR spectra of 2-cyanothiazole.

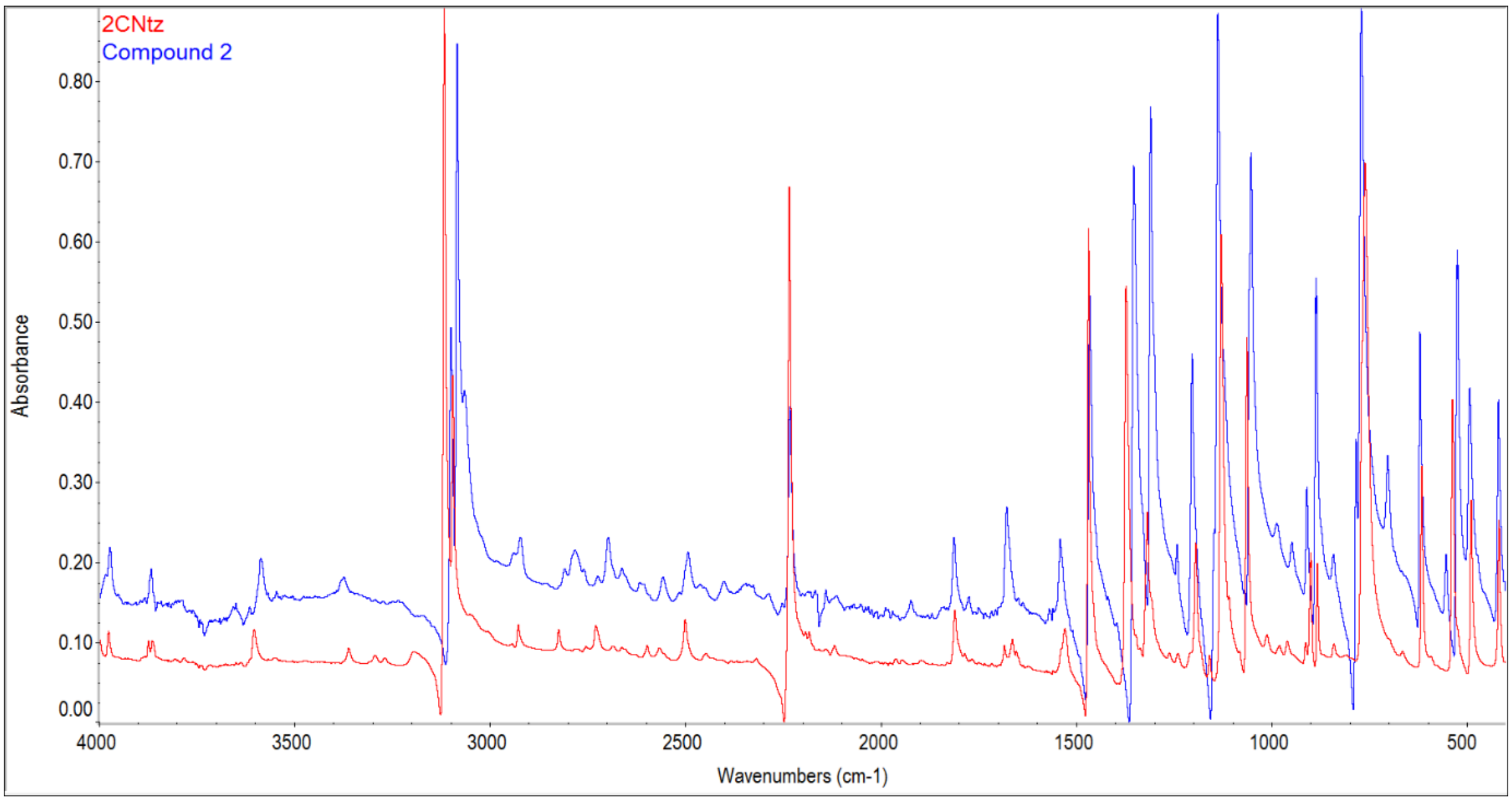

**Figure S10**. Superimposed FTIR spectra of compound **2** and **2CNtz**.

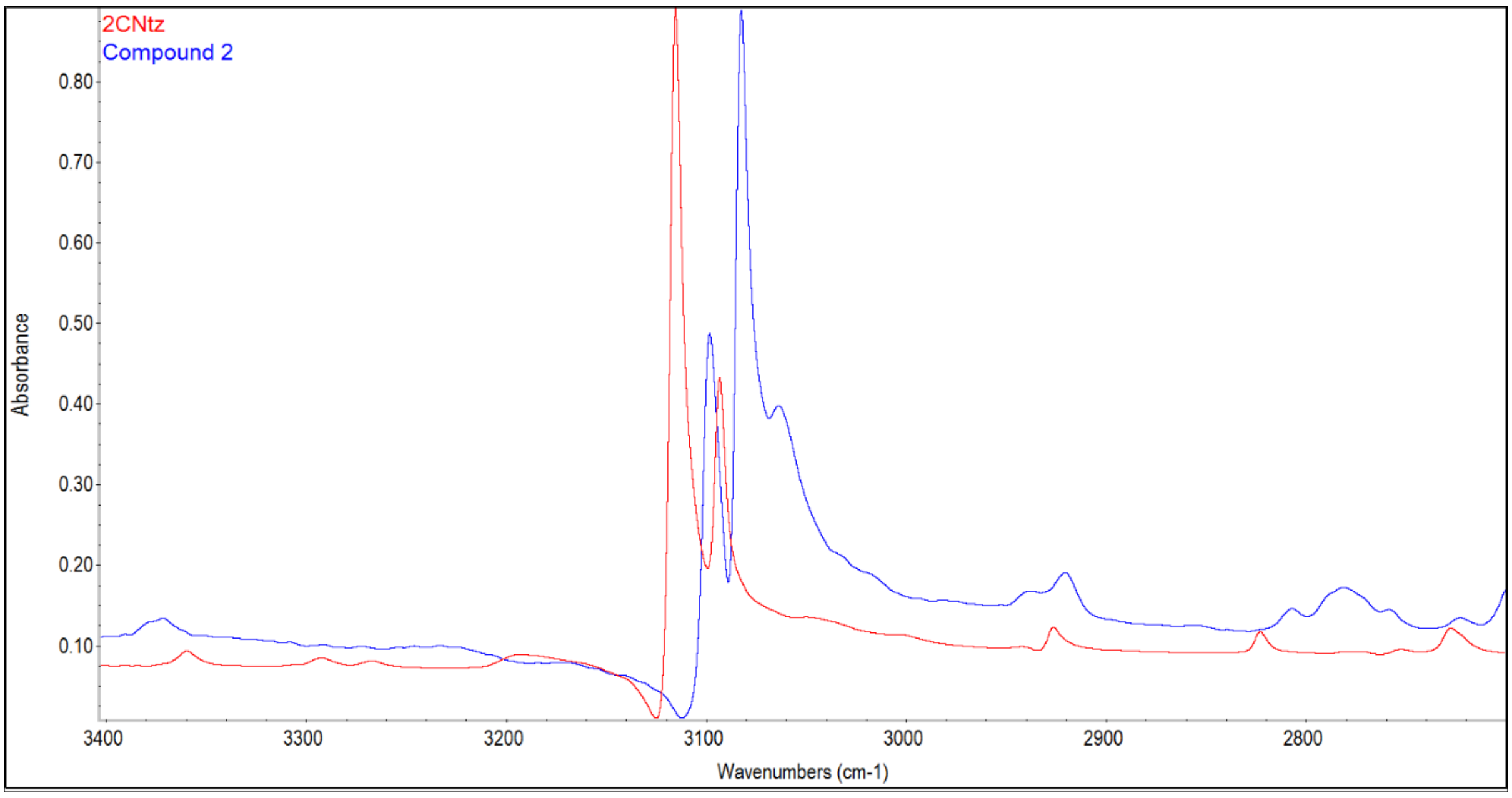

**Figure S11.** Superimposed FTIR spectra of compound **2** and **2CNtz** in the range of 3400-2700 $cm^{-1}$.

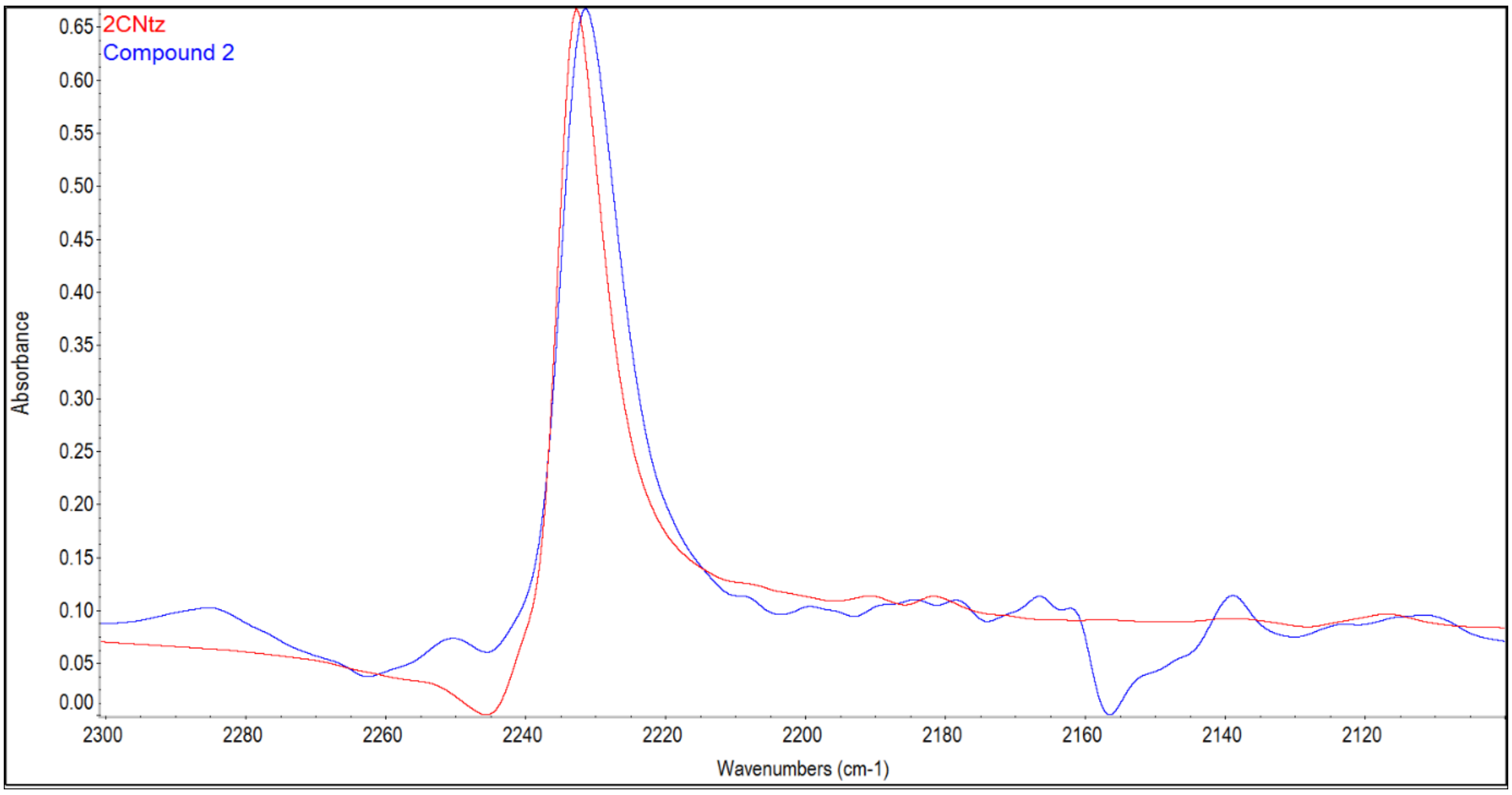

**Figure S12.** Superimposed FTIR spectra of compound **2** and **2CNtz** in the range of 2300-2100 $cm^{-1}$.

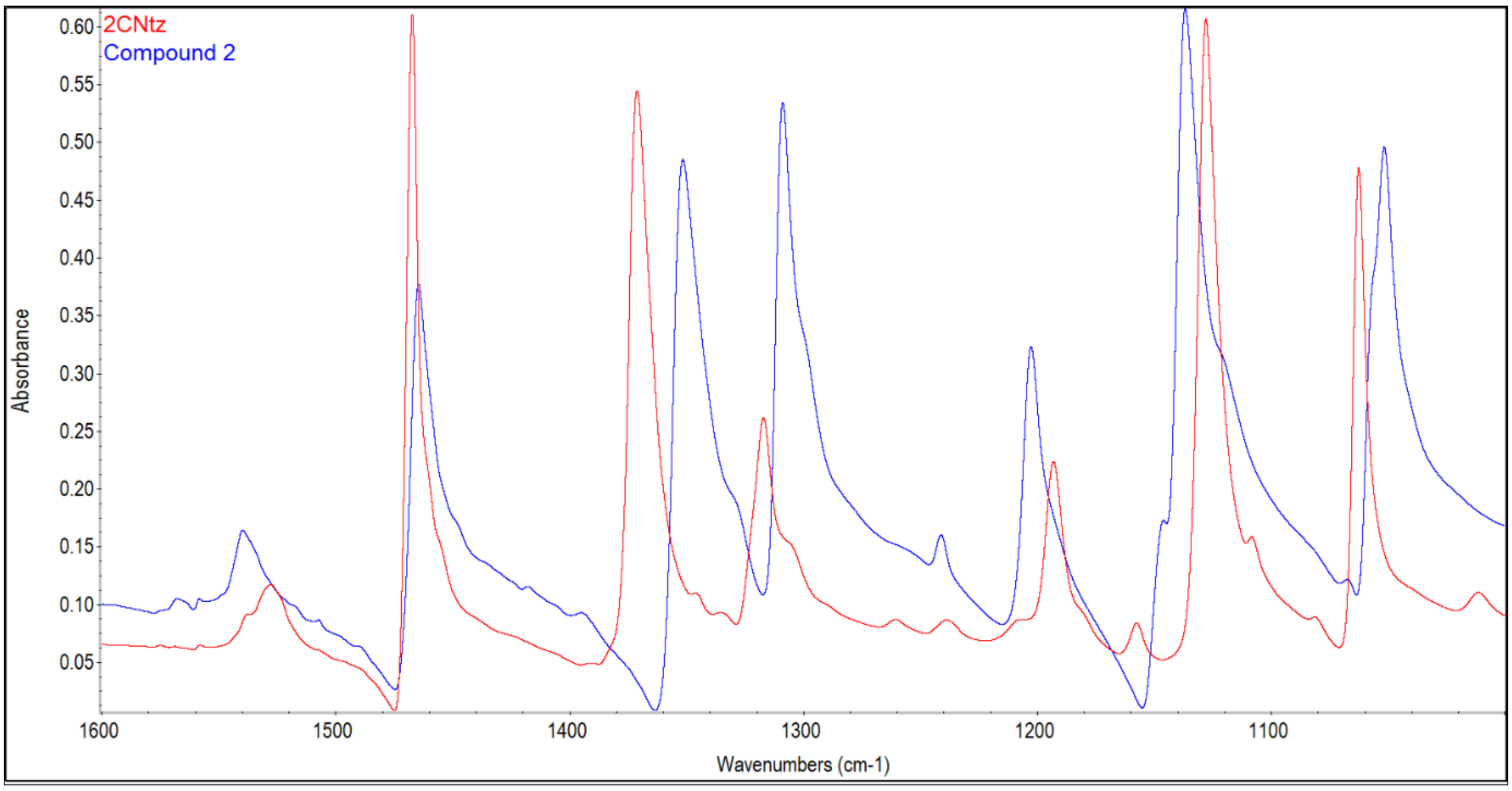

**Figure S13.** Superimposed FTIR spectra of compound **2** and **2CNtz** in the range of 1600-1000 $cm^{-1}$.

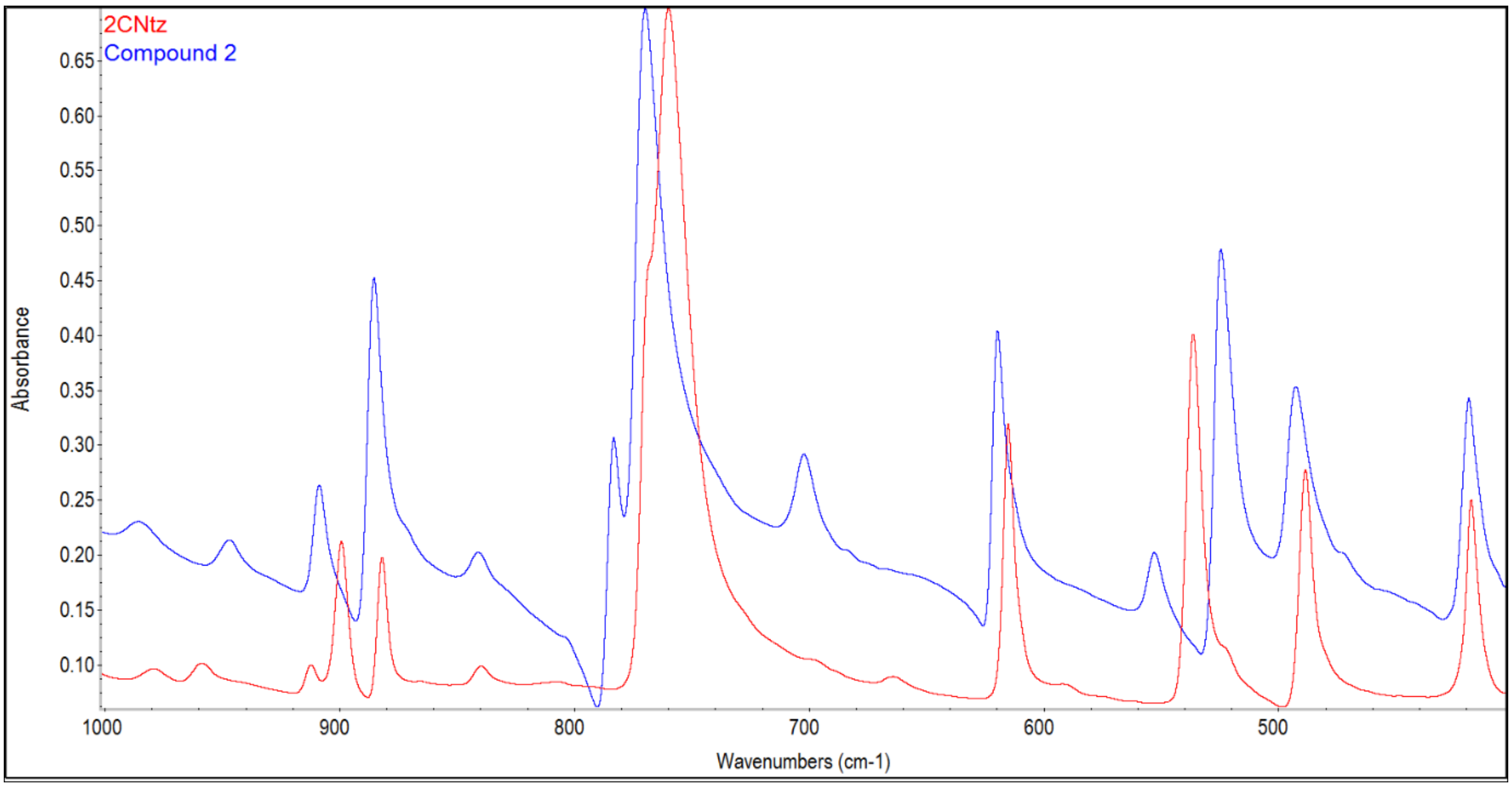

**Figure S14.** Superimposed FTIR spectra of compound **2** and **2CNtz** in the range of 1000-400 $cm^{-1}$.

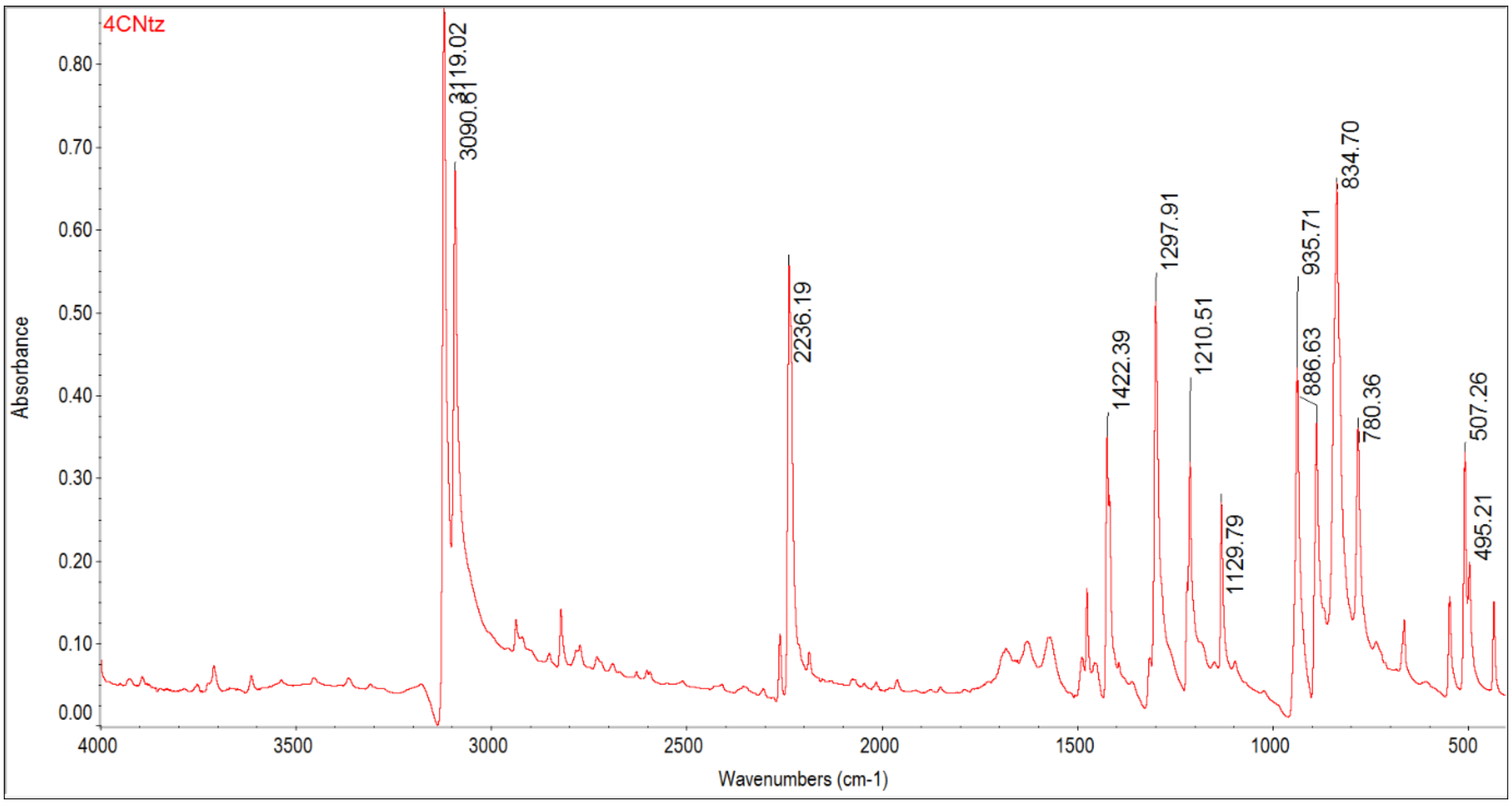

**Figure S15**. FTIR spectra of 4-cyanothiazole.

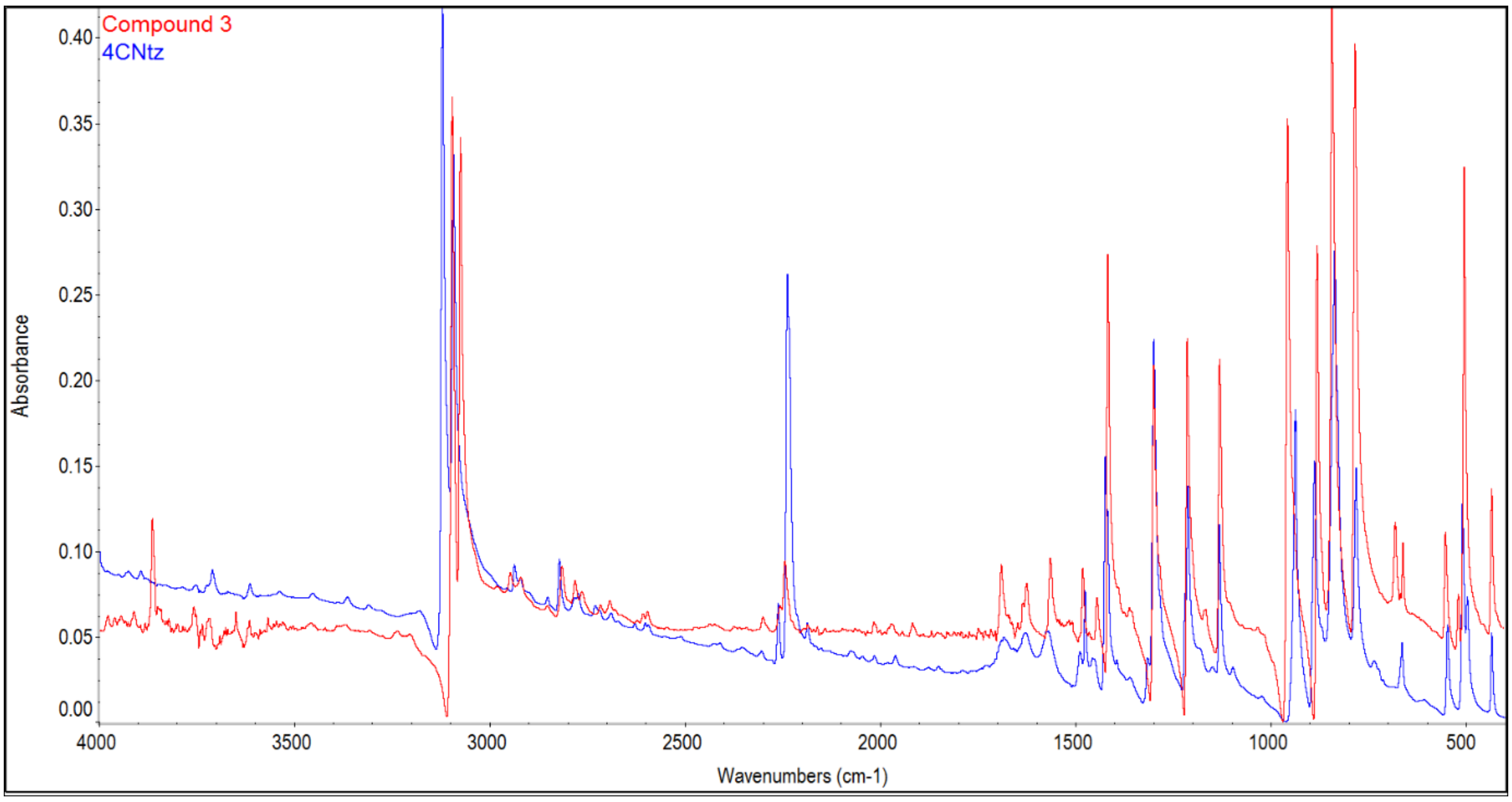

**Figure S16**. Superimposed FTIR spectra of compound **4** and **4CNtz**.

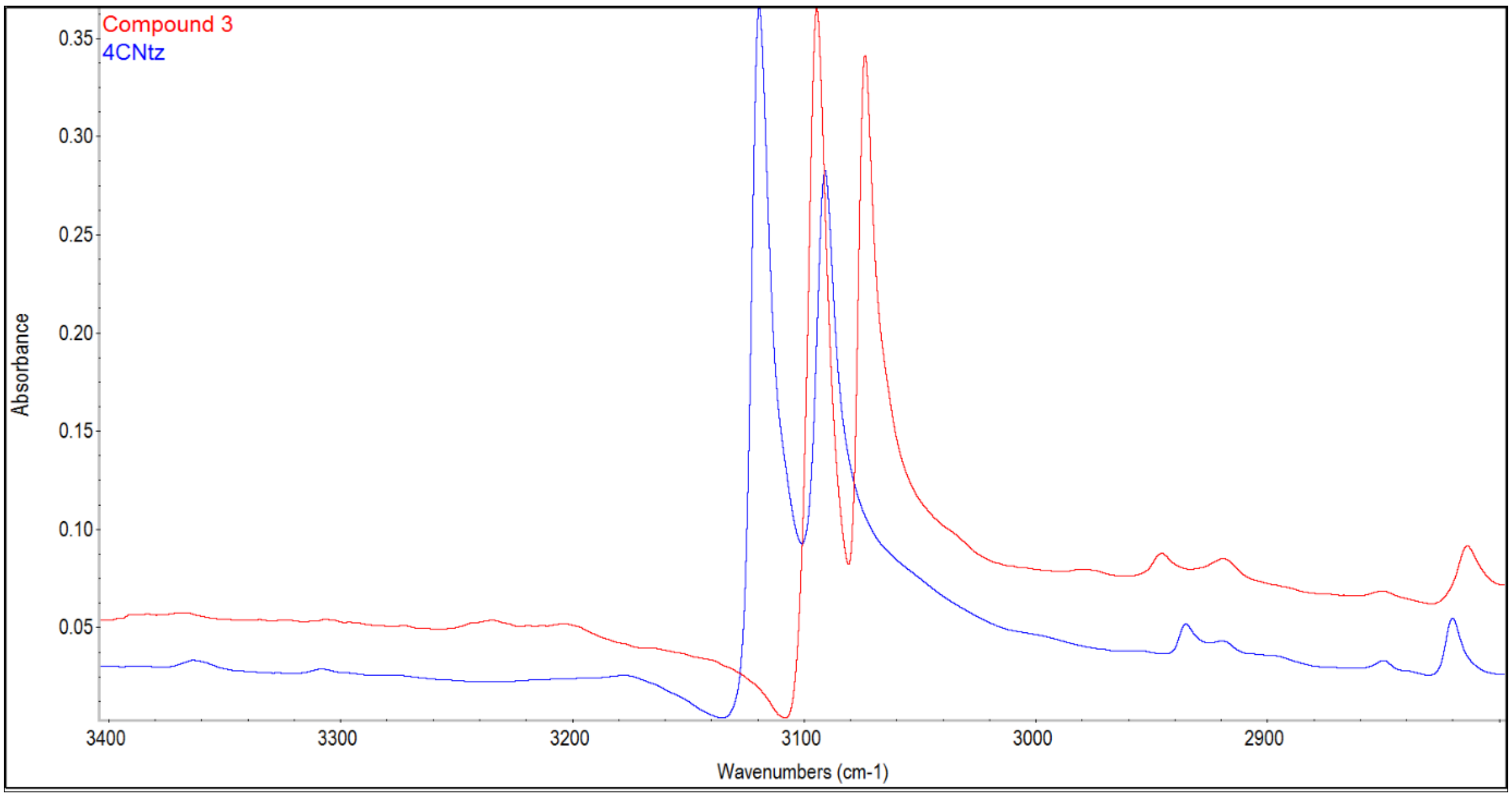

**Figure S17.** Superimposed FTIR spectra of compound **3** and **4CNtz** in the range of 3400-2800 $cm^{-1}$.

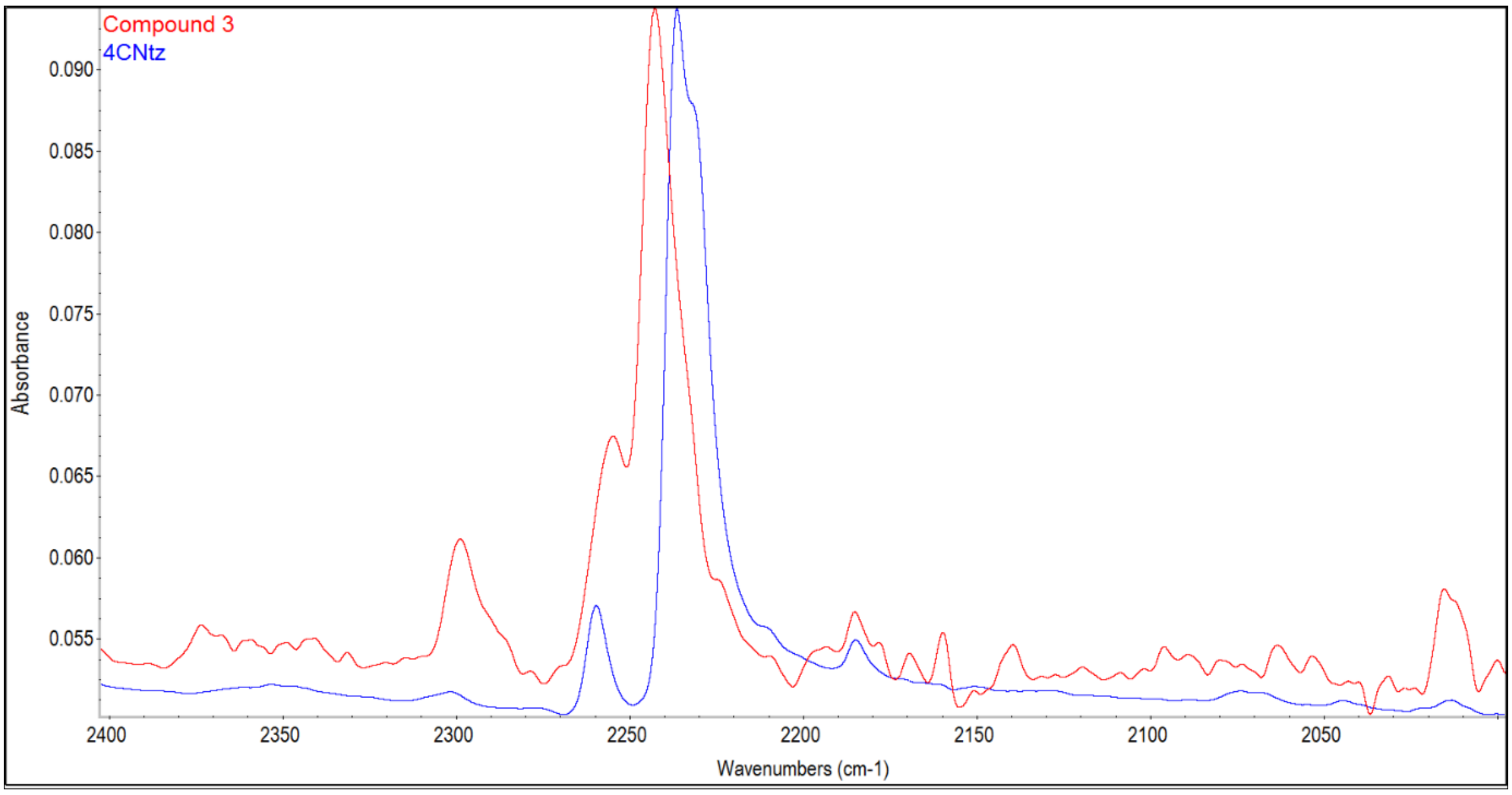

**Figure S18.** Superimposed FTIR spectra of compound **3** and **4CNtz** at the range of 2400-2000 $cm^{-1}$.

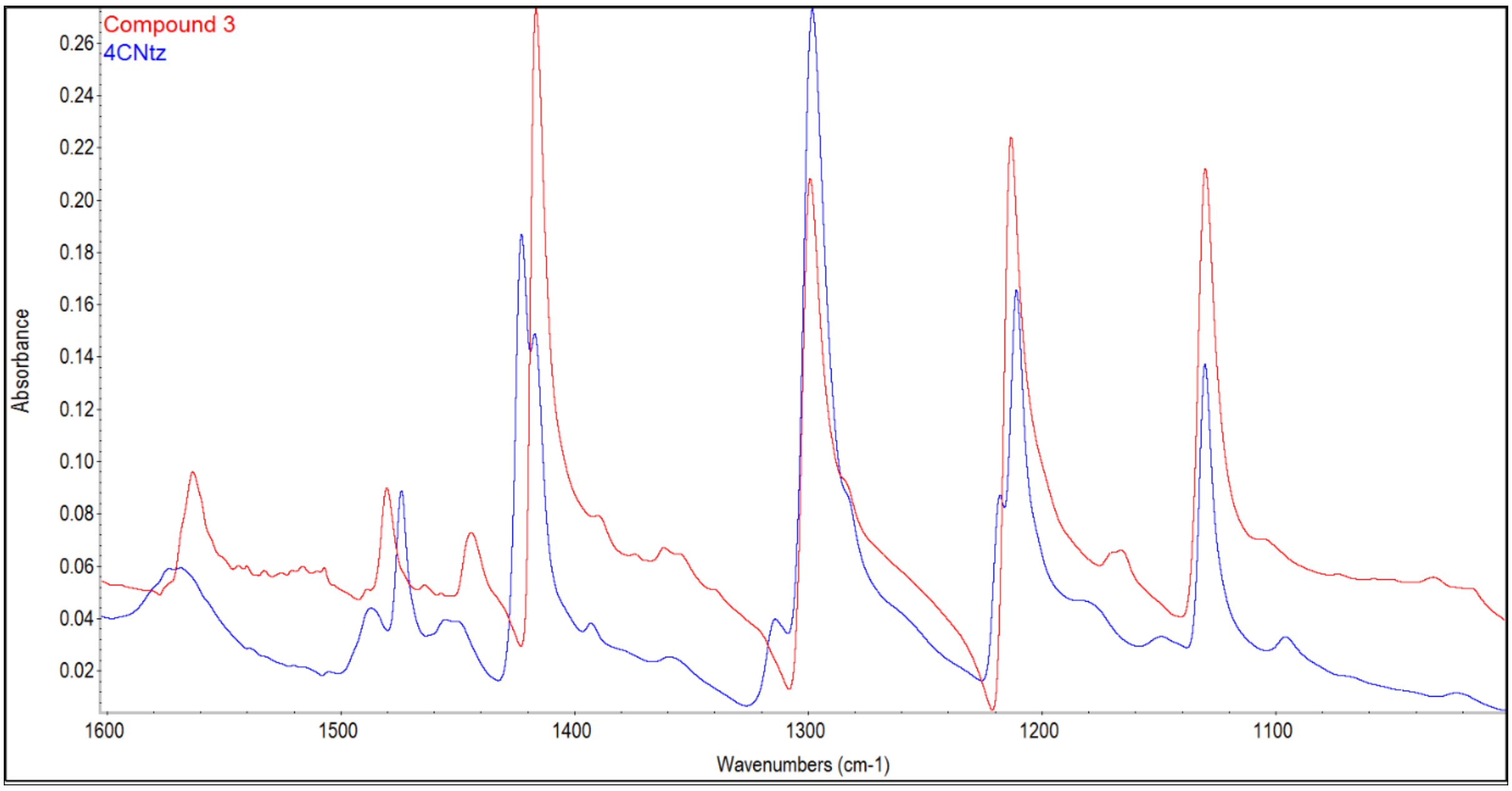

**Figure S19.** Superimposed FTIR spectra of compound **3** and **4CNtz** in the range of 1600-1000 $cm^{-1}$.

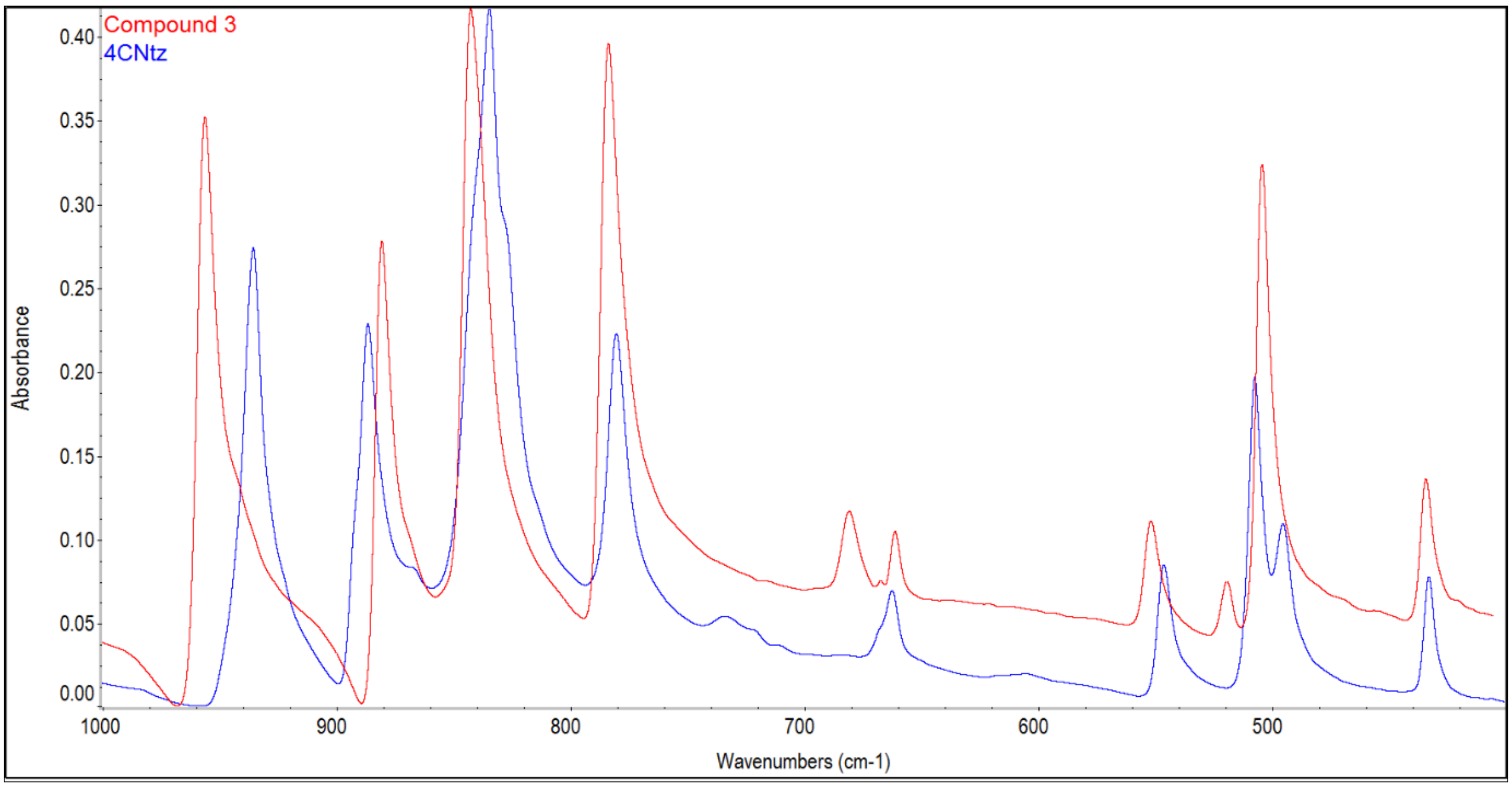

**Figure S20.** Superimposed FTIR spectra of compound **3** and **4CNtz** in the range of 1000-400 $cm^{-1}$.

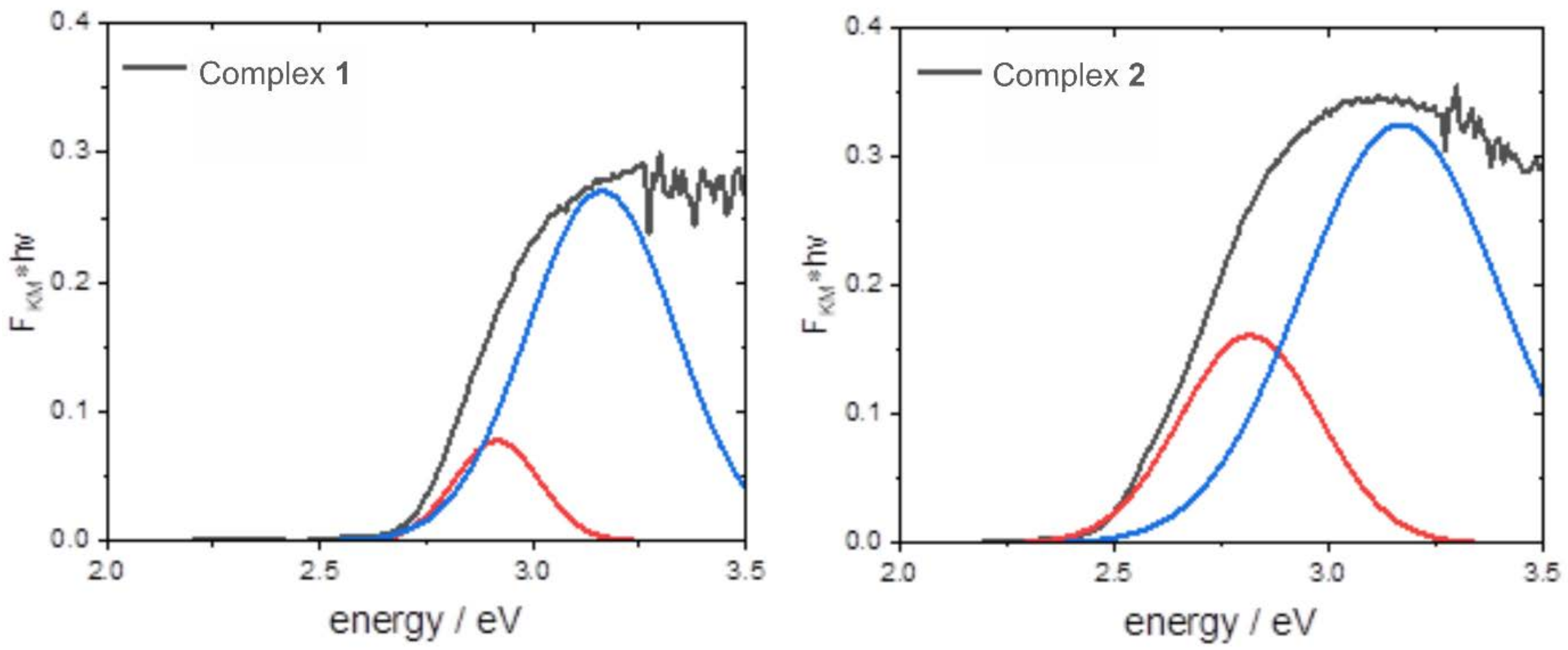

**Figure S21.** Deconvolution of absorption spectra of complexes **1** and **2** into Gaussian components.

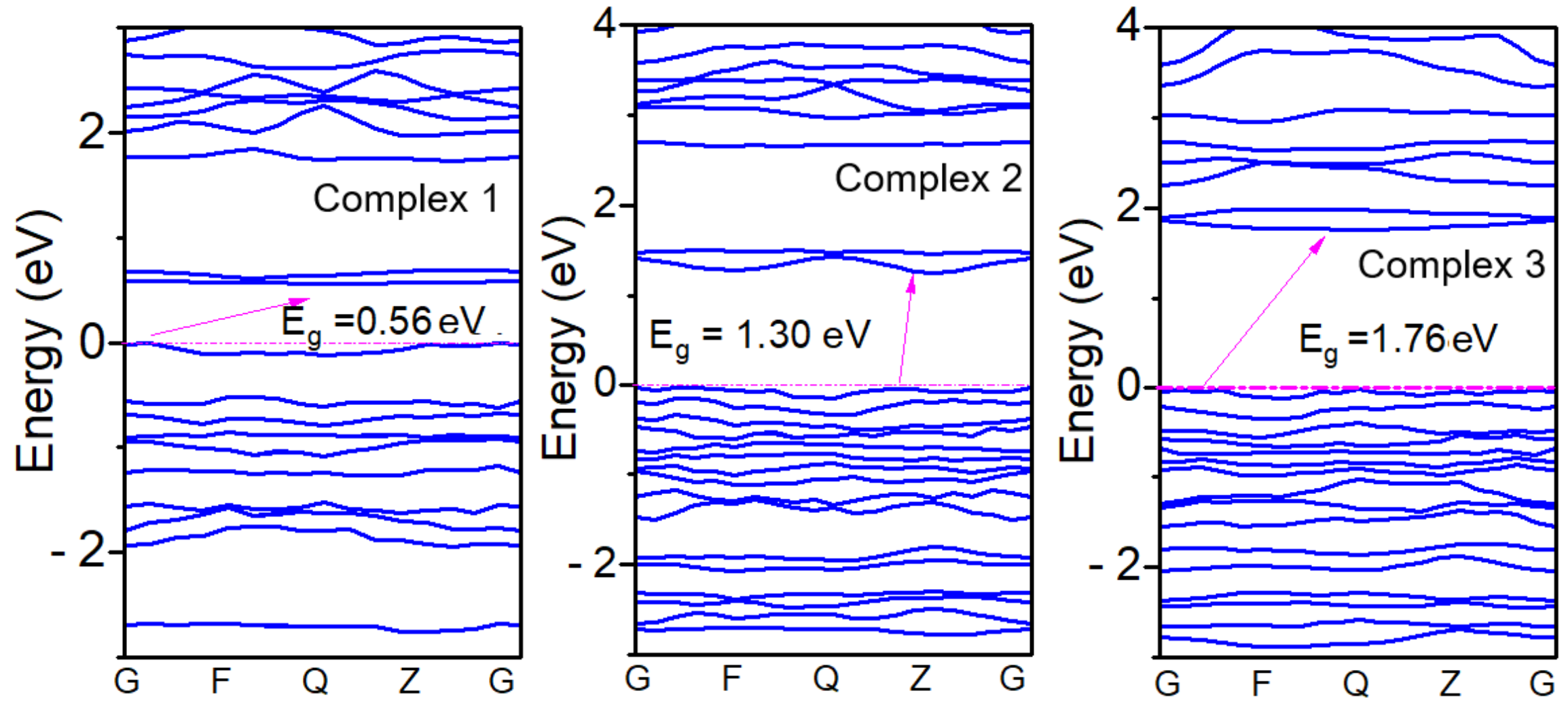

**Figure S22.** Band structure and density of states for complex **1** (**5CNtz**-CuI), complex **2** (**2CNtz**-CuI) and complex **3** (**4CNtz**CuI). (DFT-MBD+SOC)

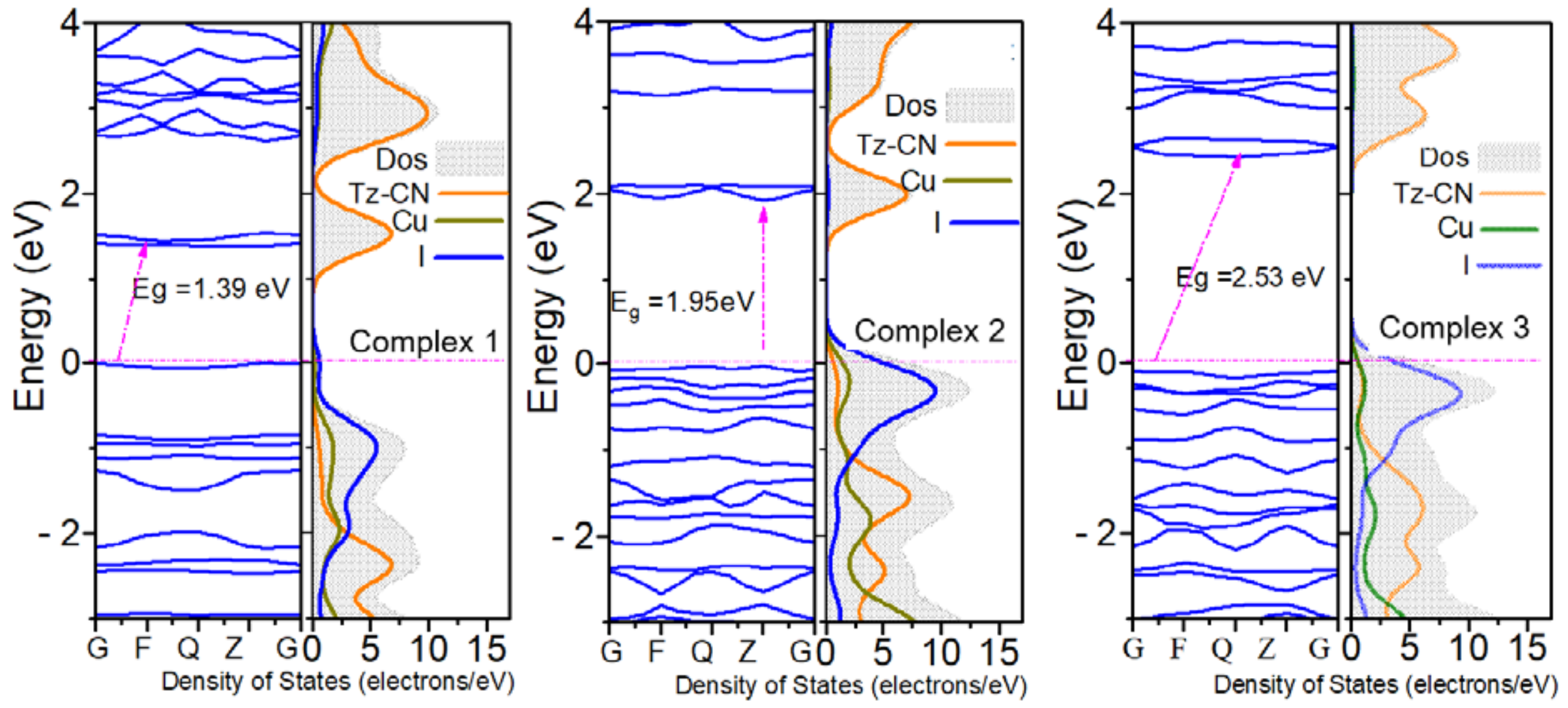

**Figure S23.** Band structure and density of states for complex **1** (**5CNtz**-CuI), complex **2** (**2CNtz**-CuI) and complex **3** (**4CNtz**-CuI). (DFT-MBD+U7.5)

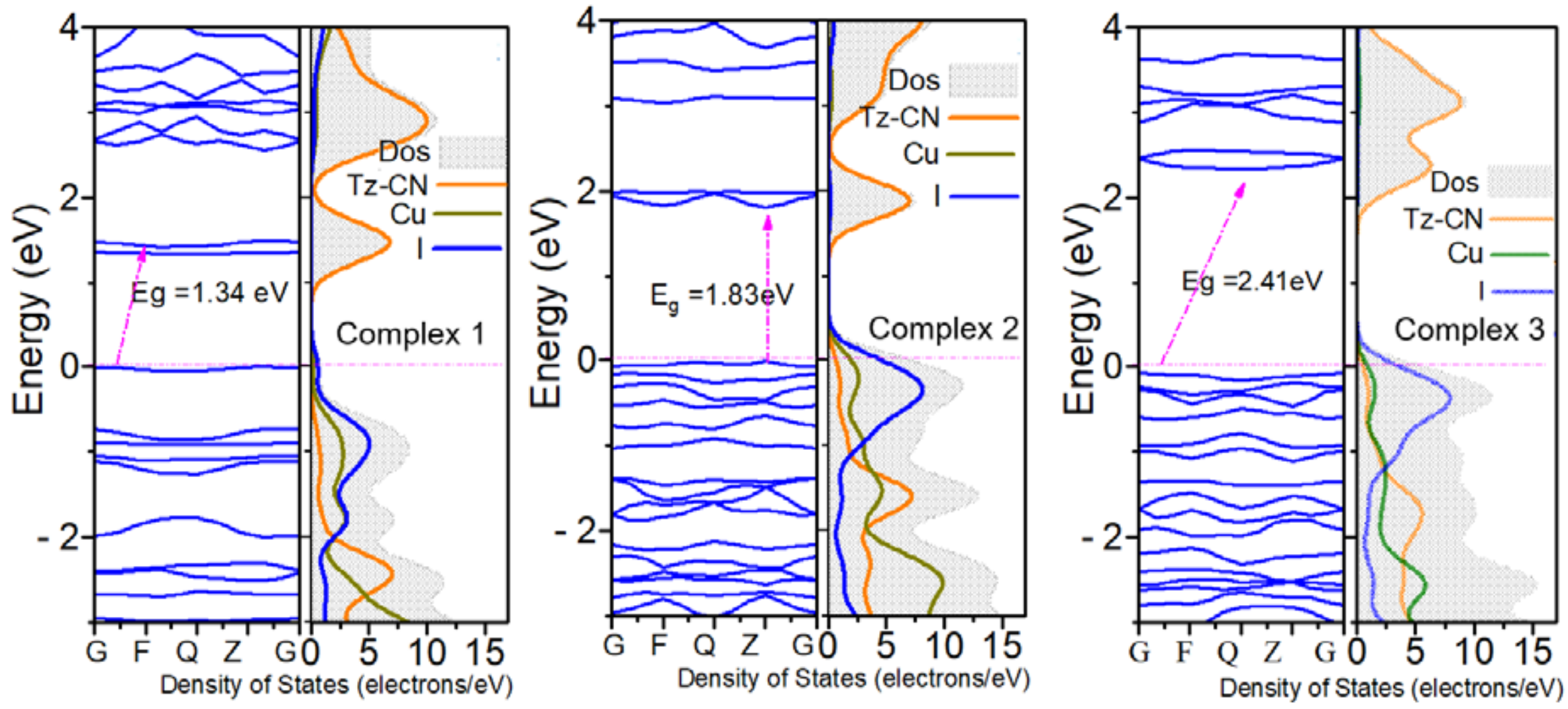

**Figure S24.** Band structure and density of states for complex **1** (**5CNtz**-CuI), complex **2** (**2CNtz**-CuI) and complex **3** (**4CNtz**-CuI). (DFT-MBD+U5.5)

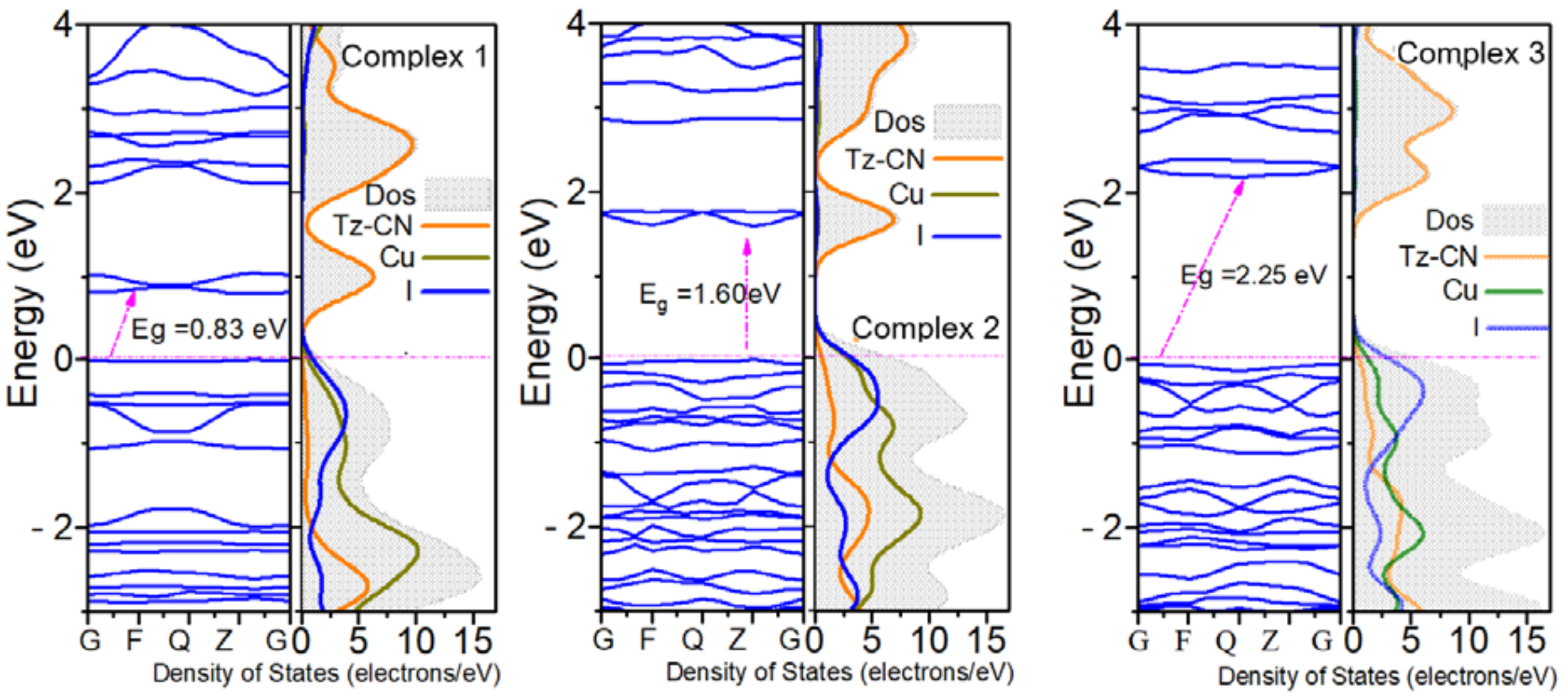

**Figure S25.** Band structure and density of states for complex **1** (**5CNtz**-CuI), complex **2** (**2CNtz**-CuI) and complex **3** (**4CNtz**-CuI). (DFT-MBD+U3.5)

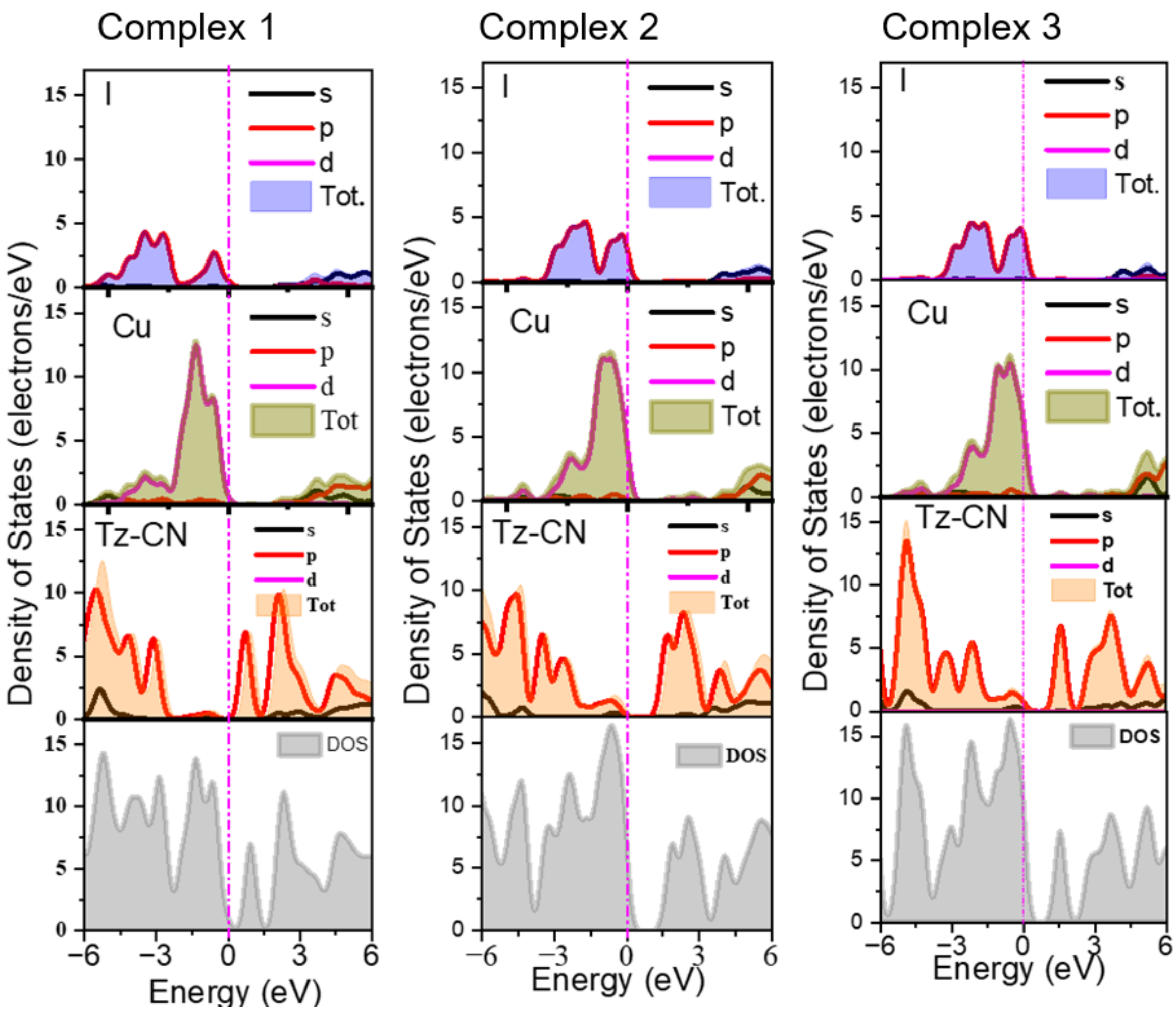

**Figure S26.** Partial density of states for complex **1** (**5CNtz**-CuI), complex **2** (**2CNtz**-CuI) and complex **3** (**4CNtz**-CuI) at DFT-MBD level of theory.

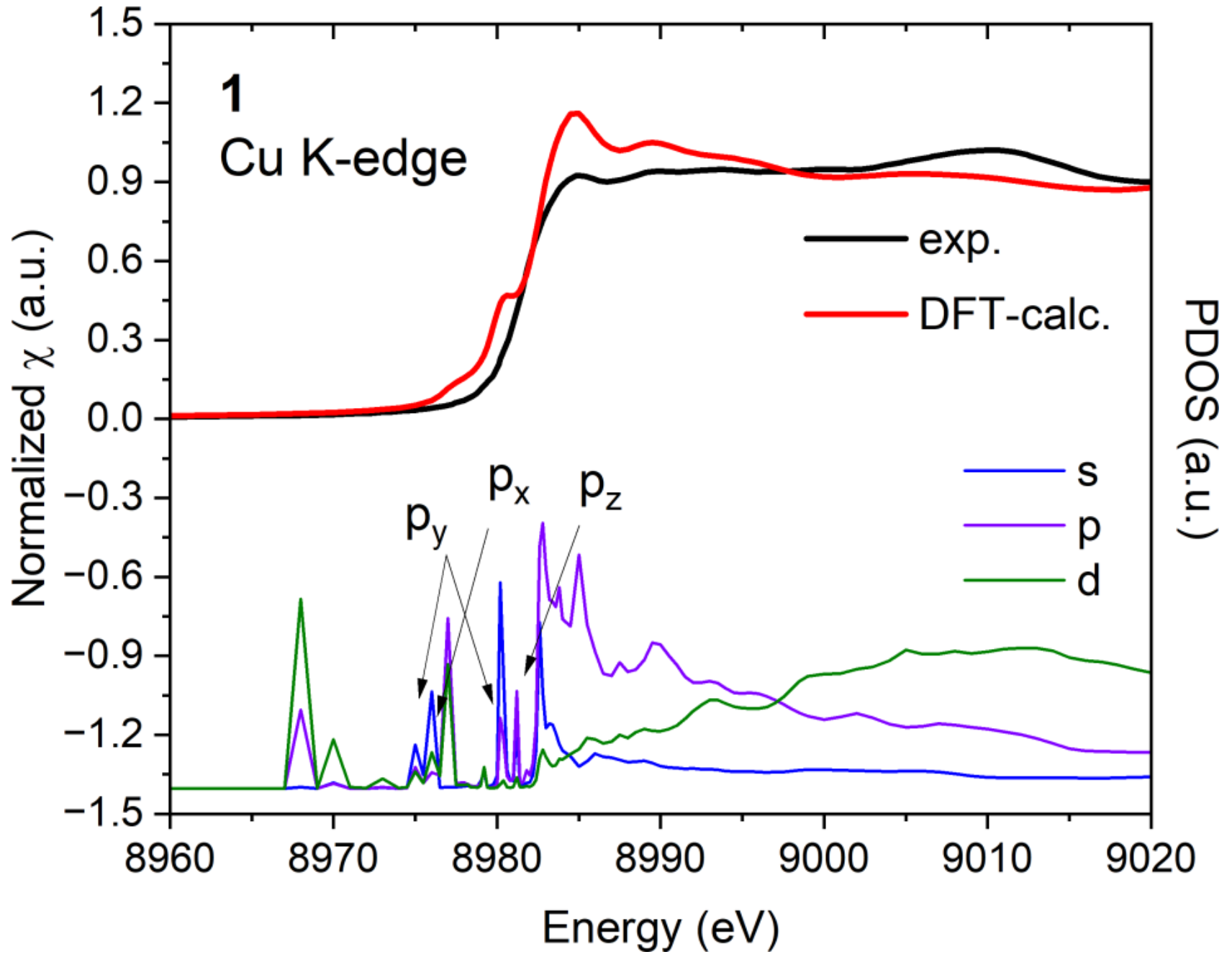

**Figure S27**. DFT-calculated XAS spectra and density for states compared with experimental data for complex **1**.

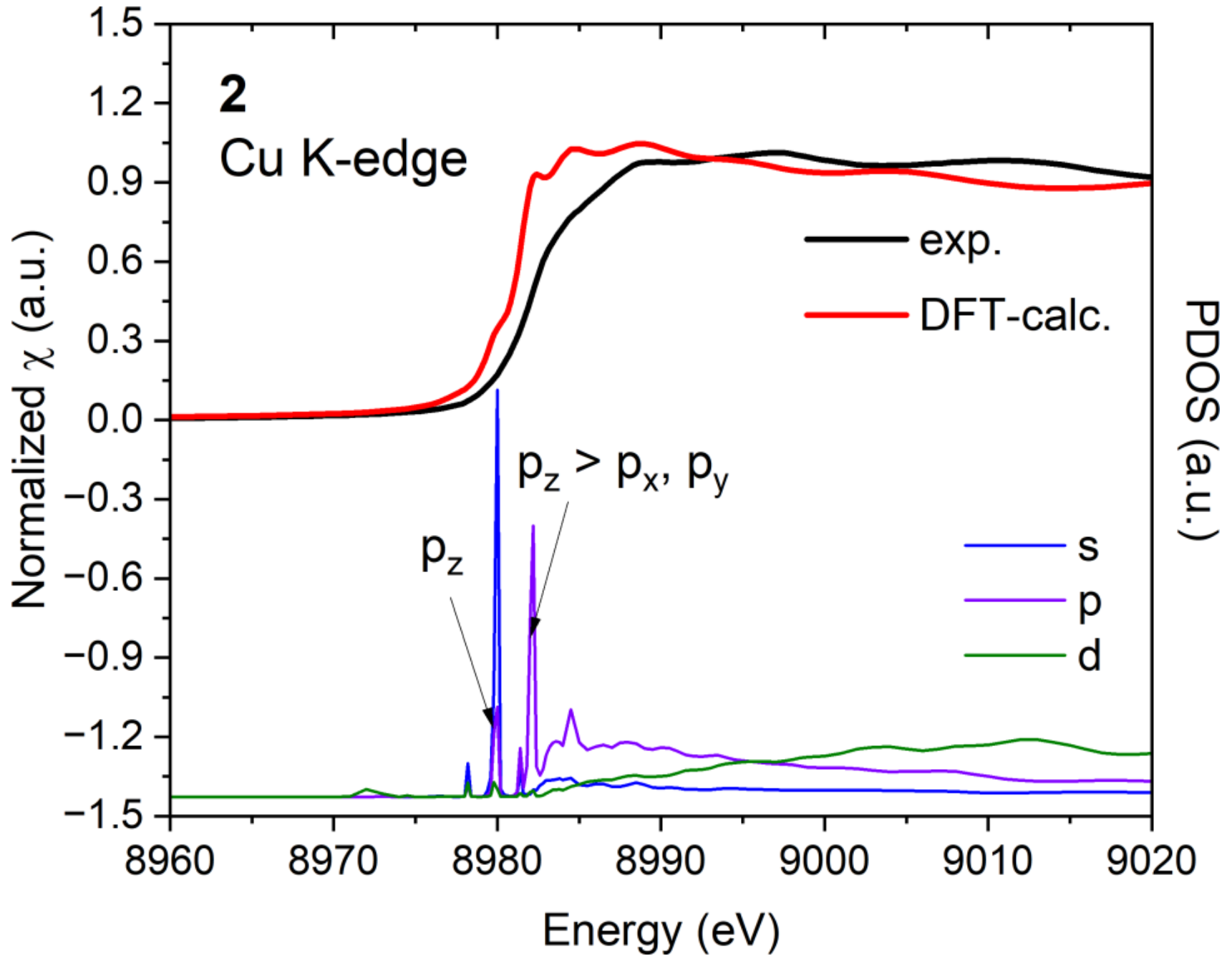

**Figure S28**. DFT-calculated XAS spectra and density for states compared with experimental data for the complex 2.

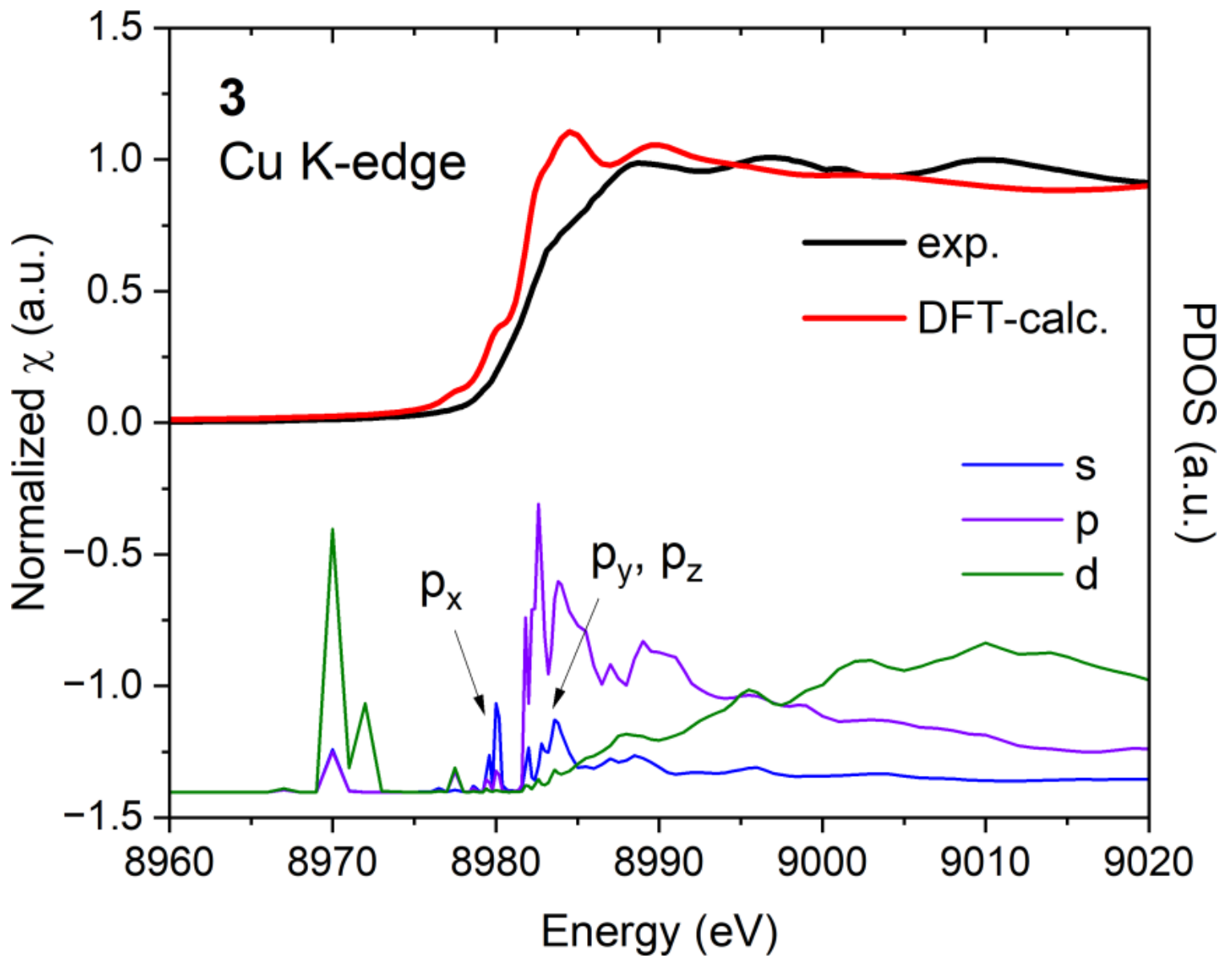

**Figure S29**. DFT-calculated XAS spectra and density for states compared with experimental data for complex **3**.

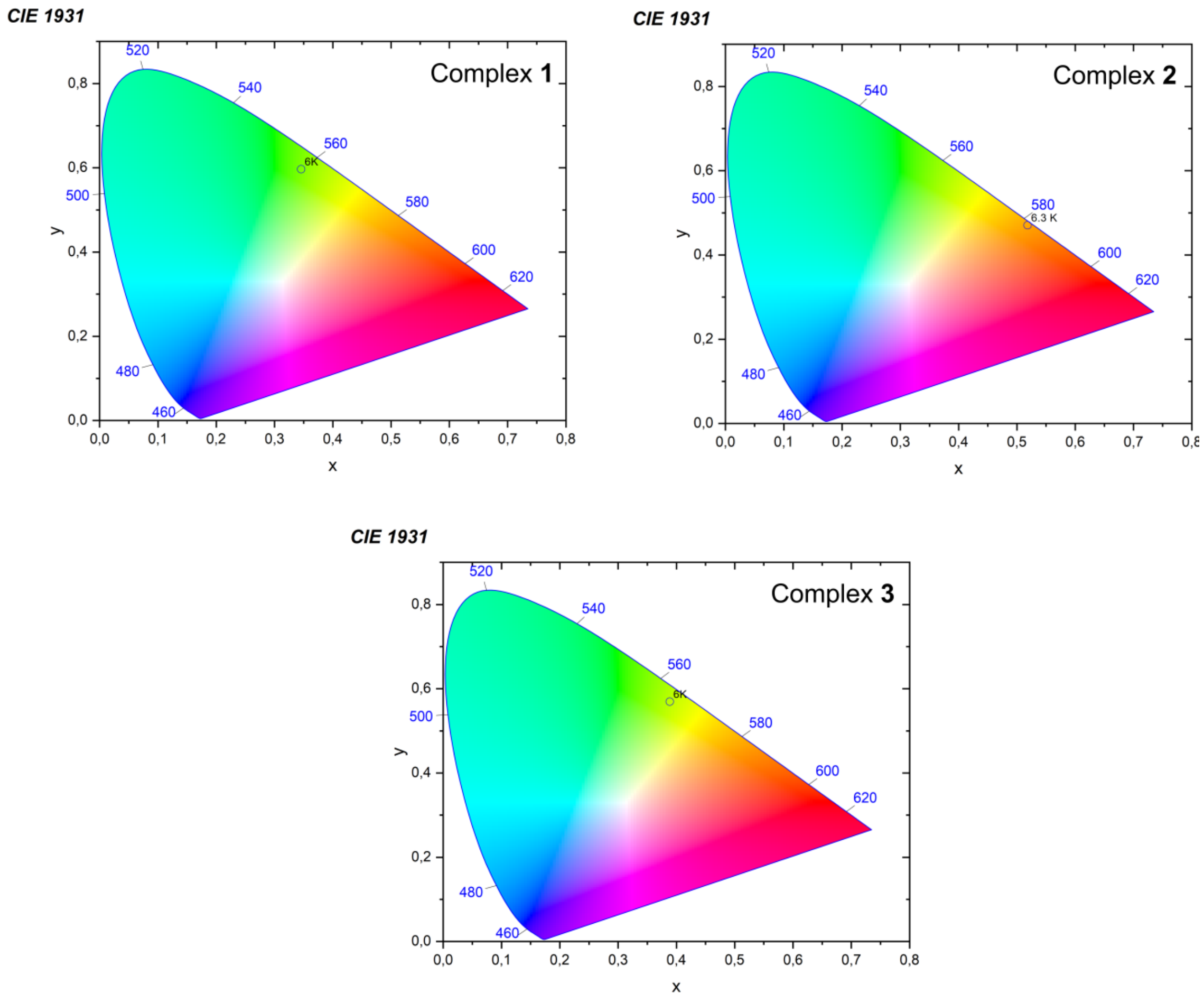

**Figure S30.** CIE1931 chromaticity diagrams calculated for emission spectra of **1** - **3** recorded at 6 K.